\documentclass[prx,twocolumn,superscriptaddress,longbibliography,aps]{revtex4-1}

\usepackage[T1]{fontenc}
\usepackage[utf8]{inputenc}
\usepackage{lmodern}
\usepackage[sumlimits]{amsmath}
\usepackage{amssymb}
\usepackage{bm}
\usepackage{yhmath}
\usepackage{graphicx}
\usepackage{epstopdf}
\usepackage{latexsym}
\usepackage{color}
\usepackage{nicefrac}
\usepackage{dsfont}
\usepackage{bbold}
\usepackage{wasysym} 
\usepackage{stmaryrd}
\usepackage{hyperref} 
\usepackage{xcolor}
\usepackage{multirow}
\usepackage{soul}
\usepackage{adjustbox}
\usepackage{cancel}

\usepackage{esint}
\usepackage{tikz}

\newcommand{\be}{\begin{equation}}
\newcommand{\ee}{\end{equation}}
\newcommand{\bea}{\begin{eqnarray}}
\newcommand{\eea}{\end{eqnarray}}

\newcommand{\idmatrix}{\text{\textbb{1}}}
\newcommand{\nn}{\nonumber}

\usepackage{verbatim}
\usepackage{dsfont} 
\usepackage{caption}
\usepackage{url}

\usepackage{subcaption}

\usepackage{sansmath}
\DeclareFontEncoding{LGR}{}{}
\DeclareSymbolFont{sfgreek}{LGR}{cmss}{m}{n}
\SetSymbolFont{sfgreek}{bold}{LGR}{cmss}{bx}{n}
\DeclareMathSymbol{\sxi}{\mathord}{sfgreek}{`x}
\DeclareMathSymbol{\stheta}{\mathord}{sfgreek}{`j}
\DeclareMathSymbol{\sepsilon}{\mathord}{sfgreek}{`e}
\DeclareMathSymbol{\sOmega}{\mathalpha}{sfgreek}{`W}
\DeclareMathSymbol{\stau}{\mathalpha}{sfgreek}{`t}

\newcommand{\mb}{\mathbf}
\newcommand{\bs}{\boldsymbol}

\newcommand{\mc}{\mathcal}
\newcommand{\ms}{\mathsf}

\renewcommand{\ul}{\underline}

\usepackage{scalerel}

\usepackage{accsupp}

\begin{document}

\title{Extrinsic contribution to bosonic thermal Hall transport}

\author{L\'eo Mangeolle}
\affiliation{Technical University of Munich, TUM School of Natural Sciences, Physics Department, 85748 Garching, Germany}
\affiliation{Munich Center for Quantum Science and Technology (MCQST), Schellingstr. 4, 80799 M{\"u}nchen, Germany}
\author{Johannes Knolle}
\affiliation{Technical University of Munich, TUM School of Natural Sciences, Physics Department, 85748 Garching, Germany}
\affiliation{Munich Center for Quantum Science and Technology (MCQST), Schellingstr. 4, 80799 M{\"u}nchen, Germany}
\affiliation{Blackett Laboratory, Imperial College London, London SW7 2AZ, United Kingdom}

\date{\today}
\begin{abstract}
  Bosonic excitations like phonons and magnons  dominate the low-temperature transport of magnetic insulators.
  Similar to electronic Hall responses, the  thermal Hall effect (THE) of charge neutral bosons has been proposed as a powerful tool for probing topological properties of their wavefunctions.
  For example, the {\it intrinsic} contribution of the THE of a perfectly clean system is directly governed by the distribution of Berry curvature, and many experiments on topological magnon and phonon insulators have been interpreted in this way.
  However, disorder is inevitably present in any material and its contribution to the THE has remained poorly understood. Here we develop a rigorous kinetic theory of the {\it extrinsic} side-jump contribution to the THE of bosons.
  We show that the extrinsic THE is always relevant for bosonic systems and can be of the same order as the intrinsic one but sensitively depends on the type of local imperfection.
  We study different types of impurities and show that a THE can even arise as a pure impurity-induced effect in a system with a vanishing intrinsic contribution.
  As a side product, we also generalize existing results for the electronic AHE to general types of impurities beyond the standard assumption of local potential scattering.
  We discuss the importance of our results for the correct interpretation of THE measurements, for example in the Kitaev magnet $\alpha$-RuCl$_3$,
  and provide a ready-to-use formula for comparison to experimental data.  
\end{abstract}

\maketitle

\section{Introduction}
\label{sec:introduction-1}
For charged particles, such as electrons, the origin of classical Hall transport from the Lorentz force has been known since the end of the 19th century.
After the discovery of the quantum Hall effect (QHE) a century later~\cite{klitzing1980new}, its origin from the intrinsic Berry curvature of the electron's wavefunctions was swiftly established~\cite{thouless1982quantized}.
Understanding the nonquantized anomalous Hall effect (AHE) in metallic materials with broken time reversal (TR) symmetry has proven much more complicated,
because it is both difficult experimentally and theoretically cumbersome to disentangle contributions from {\it intrinsic} Berry phase effects and {\it extrinsic} scattering contributions~\cite{nagaosa_anomalous_2010}.
Charge neutral excitations in solids, for example phonons, magnons or exotic fractionalized spinons, can also exhibit Hall transport but now in response to gradients in temperature (or energy density~\cite{luttinger1964theory}).
A quantized thermal Hall effect (THE) has been put forward as a powerful probe of exotic fractionalized excitations from intrinsic topological order of Mott insulators~\cite{kitaev2006anyons,kane1997quantized},
signatures of which have been arguably observed experimentally~\cite{kasahara2018majorana,paul2024topological}.
Unquantized THE in insulators can also originate from more mundane phonon and magnon excitations, as recently reported in many experiments on magnets~
\cite{onose2010observation,ideue2012effect,hirschberger2015large,hirschberger2015thermal,watanabe2016emergence,doki2018spin,
  hirokane2019,akazawa2020thermal,zhang2021anomalous,chen2022large,lefranccois2022evidence},
and particularly in the Kitaev material and spin liquid candidate $\alpha\text{-RuCl}_3$~
\cite{hentrich2019large,PhysRevB.102.220404, 10.1126/science.aay5551,
  bruin2022robustness,PhysRevB.106.L060410,PhysRevMaterials.7.114403,PhysRevMaterials.8.014402,czajka2023planar}  --- for a recent review see \cite{zhang2024thermal}.
While the theory of the {\it intrinsic} (Berry curvature) contribution to this bosonic THE has been firmly established over the last decade \cite{matsumoto_rotational_2011, qin_energy_2011, qin_berry_2012, matsumoto_thermal_2014, shitade2014heat, zhang2020thermodynamics, mangeolle_quantum_2024},
much less is known about the {\it extrinsic} contributions. Concretely, scattering contributions from disorder inevitably present in any material are still poorly understood
and are the subject of the present paper. 

The established way of calculating the THE in insulating magnets is to derive an effective quadratic magnon Hamiltonian (various ways of including effects of interactions, or not, have been explored)
and to derive from it the magnon band structure and especially the Berry curvature.
The same procedure can also be used for other types of neutral bosonic excitations, e.g.\ phonons.
In both cases, knowing the Berry curvature one may compute the thermal Hall conductivity,
 \begin{align}
  \label{eq:31}
    \kappa_{xy} ^{\rm clean} &= - \frac {1} T {\rm Tr}_+ \int_p  \int \text d \varepsilon \, \Theta (\varepsilon-  \ul{{\sf K}}_{\rm d}(p) ) \\
    &\qquad \qquad \qquad \times \varepsilon^2
                           \partial_\varepsilon n_{\rm B} \left ( \varepsilon, T\right ) \, \ms \Omega_{p_x p_y}(p)  \nonumber ,
\end{align}
where $\Theta$ is the Heaviside step function,  ${\rm Tr}_+ $ is a sum over boson bands,
$n_{\rm B}(\varepsilon,T)=1/(e^{\varepsilon/T}-1)$ is the Bose function, $\ul{{\sf K}}_{\rm d}(p)$ is the bosonic energy, and crucially $\ms \Omega_{p_x p_y}(p)$ is the bosons' Berry curvature.
This formula Eq.\eqref{eq:31} is similar to the Karplus-Luttinger formula for the intrinsic AHE of electrons.
However, in both the (bosonic) thermal and (fermionic) electrical cases, the intrinsic contribution is but part of the total Hall conductivity.
Other phenomena contribute, that are usually classified as skew-scattering \cite{PhysRevLett.113.265901, PhysRevB.105.L220301, PhysRevB.106.144111, PhysRevX.12.041031, PhysRevB.106.245139, PhysRevB.111.134405}
and side-jump mechanisms: for a review in the electronic case see \cite{nagaosa_anomalous_2010}. Here we investigate the latter.

\begin{figure*}[!t]
\centering
\includegraphics[width=0.31\textwidth]{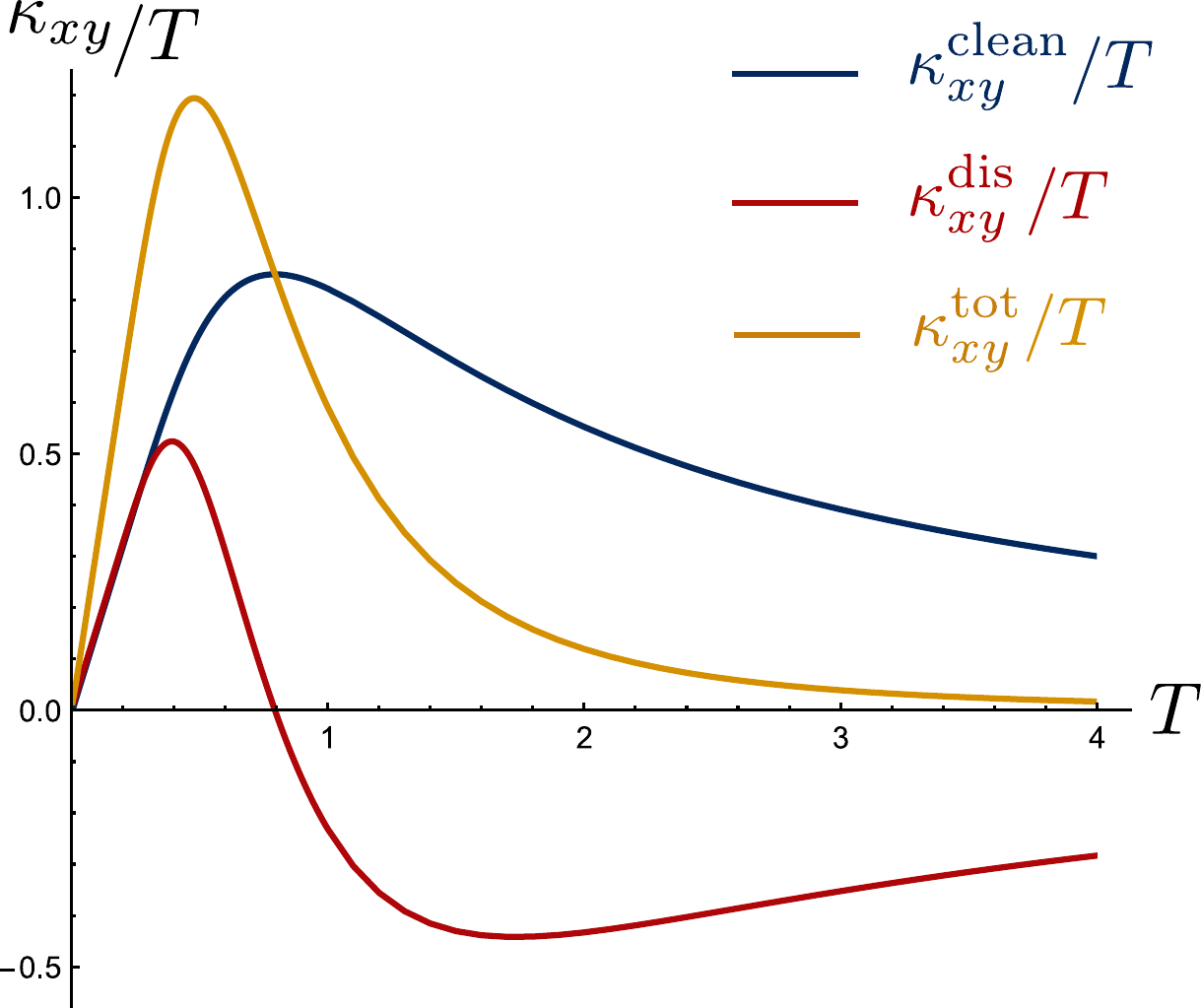} \hfill
\includegraphics[width=0.3\textwidth]{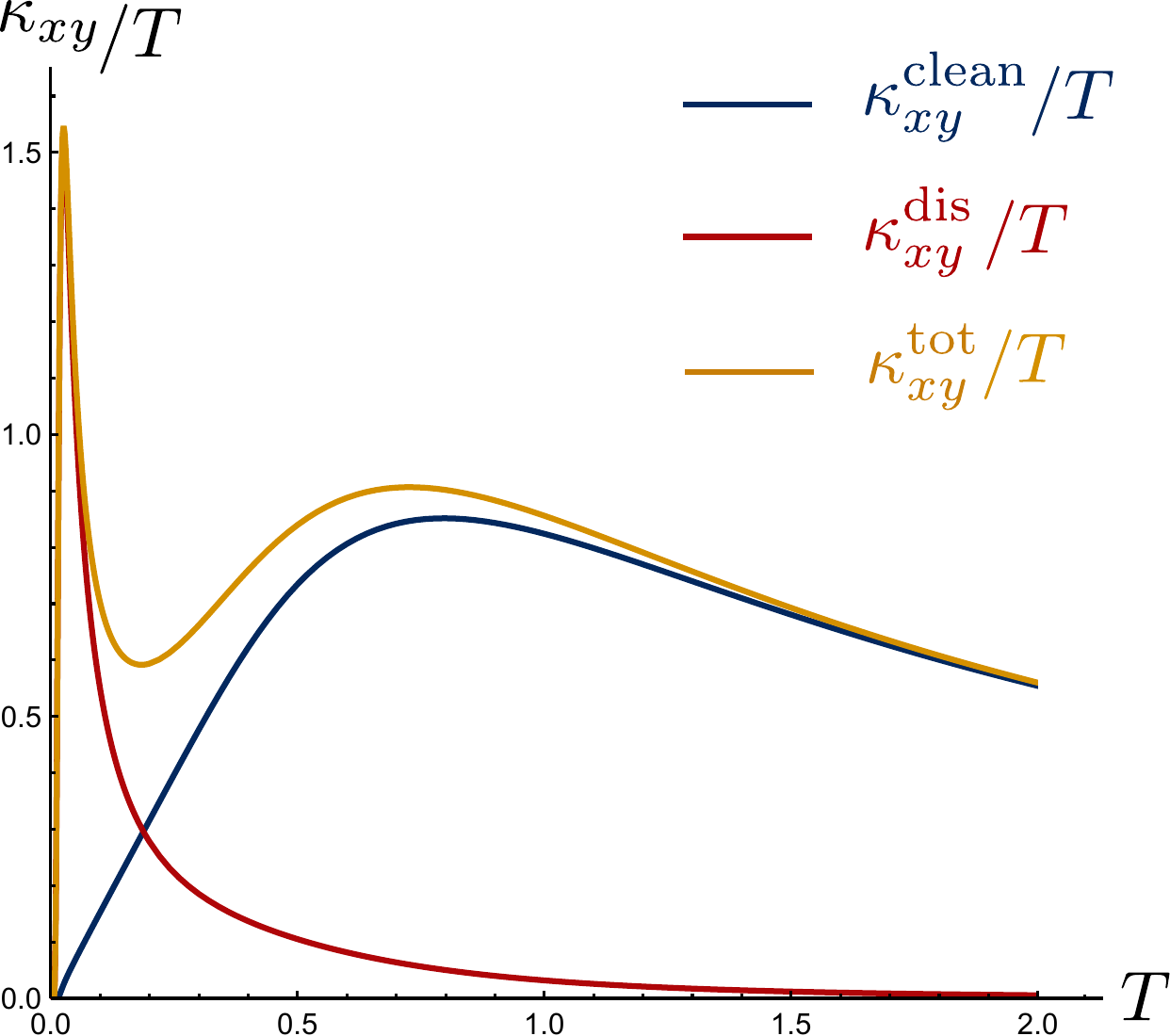}\hfill
\includegraphics[width=0.29\textwidth]{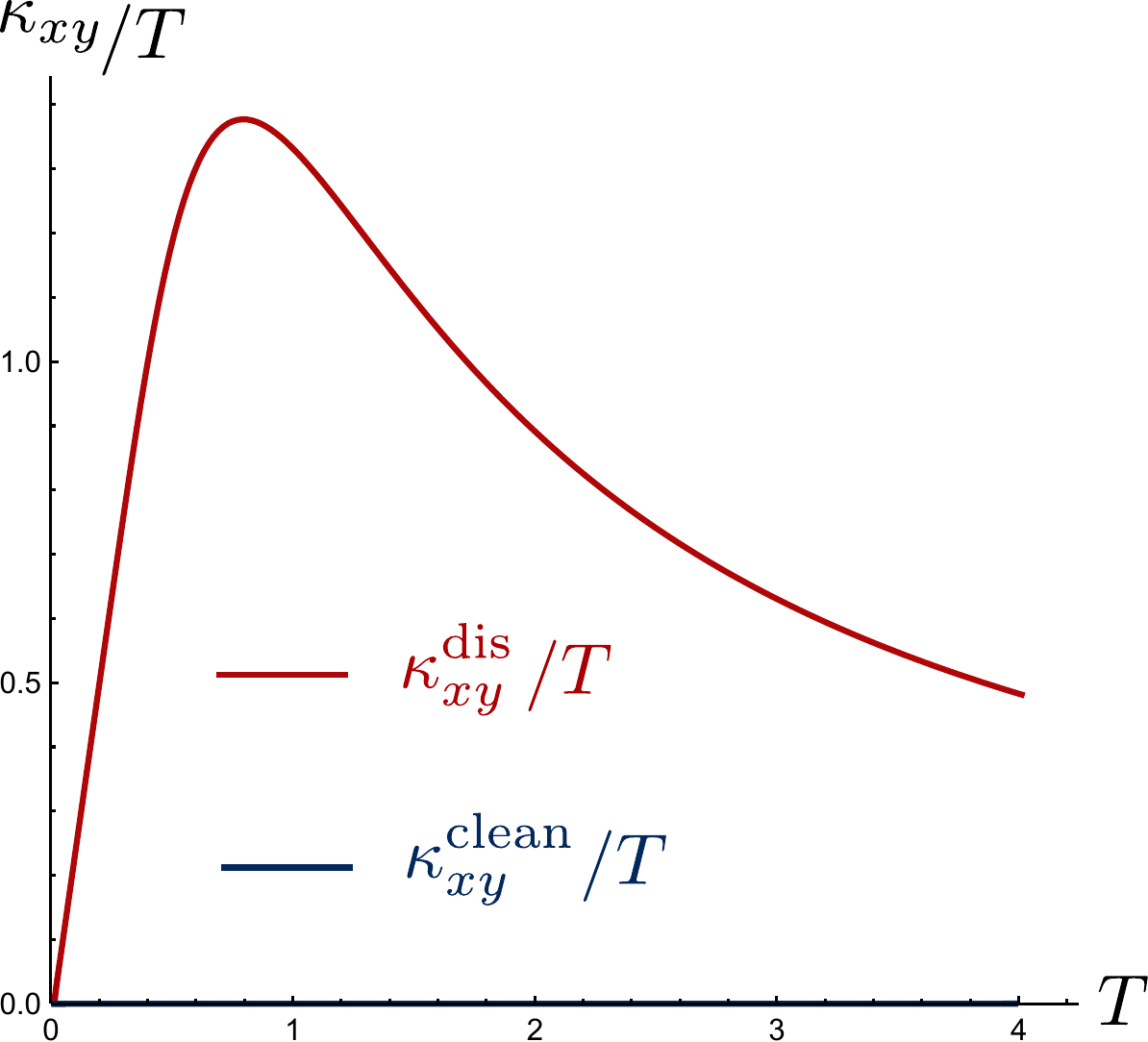}
\caption{Thermal Hall conductivity divided by temperature $\kappa_{xy}/T$ as a function of temperature $T$ for three different instances (a,b,c) of disorder and time reversal (TR) breaking parameters,
in a minimal model of bosons with viscosity (TR-breaking) terms introduced in Sec.\ \ref{sec:applications}. $\kappa_{xy}^{\rm clean}$ is the intrinsic contribution in the absence of disorder (Eq.\eqref{eq:31}), $\kappa_{xy}^{\rm dis}$ is the contribution due to disorder (Eq.\eqref{eq:32}), and $\kappa_{xy}^{\rm tot}=\kappa_{xy}^{\rm clean}+\kappa_{xy}^{\rm dis}$. Both axes are in arbitrary units, identical in all three plots. 
(a) Disorder is a local modulation of the clean Hamiltonian, with identical matrix structure. (b) Disorder is a local viscosity term, distinct from that already present in the clean theory. (c) The clean theory has zero viscosity and preserves TR, but impurities still introduce TR breaking locally: remarkably, in this case thermal Hall conductivity is a pure impurity effect. (a,b,c) The temperature dependences and distinctive features of these curves are discussed in Sec.\ref{sec:discussion}.}
\label{fig:Applications}
\end{figure*}

The general expectation is that the electronic side-jump effect is \emph{non-perturbative}, in that it does not depend on the strength of disorder (or impurity density), when inter-particle interactions can be neglected.
Yet, for electronic systems, the side-jump effect is actually not so commonly seen experimentally
\footnote{See \cite{nagaosa_anomalous_2010} and especially, in Sec. I.B.3: ``\textit{A practical approach, which is followed at present for materials in which $\sigma^{\rm AH}$ seems to be independent of $\sigma_{xx}$,
  is to first calculate the intrinsic contribution to the AHE. If this explains the observation (and it appears that it usually does), then it is deemed that the intrinsic mechanism dominates.
  If not, we can take some comfort from understanding on the basis of simple model results that there can be other contributions to $\sigma^{\rm AH}$ which are also independent of $\sigma_{xx}$
  and can for the most part be identified with the side-jump mechanism.}''},
because electron-electron interactions are usually strong especially in two dimensions~\cite{abrahams1981quasiparticle} and dominate over other scattering effects due e.g.\ to disorder.
In the presence of strong disorder, the electron liquid may also undergo localization in the bulk \cite{abrahams1979scaling},
  while mobile edge excitations account for a quantized Hall conductivity \cite{prange1987quantum}: the side-jump contribution, as a bulk effect, is again negligible in this situation.
  Meanwhile, and at odds with the electronic case, the side-jump contribution is expected to play a significant role in the thermal Hall effect of phonons or magnons.
  Indeed, in the absence of Pauli exclusion, and even in the presence of a (mobility) gap and of edge states between two bulk bands,
 bosonic states from the lower bulk band are always thermally excited, so that they \emph{always} contribute to thermal transport. Thus, side-jump effects are always present.
 Moreover, the latter are presumably much less dominated by interaction effects than in electron liquids,
insofar as interactions between collective bosonic modes like phonons or magnons are usually small.
For instance, they are due to lattice anharmonicity for phonons, or to higher-order terms in the $1/S$ expansion for magnons,
and they are usually local (as opposed to long-range like the Coulomb interaction between electrons).
\footnote{In magnetic systems, long-range interactions also exist such as dipolar exchange,
  though instances where these dominate over other (local) magnetic interactions governing the physical properties are rare, especially for bulk 3D insulating magnets.}
 Possible experimental evidence for significant side-jump effects may be found in the kagome antiferromagnet Volborthite \cite{watanabe2016emergence},
  in pyrochlore terbium oxides \cite{hirschberger2015large,hirokane2019}, and in the Kitaev compound $\alpha \text{-RuCl}_3$ \cite{PhysRevMaterials.7.114403} 
  -- see Sec. \ref{sec:experimental-aspects}.

Motivated by many reports of THE in insulating materials \cite{onose2010observation,ideue2012effect,hirschberger2015large,hirschberger2015thermal,watanabe2016emergence,doki2018spin,
  hirokane2019,akazawa2020thermal,zhang2021anomalous,chen2022large,lefranccois2022evidence,hentrich2019large,PhysRevB.102.220404, 10.1126/science.aay5551,
  bruin2022robustness,PhysRevB.106.L060410,PhysRevMaterials.7.114403,PhysRevMaterials.8.014402,czajka2023planar,zhang2024thermal}, we derive rigorously the side-jump effect in the thermal Hall conductivity of neutral bosons.
Employing a kinetic equation formalism (see below) for systems of non-interacting bosons, we include the presence of generic disorder fields,
i.e. not only \emph{potential} but a fully-fledged local deformation of the Hamiltonian as the most genereal form of disorder. 
Within the semiclassical approximations which we discuss later, we find that disorder scattering induces new contributions in the boson thermal Hall conductivity which are carried by \emph{bulk} currents.
As our main result we derive the corresponding disorder-induced thermal Hall conductivity reading
 \begin{align}
   \label{eq:32}
   \kappa_{xy}^{\rm dis}
  &= - \frac 1 {T} {\rm Tr}_+ \int_p   \int \text d \varepsilon \, \delta (\varepsilon-  \ul {\sf K}_{\rm d}(p) ) \,\varepsilon^2
    \partial_\varepsilon n_{\rm B} \left ( \varepsilon, T\right )  \\
   & \qquad \times p_\lambda  \left [ {\ms \Omega}^{\ms W}_{p_y p_\lambda}(p)\, \partial_{p_x}\ul {\ms K}_{\rm d}^{}(p)
                          - {\ms \Omega}^{\ms W}_{p_x p_\lambda}(p)\,  \partial_{p_y}\ul {\ms K}_{\rm d}^{}(p) \right ]\nonumber   ,
 \end{align}
 where ${\ms \Omega}^{\ms W}_{p_\mu p_\mu}(p)$ is a band-dependent and disorder-dependent quantity,
 generally distinct from the Berry curvature and which reduces to the latter in specific cases.
 Its expression is provided in Eq.\eqref{eq:24}, and a concrete numerical recipe how to compute it can be found in Appendix \ref{sec:expl-form-analyt}.
 These show that ${\ms \Omega}^{\ms W}_{p_\mu p_\mu}$ depends on the detailed matrix structure of impurities:
 we discuss how the latter may be obtained for any given physical model in Sec.\ref{sec:disord-matr-struct}.
 Interestingly, our derivation translates with very little change to the case of electrical Hall transport by electrons.
This provides a generalized formula for the side-jump AHE where impurities, as local random Hamiltonians,
have a general matrix form in the space of fermion flavors and a general momentum dependence.
We provide this electronic AHE contribution, together with a simple derivation, in Eq.\eqref{eq:74analog}.
 
As a concrete application, in Sec.\ref {sec:concr-appl-honeyc} we consider the honeycomb-lattice $K\Gamma\Gamma'$ spin model in a magnetic field.
  This hamiltonian was proposed as a minimal model for the Kitaev material $\alpha\text{-RuCl}_3$ in \cite{PhysRevLett.126.147201},
  where qualitative agreement was found with experimental measurements of $\kappa_{xy}$ in this compound.
  We investigate the role of disorder, in the form of random fluctuations of the model's parameters that may be caused by chemical variations.
  We find that the side-jump term is a sizable contribution that leads to clear differences, both quantitative and qualitative, from the intrinsic THE,
  in accordance with experiments \cite{czajka2023planar}.

  These observations appear to be quite generic.
  To illustrate this somewhat formally, in Sec.\ref{sec:gener-boson-model} we investigate a generic low-energy theory for bosonic fields with broken time reversal symmetry.
We find again that the side-jump thermal Hall effect is comparable in magnitude to the intrinsic thermal Hall effect (see Fig.\ \ref{fig:Applications}).
This makes Eq.\eqref{eq:32} an indispensable contribution alongside Eq.\eqref{eq:31}.
Since for bosonic systems the intrinsic and extrinsic contributions are not clearly seperated in different temperature regimes,
   i.e. they generically coexist, we argue that both necessarily need to be treated on equal footing for a correct interpretation of experimental data.
Because ${\ms \Omega}^{\ms W}_{p_\mu p_\mu}(p)$ depends on the form of disorder considered,
the above formula shows that $\kappa_{xy}^{\rm dis}$ is non-universal and depends on the specific instance of disorder that we consider, as  illustrated in Fig.\ \ref{fig:Applications}.
We highlight in particular case (c), which shows that even with vanishing intrinsic Berry curvature $\ms \Omega_{p_x p_y}$ there may still exist sizeable disorder-induced thermal Hall conductivity.

While our paper discusses the physics of Eqs.\eqref{eq:31},\eqref{eq:32}, both in generic and in concrete applications,
  a significant part of it is devoted to actually \emph{deriving} Eq.\eqref{eq:32} rigorously. Our proof uses the language of kinetic theory,
  thereby satisfying formal constraints (of systematicity and physical intuition) that we introduce next with some historical context.
The general formula Eq.\eqref{eq:31} for the \emph{intrinsic} thermal Hall conductivity of neutral bosons was first derived rigorously using the Kubo-Streda formalism 
for phonons \cite{qin_berry_2012} and magnons \cite{matsumoto_thermal_2014}.
Its intuitive picture as a semiclassical current carried by particles obeying the semiclassical equations of motion \cite{matsumoto_rotational_2011}
was recently provided rigorous support in Ref.~\cite{mangeolle_quantum_2024}, where one of the authors took part in deriving Eq.\eqref{eq:31} directly from the inhomogeneous kinetic equation.
Except for the now well-known complications introduced by the energy magnetization \cite{qin_energy_2011},
this development is very similar to that of the intrinsic anomalous Hall effect (AHE) of electrons,
which was first obtained from the Kubo formula \cite{karplus_hall_1954} before it was given a semiclassical picture using the semiclassical equations of motion of the electron wavepacket \cite{sundaram_wave-packet_1999}, 
later derived rigorously within the kinetic language using the Wigner-Weyl formalism \cite{wickles_effective_2013}.
The effect of disorder on the \emph{electronic} AHE, and especially the side-jump effect,
was originally computed in specific electron models using Kubo's formula \cite{adams_energy_1959, fivaz_transport_1969, nozieres_simple_1973}, 
but can also be derived in the language of semiclassical dynamics. This was shown much later, based on the wavepacket analysis \cite{sinitsyn_disorder_2005, sinitsyn_coordinate_2006},
and was also derived rigorously in a particular model within the Schwinger-Keldysh kinetic formalism \cite{konig_quantum_2021}. At least in one particular case, a formal connection can be drawn between both approaches \cite{sinitsyn2007anomalous}. In the case of the thermal Hall conductivity of neutral bosons, the side-jump effect was seldom studied
(for a rare exception see Ref.~\cite{guo2022resonant} within the linear response formalism \cite{kapustin_thermal_2020}),
and a general, intuitive, semiclassical picture based on kinetic theory was still missing
 -- this is the problem we address in Sections \ref{sec:setup} to \ref{sec:side-jump-effect}.

The rest of the paper is organized as follows. In Section \ref{sec:setup}, we summarize the derivation of Ref.\cite{mangeolle_quantum_2024} of the inhomogeneous kinetic theory without disorder,
which forms the basis for the inclusion of disorder in a real-space, real-time picture. We then show how disorder scattering can be included in this formalism.
In Section \ref{sec:deriv-kinet-equat} we expand the time evolution equation in phase space
and show how disorder-induced coordinate shifts naturally appear in the full gauge-invariant kinetic equation. In Section \ref{sec:side-jump-effect} we show how these coordinate shifts
lead to a side-jump current operator, which we derive, and to a displaced equilibrium distribution, which together lead to Eq. \eqref{eq:32}. 
Finally, in Section \ref{sec:applications} we apply our results, first very concretely to a spin model with direct experimental implications
  for the Kitaev compound $\alpha \text{-RuCl}_3$ \cite{PhysRevLett.126.147201}  (Sec.\ref {sec:concr-appl-honeyc}),
then to a more generic low-energy theory of chiral bosons (Sec.\ref{sec:gener-boson-model}), in both cases with different instances of disorder.

\section{Setup}
\label{sec:setup}

\subsection{Reminders about the clean case}
\label{sec:remind-about-clean}

We start from the general formulation introduced in Ref.\ \cite{mangeolle_quantum_2024},
of a free boson system described by $2N$ hermitian fields $\Phi_a(r)$, $r\in\mathbb{R}^d$, $a=1,..,2N$,
with commutation relation $[{\Phi}_a(r_1),{\Phi}_b(r_2)]\equiv\hbar\,\hat{\sf \Gamma}_{ab}(r_1,r_2)$
a $2N\times2N$ matrix with pure imaginary $c$-number entries.
The free theory \emph{without disorder} is defined by an arbitrary quadratic hamiltonian,
$ \textrm{H}=\frac{1}{2} {\rm Tr}\left [ \Phi \circ \hat{\sf H} \circ \Phi \right ]$
where the convolution $\circ$ and the trace Tr run over both flavor and spatial indices, explicitly
\begin{equation}
  \label{eq:1}
  \textrm{H}=\frac{1}{2}\int_{r_1,r_2}\sum_{a,b}\Phi_a(r_1)\hat{\sf H}_{ab}(r_1,r_2)\Phi_b(r_2),
\end{equation}
where $\hat{\sf H}$ is a $2N\times2N$ matrix of pure real $c$-number entries.
The kinetic theory describes the evolution of the (real) density matrix
\begin{equation}
  \label{eq:2}
  \hat{\sf  F}_{ab}(r_1,r_2)\equiv\tfrac{1}{2}\left\langle\left\{\Phi_a(r_1),\Phi_b(r_2)\right\}\right\rangle.
\end{equation}
All three $\hat{\sf H}, \hat{\sf  F}, \hat{\sf \Gamma}$, regarded as matrices in the joint
flavor ($a,b$) and coordinate ($r_1,r_2$) space, are hermitian.

The theory can be formulated in phase space, with $(X,p)$ coordinates, by means of the Wigner transform,
\begin{align}
  \label{eq:3}
   {\sf  O} (X,p)& \equiv\int_xe^{-\frac{i}{\hbar}px} \; \hat{\sf  O} \left(X+\tfrac{x}{2},X-\tfrac{x}{2}\right),
\end{align}
where $p,X \in\mathbb{R}^d$ and $\int_x \equiv \int d^dx$.
Then $\mathsf{F},\mathsf{H},\mathsf{\Gamma}$ are functions on phase
space ($X,p$) and are all hermitian as $2N\times2N$ matrices at fixed $(X,p)$.
The equation of motion for the density matrix is
\begin{equation}
  \label{eq:4}
     \partial_t{\sf F}(X,p)=-\frac{i}{\hbar}\left({\sf K}\star{\sf F}-{\sf F}\star{\sf K}^\dagger\right),
\end{equation}
where $\star$ is the Moyal product (with sign convention as in \cite{mangeolle_quantum_2024, kamenev2011field})
and $ {\sf K}(X,p)=\hbar\,{\sf \Gamma}\star{\sf H}$ is the dynamical matrix.

\{ At this stage, it is worth noticing that the same procedure applied to fermions instead of bosons,
with $ \textrm{H}= {\rm Tr}\left [ \Psi^\dagger \circ \hat{\sf H} \circ \Psi \right ]$,
yields the very same kinetic equation Eq.\eqref{eq:4}, for the density matrix now defined as
$\hat{\sf  F}_{ab}(r_1,r_2)\equiv \left\langle \Psi_a^\dagger(r_1) \Psi_b(r_2) \right\rangle$
and where the dynamical matrix identifies as the hamiltonian, $ {\sf K}=\hbar\,{\sf H}$. \}

Back to bosonic systems, the energy current density, obtained from the continuity equation \cite{mangeolle_quantum_2024},
reads (to the first order in $\hbar$)
\begin{align}
  \label{eq:5}
  J_\mu(X)=\frac{1}{2} \int_p {\rm Re}\, {\rm  Tr}\left(\partial_{k_\mu}{\sf K}({\sf F}\star{\sf H})\right),
\end{align}
where $\mu \in \{1,\dots,d\}$ and $\int_p = \int d^dp/(2\pi\hbar)^d$.
We will come back to this in Sec.\ \ref{sec:side-jump-effect}.

\subsection{Introducing disorder}
\label{sec:form-with-disord}

We now consider the similar problem of an inhomogeneous theory (allowing for a possibly spatially-dependent lagrangean) of free bosons with disorder.
While, in principle, frozen disorder may be treated as a specific instance of spatial dependence in the hamiltonian Eq.\eqref{eq:1},
we will instead consider disorder as a distinct \emph{random} contribution to the hamiltonian. 
This means replacing Eq.\eqref{eq:1} with
\begin{align}
  \label{eq:6}
{\rm H} &= \frac 1 2 {\rm Tr}\left [\Phi \circ (\hat{\sf H}+\hat{V}) \circ \Phi \right ]
\end{align}
where $\hat{V}_{ab}(r_1,r_2)$ is a random matrix variable.
Since ${\rm H} $, together with the commutation relation, defines a free theory for the fields $\Phi$,
one can simply use the classical equations of motion to derive the kinetic equation,
\begin{align}
  \label{eq:7}
  \partial_t\hat{\sf F}
  &=-\frac{i}{\hbar}\big ( (\hat{\sf K}+\hat{\sf V}) \circ\hat{\sf F}-\hat{\sf F}\circ (\hat{\sf K}+\hat{\sf V})^\dagger \big ),
\end{align}
where now $\hat{\sf K}= \hbar\,\hat{\sf \Gamma}\circ\hat{\sf H}$ and $\hat{\sf V}= \hbar\,\hat{\sf \Gamma}\circ \hat{V}$.
\{ In the fermionic case, $\hat{\sf K}=\hbar\,\hat{\sf H}$ and $\hat{\sf V}= \hbar\, \hat{V}$. \}

The new ingredient here is $\hat{V}$, accounting for disorder.
Generally, $\hat{V}_{ab}(r_1,r_2)$ has an arbitrary matrix structure, only constrained by the same symmetry properties as $\hat{\sf H}_{ab}(r_1,r_2)$.
In the existing treatments of disorder in (fermionic) anomalous Hall effects \cite{adams_energy_1959, fivaz_transport_1969, nozieres_simple_1973,
  sinitsyn_disorder_2005, sinitsyn_coordinate_2006,sinitsyn2007anomalous,konig_quantum_2021},
a diagonal scalar potential $\hat{V}_{ab}(r_1,r_2)= \delta_{a,b}\, \delta(r_1-r_2)\,V(r)$ is used,
which corresponds effectively to an on-site shift of the electronic chemical potential 
\footnote{In \cite{konig_quantum_2021} an extra valley index is introduced, but the $(a,b)$ and $(r_1,r_2)$ indices remain decoupled,
  the spatial dependence is still of the $\delta(r_1-r_2)$ type, and within one valley disorder is still $\hat V \propto \idmatrix$.};
in the following we will refer to this as the ``case of fermions in a random scalar potential''.
For real bosons this would be unreasonable in general: because of the relation $\hat{\sf V}= \hbar\,\hat{\sf \Gamma}\circ \hat{V}$,
an on-site chemical potential (i.e.\ a term $\hat{\sf V} \propto \idmatrix$) corresponds in general to a nontrivial and fine-tuned matrix structure for $\hat{V}$.
We also drop the assumption $\hat{V}_{ab}(r_1,r_2)\propto\delta(r_1-r_2)$ and instead consider a more general spatial dependence.
This allows us to describe structural disorder, bond disorder, impurities interacting or mediating interactions at longer range, and more generally any local disturbance of \emph{momentum-dependent} features of the clean theory (as opposed to just an on-site scalar potential).

Explicitly, we consider $\hat{V}_{ab}(r_1,r_2) $ a gaussian random variable with zero average, $\langle \hat{V}_{ab}(r_1,r_2) \rangle =0$,
interpreted as a fixed (i.e.\ non-random) function of $(r_1-r_2)$ centered at a position $\tfrac 1 2 (r_1+r_2)$ which is random. 
This means, mathematically,
\begin{align}
  \label{eq:8}
 &  \langle \hat{V}_{ab}(r_1,r_2) \hat{V}_{cd}(r_3,r_4)\rangle \nonumber \\
  &= \hat{W}_{ab}(r_1-r_2) \hat{W}_{cd}(r_3-r_4) \; w(\tfrac{r_1+r_2}{2}- \tfrac{r_3+r_4}{2}),
\end{align}
where the first two factors describe the fixed matrix structure of single impurities,
and the function $w$ describes the impurities' spatial correlation -- we assume $w^*(r-r')=w(r-r')=w(r'-r)$.
We choose the normalization $w(0)=n_{\rm imp}$, the impurity density, to remove ambiguity.
For future convenience, we also define $\hat{\sf W}= \hat{\ms \Gamma}\circ \hat{W}$.
We note that the general disorder we consider here is similar to that of \cite{culcer_interband_2017}, where a homogeneous problem is considered
in a full momentum representation -- at odds with our approach of an inhomogeneous problem in a phase-space representation,
allowing for a semiclassical identification of wavepacket coordinate shifts.

We provide more intuition on Eq.\eqref{eq:8}, together with a concrete recipe to compute $\hat{\sf W}$,
in the more convenient language of phase space coordinates in Sec.\ref{sec:disord-matr-struct}.
For now, it suffices to say that $\hat{W}_{ab}(x-x')$, the ``disorder matrix'', accounts for the local matrix structure of an impurity
in the basis of the field components $\Phi_a(x), \Phi_b(x')$.

\begin{figure}[htbp]
\centering
\includegraphics[width=.95\columnwidth]{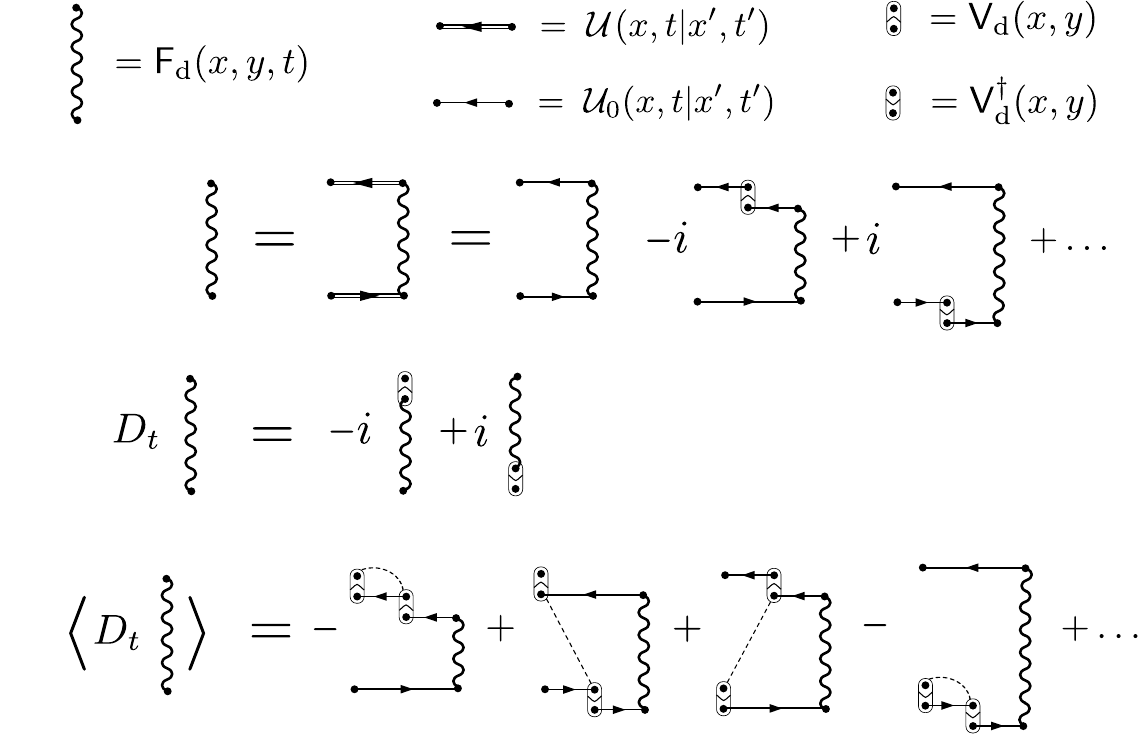}
\caption{Graphical representation of our equal-time procedure, inspired by \cite{rammerbook}. The horizontal direction represents time (with later times to the left),
  and the dashed lines represent a statistical contraction as per Eq.\eqref{eq:8}.
  Symbols are defined in the first line, and the other three lines represent Eqs.\eqref{eq:10}-\eqref{eq:11}, Eq.\eqref{eq:12} and Eq.\eqref{eq:13}, respectively.
 We use the shorthand $D_t  \ms F_{\rm d} = \partial_t +i  [{\sf K}_{\rm d}\,\overset{\circ},\,{\sf F}_{\rm d}]$, and for notational clarity $\hbar=1$.}
\label{fig:rammer}
\end{figure}

\subsection{Equal-time formalism}
\label{sec:equal-time-formalism}

The next step is to diagonalize the clean problem: here we just outline in two sentences the method described in \cite{mangeolle_quantum_2024}.
Explicitly the diagonalization is realized by solving for a matrix $\ms S$ such that $\ms K_{\rm d} = \ms S^{-1}\circ \ms K \circ \ms S$ is diagonal (one can also prove that its entries are real):
this defines the basis of bosonic bands.
Then, we defime $\ms F = \ms S \circ \ms F_{\rm d} \circ \ms S^\dagger $ and $\ms V_{\rm d} = \ms S^{-1}\circ \ms V \circ \ms S$ --
the index ``d'' indicates that a quantity is expressed in the basis where $\ms K_{\rm d}$ is made diagonal, but note that neither $\ms V_{\rm d}$ nor $\ms F_{\rm d}$ is diagonal a priori.
Then the kinetic equation reads
\begin{align}
  \label{eq:9}
  i \hbar  \partial_t{\sf F}_{\rm d}- [{\sf K}_{\rm d}\,\overset{\circ},\,{\sf F}_{\rm d}] = \ms V_{\rm d} \circ {\sf F}_{\rm d} - {\sf F}_{\rm d} \circ \ms V_{\rm d}^\dagger ,
\end{align}
with the shorthand $ [{\sf K}_{\rm d}\,\overset{\circ},\,{\sf F}_{\rm d}] = {\sf K}_{\rm d}\circ{\sf F}_{\rm d} - {\sf F}_{\rm d}\circ{\sf K}_{\rm d}$.
Note that $\ms V_{\rm d} $ is a random variable, and consequently ${\sf F}_{\rm d}$, as a solution of Eq.\eqref{eq:9}, is also random:
to extract physical quantities, a disorder average of Eq.\eqref{eq:9} must be taken. 

Because the disorder we consider is correlated in space but not in time, scattering is elastic and an equal-time formalism
(as opposed to the full Schwinger-Baym-Kadanoff-Keldysh treatment) is sufficient. We write
\begin{align}
  \label{eq:10}
   \ms F_{\rm d}(x,y,t) &= \int_{x',y'}\, \mc U(x,t|x',t') \, \ms F_{\rm d}(x',y',t') \,\mc U^\dagger(y,t|y',t') ,
\end{align}
in terms of the evolution operator 
\begin{align}
  \label{eq:11}
   \mc U(x,t|x',t') &\equiv \langle x|  {\mathbb T} \exp \left [ - \frac i \hbar \int_{t'}^t \text d \tau (\ms K_{\rm d}+\ms V_{\rm d})(\tau) \right ] |x'\rangle ,
\end{align}
where $ {\mathbb T} $ stands for time ordering, and note that in Eq.\eqref{eq:10} one integrates over $x',y'$ but not $t'$ which is an \emph{arbitrary, but fixed} ``initial'' time.
Then the next steps, following the standard procedure (see for instance \cite{rammerbook}), are to replace Eq.\eqref{eq:10} into the kinetic equation,
expand in powers of $\ms V_{\rm d}$, and average over disorder using Eq.\eqref{eq:8}: this is represented diagrammatically on Fig.\ref{fig:rammer},
using the standard intuitive pictorial language of e.g.\ \cite{rammerbook}.

\begin{widetext}
  To the lowest diagrammatic order, one uses the expansion 
  \begin{align}
    \label{eq:12}
   \mc U(x,t|x',t') &= \mc U_0(x,t|x',t') - \frac i \hbar \int_{t'}^t \text d \tau \int_{\xi,\zeta} 
                     \mc U_0 (x,t|\xi,\tau)    \,\ms V_{\rm d}(\xi,\zeta) \, \mc U_0(\zeta,\tau|x',t') + \dots,
  \end{align}
  where $ \mc U_0 (x,t|x',t')   \equiv \langle x| \exp \left [ - \frac i \hbar (t-t') \ms K_{\rm d} \right ] |x' \rangle$ is the clean evolution operator.
  In the last line of Fig.\ref{fig:rammer}, the rightmost part of each diagram can be contracted using again Eq.\eqref{eq:10}
  (to the lowest order in powers of the disorder potential), yielding the disorder-averaged kinetic equation
\begin{align}
  \label{eq:13}
   &\langle  i \hbar  \partial_t({\sf F}_{\rm d})_{n,n}(x,y,t) - [{\sf K}_{\rm d}\,\overset{\circ},\,{\sf F}_{\rm d}]_{n,n}(x,y,t) \rangle \\
  &= - \frac i \hbar \bigg (  \int_{t'}^t \text d \tau \int_{\{\xi_i\}} \,
     ( \mc U_0)_{n',n'}(\xi_1,t|\xi_2,\tau)   \langle (\ms V_{\rm d})_{n,n'}(x,\xi_1)  (\ms V_{\rm d})_{n',n}(\xi_2,\xi_3) \rangle (\ms F_{\rm d})_{n,n}(\xi_3,\xi_4,\tau)  ( \mc U_0)_{n,n}(\xi_4,\tau|y,t) \nonumber \\
  &  - \int_{t'}^t \text d \tau \int_{\{\xi_i\}} \,
     ( \mc U_0)_{n,n}(x,t|\xi_1,\tau)   \langle (\ms V_{\rm d})_{n,n'}(\xi_1,\xi_2)  (\ms V_{\rm d}^\dagger)_{n',n}(\xi_4,y) \rangle (\ms F_{\rm d})_{n',n'}(\xi_2,\xi_3,\tau)  ( \mc U_0)_{n',n'}(\xi_3,\tau|\xi_4,t)  \nonumber \\
  & - \int_{t'}^{t} \text d \tau \int_{\{\xi_i\}} \,  ( \mc U_0)_{n',n'}(\xi_1,t|\xi_2,\tau) (\ms F_{\rm d})_{n',n'}(\xi_2,\xi_3,\tau)
    \langle (\ms V_{\rm d})_{n,n'}(x,\xi_1)   (\ms V_{\rm d}^\dagger)_{n',n}(\xi_3,\xi_4) \rangle   ( \mc U_0)_{n,n}(\xi_4,\tau|y,t)   \nonumber \\
  & + \int_{t'}^{t} \text d \tau \int_{\{\xi_i\}}\,  ( \mc U_0)_{n,n}(x,t|\xi_1,\tau) (\ms F_{\rm d})_{n,n}(\xi_1,\xi_2,\tau)
    \langle (\ms V_{\rm d}^\dagger)_{n,n'}(\xi_2,\xi_3) (\ms V_{\rm d}^\dagger)_{n',n}(\xi_4,y)   \rangle   ( \mc U_0)_{n',n'}(\xi_3,\tau|\xi_4,t)  \bigg ) + \dots \nonumber
\end{align}
where the integral is over all four $\xi_i$, $i=1,\dots, 4$.
In Eq.\eqref{eq:13}, disorder now enters solely in the form of statistical correlations
\begin{align}
  \label{eq:14}
&\langle (\ms V_{\rm d}^s)_{n,n'}(z_1,z_2)  (\ms V_{\rm d}^{s'})_{n',n}(z_3,z_4) \rangle 
 = \int_{\{\lambda_i\}} w \Big (\tfrac 1 2 (\lambda_1+\lambda_2) - \tfrac 1 2 (\lambda_3+\lambda_4) \Big ) \\
  &\qquad \times \left [ \ms S^{-1}(z_1,\lambda_1) \ms W(\lambda_1,\lambda_2) \ms S(\lambda_2,z_2) \right ]^s_{n,n'}\;
   \left [ \ms S^{-1}(z_3,\lambda_3) \ms W(\lambda_3,\lambda_4) \ms S(\lambda_4,z_4) \right ]^{s'}_{n',n}  , \nonumber
\end{align}
where the integral is over all four $\lambda_i$, $i=1,\dots, 4$ and the upper index $s,s'=-$ (resp. $+$) on an operator means hermitian conjugation (resp. absence thereof).
\end{widetext}

Taking $t'\rightarrow -\infty$, Eq.\eqref{eq:13} gives ${\sf F}_{\rm d}(t)$ as a function of ${\sf F}_{\rm d}(\tau)$ at all earlier times $\tau < t$.
Considering ${\sf F}_{\rm d}(-\infty)$ to be disorder-independent, then ${\sf F}_{\rm d}(\tau)$ on the right-hand side of Eq.\eqref{eq:13},
as a solution to the equation at an earlier time, is no longer a random variable but a deterministic function.
Therefore we can drop the disorder average $\langle \cdot \rangle$ on the left-hand side,
and henceforth treat Eq.\eqref{eq:13} as a \emph{deterministic} integro-differential equation for ${\sf F}_{\rm d}(t)$.

\section{Derivation of the kinetic equation}
\label{sec:deriv-kinet-equat}

\subsection{Phase space formulation and semiclassical approximations}
\label{sec:phase-space-form}

In order to obtain the kinetic equation in terms of phase space variables $(X,p)$,
we now perform the Wigner transformation (see Eq.\eqref{eq:3}) on the kinetic equation.
We provide some technical details about the intermediate steps in Appendix \ref{sec:deriv-kinet-equat-calc-det},
and here we only outline a few salient aspects of the calculation.

First, our semiclassical derivation relies on an expansion in powers of $\hbar$.
Throughout the paper, this is referred to as the \emph{semiclassical} expansion \cite{footnote_perturbative}.
Scattering of bosons on disorder affects the phase-space coordinates $X$ and $p$ of the wavepackets, which by convention are $O(\hbar^0)$ in the free case.
Upon scattering, the particle momentum is modified, $p \rightarrow p'$, by an amount $p'-p=O(\hbar^0)$
governed by the Fourier-transformed disorder correlation $\tilde w(k)=\int_\xi e^{-i k \xi/\hbar} w(\xi)$: since the latter can hardly be known with sufficient
accuracy, any $O(\hbar^1)$ semiclassical correction is irrelevant, and we neglect scattering shift effects on the \emph{momentum} of excitations.
Meanwhile, we focus on the \emph{position} shift effects induced by disorder which will occur at order $O(\hbar^1)$ -- in other words,
we consider semiclassical corrections on the $X$ but not the $p$ variable of ${\sf F}_{\rm d}(X,p), {\sf K}_{\rm d}(X,p)$ in our derivation.

Second, we use the classical limit of the time evolution, given by the limit $\hbar\rightarrow 0$ and the identity
\begin{align}
  \label{eq:15}
  \lim_{\hbar \rightarrow 0} \frac 1 \hbar \theta(t)  e^{-it(E-i0)/\hbar}
  &= - i \delta(t) \frac 1 {E-i0} \\
  & = -i \delta(t) \mc P \frac 1 E + \pi \delta(t) \delta(E) \nonumber ,
\end{align}
where $\mc P$ stands for principal value.
Physically, this amounts to assuming destructive intereferences between all contributions in Eq.\eqref{eq:13} at times $\tau$ earlier than $t$,
yielding an \emph{equal-time} differential equation for the time evolution of the density matrix.
Equivalently, the $\hbar \rightarrow 0$ limit selects the classical path in the functional integral representation of time evolution.
In the following we also discard the principal value term, which accounts for the perturbative \cite{footnote_perturbative} effect of disorder on the bosonic quasiparticle energies,
and keep only the dirac delta term, which accounts for elastic scattering of bosonic waves.

Third, we assume adiabaticity in the sense that the disorder correlation length $\lambda_{\rm dis}$ is much larger than the typical wavelength of bosonic excitations
$\lambda_{\rm bos}$ (at a given temperature $T$), $\lambda_{\rm dis} \gg \lambda_{\rm bos}$, equivalently that the disorder correlation function
$\tilde w(k)$ is a narrow enough peak centered around zero momentum: the momentum transfer $k$ during a single scattering event is then small
in comparison to a typical bosonic excitation's momentum $p$.
Based on this, as a crucial step we perform a small-$k$ expansion 
\begin{align}
  \label{eq:16}
&  (\ms S^{-1})_{n,a} (p+k/2)  \,\ms W_{ab}(\tfrac{p+p'}2+k/2)\,\ms S_{b,n'}(p'+k/2) \nonumber \\
   & \quad \approx   \ms M_{n,n'}(p,p') \,  \exp \left [ {i (k/2) \, \delta \ms r_{n,n'}(p,p')/\hbar} \right ],
\end{align}
where we defined
\begin{subequations}
  \label{eq:17}
  \begin{align}
  \label{eq:17a}
    \ms M_{n,n'}(p,p')&=  (\ms S^{-1})_{n,a} (p) \ms W_{ab}(\tfrac{p+p'}2) \ms S_{b,n'}(p') ,\\
      \label{eq:17b}
 \delta \ms r_{n,n'}(p,p') &= \hbar \,(\partial_p+\partial_{p'})\,{\rm Arg} \left [ \ms M_{n,n'}(p,p') \right ].
\end{align}
\end{subequations}
Contrary to the previous two approximations, Eq.\eqref{eq:16} is in principle an uncontrolled assumption,
which amounts to neglecting variations of the moduli of eigenvectors, focussing on their \emph{phase} variations.
In the wavepacket interpretation of phase-space quantum dynamics, this means we are focussing on the position and momentum of particle wavepackets (contained in the phase),
neglecting variations of their extension or spreading (contained in the amplitude): this is again a semiclassical approximation in the traditional sense \cite{gutzwillerbook}. 

\{ In the case of fermions in a random scalar potential, because $\ms W \propto \idmatrix$ and $\ms S^\dagger \circ \ms S = \idmatrix$,
Eq.\eqref{eq:16} is strictly equivalent to $k \ll p{}^({}'{}^)$ ; in our general bosonic case there is no such equivalence. \}

Note also that in Eqs.\eqref{eq:16},\eqref{eq:17} we dropped the $X$ variable of all quantities ($\ms S, \ms M, \delta \ms r$, etc):
while these may indeed vary spatially in our inhomogeneous theory,
this dependence is irrelevant to the order of $\hbar$ considered here,
and we will keep omitting the $X$ variable in the following wherever this cannot possibly cause any confusion.

\begin{widetext}
  The upshot of the procedure just described is Eq.\eqref{eq:49} in Appendix \ref{sec:deriv-kinet-equat-calc-det},
which, after one extra (minor) simplification discussed there, becomes
  \begin{align}
    \label{eq:97}
    & i \hbar  \partial_t ({\sf F}_{\rm d})_{n,n}(X,p,t) - [{\sf K}_{\rm d}\,\overset{\star},\,{\sf F}_{\rm d}]_{n,n}(X,p,t) \\
                         &= -i2 \pi \sum_{n'}\int_{p'} \tilde w(p-p')      \, \delta \left ( (\ms K_{\rm d})_{n',n'}(X + \delta \ms r_{n,n'}^{-}(p,p') , p') - (\ms K_{\rm d})_{n,n} (X , p) \right )\nonumber \\
                         &  \qquad \times \bigg \{ (\ms F_{\rm d})_{n,n}(X ,p ,t)\; {\rm Re} \left [ \ms M_{n,n'}(p,p')\, \ms M_{n',n}(p',p) \right ]
                     - (\ms F_{\rm d})_{n',n'} (X +\delta \ms r_{n,n'}^{-}(p,p') , p',t) \;  {\rm Re} \left [ \ms M_{n,n'}(p,p') \,\ms M_{n,n'}(p,p')^* \right ]  \bigg \} ,\nonumber
  \end{align}
where $ [ \ms O_1 \overset{\star}, \ms O_2] =  \ms O_1 \star \ms O_2 - \ms O_2 \star \ms O_1 $ is the Moyal bracket,
and $\delta \ms r_{n,n'}^{-}(p,p') =  \tfrac 1 2 [\delta \ms r_{n,n'}(p,p') - \delta \ms r_{n',n}(p',p)] $ appears as a coordinate shift. We note that $\tilde w^*(k)=\tilde w(k)=\tilde w(-k)$.
\end{widetext}

\subsection{Gauge-invariant coordinates and shift}
\label{sec:gauge-invar-form}

When defining eigenfunctions of a given operator (here, $\ms K$), one is faced with a gauge choice:
after fixing the eigenvector normalization $\ms S^\dagger \star \ms H \star \ms S = \idmatrix$
there remains a $U(1)$ phase redundancy
\begin{align}
  \label{eq:114}
  \ms S \rightarrow \ms S \star e^{i\stheta}
\end{align}
at any point $(X,p)$ in phase space ($\stheta$ a diagonal matrix of phases, function of $X$ and $p$). 
As explained e.g.\ in \cite{mangeolle_quantum_2024}, ${\sf K}_{\rm d}(X,p)$ and ${\sf F}_{\rm d}(X,p)$ themselves are not invariant under such a gauge transformation Eq.\eqref{eq:114}.
Instead, $\ul {\sf K}_{\rm d}(X,p) =  {\sf K}_{\rm d}(\ul X,\ul p)$ and $\ul {\sf F}_{\rm d}(X,p) =  {\sf F}_{\rm d}(\ul X,\ul p)$,
functions of the ``kinetic'' coordinates
\begin{subequations}
      \label{eq:18}
  \begin{align}
    \ul X_\mu &=  X_\mu  + \hbar \ms A_{p_\mu},\\
     \ul p_\mu &=  p_\mu  - \hbar \ms A_{X_\mu}  ,         
  \end{align}
\end{subequations}
are gauge-invariant (to the first order in $\hbar$).

Here $\ms A_\gamma = {\rm Im}\left (\ms S^{-1}\partial_\gamma \ms S\right )^{(\rm d)}$, for $\gamma=X_\mu,p_\mu$,
where Im selects the imaginary part and (d) the diagonal part,
is the generalized Berry connection in phase space (up to $O(\hbar^1)$ corrections).

In terms of these new coordinates and physical quantities, one can show (see \cite{wickles_effective_2013, konig_quantum_2021, mangeolle_quantum_2024})
that the time derivative operator appearing on the left-hand side of the gauge-invariant kinetic equation is
\begin{align}
  \label{eq:19}
{\rm D}_t = \partial_t + \dot {\ms X}_\mu\partial_{X_\mu}  + \dot {\ms p}_\mu\partial_{p_\mu}  ,
\end{align}
with what can be identified as the Hamilton equations,
\begin{subequations}
   \label{eq:192}
  \begin{align}
  \label{eq:192a}
    \dot {\ms X}_\mu&=\partial_{p_\mu} \ul{\sf K}_d-\hbar\left({\sf \Omega}_{p_\mu p_\nu}\partial_{X_\nu}\ul{\sf K}_d
                      -{\sf  \Omega}_{p_\mu X_\nu}\partial_{p_\nu}\ul{\sf K}_d\right),\\
    \label{eq:192b}
    \dot {\ms p}_\mu&=- \partial_{X_\mu}\ul{\sf K}_d+\hbar\left({\sf \Omega}_{X_\mu p_\nu}\partial_{X_\nu}\ul{\sf K}_d
                           -{\sf \Omega}_{X_\mu X_\nu}\partial_{p_\nu}\ul{\sf K}_d\right),
\end{align}
\end{subequations}
in terms of the generalized Berry curvature in phase space,
\begin{align}
  \label{eq:20}
  {\sf \Omega}_{\alpha\beta}&\equiv \partial_\alpha \ms A_\beta - \partial_\beta \ms A_\alpha ,
\end{align}
with $\alpha,\beta \in \{X_\mu,p_\nu\}$, and always up to $O(\hbar^2)$ corrections.
We note that all these quantities $ {\sf \Omega}_{\alpha\beta}, \ms A_\gamma ,  \dot {\ms X}_\mu,  \dot {\ms p}_\mu, {\rm D}_t $
must be understood as diagonal matrices in the eigenbasis of energy bands (where $\ms K_{\rm d}$ is diagonal).
\{ We also note that this discussion translates straightforwardly to the language of fermions \cite{wickles_effective_2013}. \}

The above results hold for a general inhomogeneous theory.
In the following, we focus on physical problems where the ``magnetic field'' $\ms \Omega_{X,X} $, the mixed Berry curvatures $\ms \Omega_{X,p}, \ms \Omega_{p,X}$,
and the real-space Berry connection $\ms A_{X_\mu}$ play no role (which includes not only homogeneous theories but also,
for instance, the ``gravitational potential'' or ``separable'' case \cite{luttinger1964theory,mangeolle_quantum_2024}).

Then, performing the change of coordinates Eq.\eqref{eq:18} in Eq.\eqref{eq:97}, one obtains the gauge-invariant kinetic equation Eq.\eqref{eq:22},
whose left-hand side is ${\rm D}_t  \,\ul {\sf F}_{\rm d}(X,p) $
and whose right-hand side is now expressed fully in terms of $\ul {\sf K}_{\rm d}(X,p), \ul {\sf F}_{\rm d}(X,p) $,
with the corresponding position coordinates
\begin{align}
\text{for}\quad   (\ms K_{\rm d})_{n,n} ,(\ms F_{\rm d})_{n,n} \quad& : \quad X_\mu , \nonumber \\
\text{for}\quad    (\ms K_{\rm d})_{n',n'} ,(\ms F_{\rm d})_{n',n'} \quad& : \quad X_\mu +  \delta {\sf X}_{n,n'}(p,p') , \nonumber 
\end{align}
where we have defined the gauge-invariant position shift
\begin{align}
  \label{eq:21}
  \delta {\sf X}^\mu_{n,n'}(p,p') &= \delta \ms r_{n,n'}^{-,\mu} (p,p') \\
  &\; + \hbar (\ms A_{p_\mu})_{n,n}(p) - \hbar (\ms A_{p_\mu})_{n',n'} (p') \nonumber .
\end{align}

We note that all three terms in Eq.\eqref{eq:21} are order $O(\hbar^1)$,
and that although $\delta \ms r_{n,n'}^- (p,p') $ is not gauge-invariant, the full position shift $\delta {\sf X}^\mu_{n,n'}(p,p')$ is.

\begin{widetext}
  We have obtained:
  \begin{align}
  \label{eq:22}
  \big ( \partial_t &+ \dot {\ms X}_\mu  \partial_{X_\mu}  + \dot {\ms p}_\mu\partial_{p_\mu}  \big ) \ul{\ms F}_{\rm d} (X,p,t) \big |_{n,n}   \\
  &= - \frac {2\pi} {\hbar} \,\sum_{n'} \int_{p'} \tilde w(p-p')  \times \delta \Big ( (\ul{\ms K}_{\rm d})_{n',n'}(X+\delta {\sf X}_{n,n'}(p,p') , p') - (\ul{\ms K}_{\rm d})_{n,n} (X, p) \Big ) \nonumber \\
  & \times   \bigg \{ (\ul{\ms F}_{\rm d})_{n,n}(X ,p ,t)\; {\rm Re} \left [ \ms M_{n,n'}(p,p') \, \ms M_{n',n}(p',p) \right ]
    - (\ul{\ms F}_{\rm d})_{n',n'} (X+ \delta {\sf X}_{n,n'}(p,p') , p',t) \; {\rm Re} \left [ \ms M_{n,n'}(p,p') \,\ms M_{n,n'}(p,p')^* \right ] \bigg \} .  \nonumber 
  \end{align}
  Note that the position arguments of the functions $\ul{\ms F}_{\rm d}$ and $\ul{\ms K}_{\rm d}$ in Eq.\eqref{eq:22} are derived systematically
  to semiclassical order $O(\hbar^1$), and they are crucial in the following.
  \end{widetext}

\subsection{Corrected kinetic equation: practical implementation and further assumptions}
\label{sec:corr-kinet-equat}

In order to make quantitative predictions in specific applications, it is possible to use Eq.\eqref{eq:22} as a starting point,
and to solve it for instance numerically, using variational methods \cite{zimanbook} and/or decomposing the distribution function onto a basis of harmonics \cite{PhysRevB.17.3725}.
Here, we will remain general and proceed with further approximations to make analytical progress.

First we will replace $\ms M_{n,n'}(p,p')$ by its modulus: while it is clear that this does not affect the semiclassical physics of coordinate shifts but merely modifies the scattering rates, whose precise form ultimately \emph{does not} affect our final results, it is also possible to justify this assumption more rigorously based on a small-momentum expansion (see Appendix \ref{sec:repl-msm-absmsm}).
Second, contrary to the case of fermions in a random scalar potential, where it is always true,
\begin{equation}
  \label{eq:89}
  \big |\ms M_{n,n'}(p,p') \big | \overset ? = \big |\ms M_{n',n}(p',p) \big |
\end{equation}
does not hold \emph{a priori} for the problem at hand. We discuss the validity of this assumption in Appendix \ref{sec:valid-appr-eq.eqr},
where we show that it is valid in a particular (yet still reasonably general) case.
Henceforth we will assume Eq.\eqref{eq:89} to hold, so that $ \big |\ms M_{n,n'}(p,p') \big |^2$ may be factorized in the right-hand side of Eq.\eqref{eq:22}.
The latter then takes a simplified form, 
\begin{align}
  \label{eq:103}
  {\rm D}_t \,  \ul{\ms F}_{\rm d} (X,p,t) \big |_{n,n}
  &= - \frac {2\pi} {\hbar} \, \sum_{n'}\int_{p'} \tilde w(p-p')  \times \delta \big ( \text{energy} \big ) \nonumber \\
  & \times \big |\ms M_{n,n'}(p,p') \big |^2 \times \delta F _{n,n'}(p ,p'),
\end{align}
where one defined the distribution difference
\begin{align}
  \label{eq:105}
  \delta F _{n,n'}(p ,p') &\equiv  (\ul{\ms F}_{\rm d})_{n,n}(X ,p ) \nonumber \\
  & - (\ul{\ms F}_{\rm d})_{n',n'} (X+ \delta {\sf X}_{n,n'}(p,p') , p')
\end{align}
and the energy conservation delta-function is the same as in Eq.\eqref{eq:22}.
Besides, in the following we neglect $n\neq n'$ contributions to the collision integral.
An argument based on the relative magnitude of matrix elements $\ms M_{n,n'}(p,p')$ is given in App.\ref{sec:argument-2},
but here we simply notice that for non-degenerate bands and small deviations $p'-p$,
the energy conservation delta strongly favors $n'=n$ terms, which we henceforth focus on.

\subsection{The disorder-induced curvature}
\label{sec:disord-berry-curv}

We now consider the band-diagonal position shift $\delta {\sf X}_{n,n}$.
An expansion at small deviations $p'-p$ (see details in Appendix \ref{sec:kinet-coord-comm}) yields
\begin{align}
  \label{eq:23}
  \delta \ms X^\mu_{n,n}(p,p')  &\approx  (p'-p)_\nu \, (\ms \Omega^{\ms W}_{p_\mu p_\nu})_{n,n} ,
\end{align}
where we defined a gauge-invariant, disorder-dependent quantity,
\begin{align}
  \label{eq:24}
\ms \Omega^{\ms W}_{p_\mu p_\nu}(p) &\equiv -  \partial_{p_\nu} \,  {\rm Im} \left [ \ms S^{-1}(p) \partial_{p_\mu}\ms S(p) \right ]^{(\rm d)}  \\
                                           &  + \partial_{p_\mu} \,  {\rm Im} \left \{ [\ms S^{-1}(p) \ms W(p) \partial_{p_\nu}\ms S(p)]^{(\rm d)} {\ms D}_{\rm d}^{-1}(p) \right \} \nonumber \\
                                           &   + \frac 1 2   \partial_{p_\mu} \,  {\rm Im} \left \{ [\ms S^{-1}(p) \partial_{p_\nu} \ms W(p) \ms S(p)]^{(\rm d)} {\ms D}_{\rm d}^{-1}(p) \right \} \nonumber ,
\end{align}
where we used the shorthand
\begin{align}
  \label{eq:113}
  {\ms D}_{\rm d}(p) &\equiv  [\ms S^{-1}(p) \ms W(p) \ms S(p)]^{(\rm d)}  ,
\end{align}
and we note that a necessary condition for this small-deviation expansion to be well grounded
is that $ ({\ms D}_{\rm d})_{n,n} (p)\neq 0\;\forall (n,p)$, i.e.\ that the diagonal elements of $\ms S^{-1}\ms W \ms S$ be non-vanishing.
To the order we consider in $\lambda_{\rm bos}/\lambda_{\rm dis}\ll 1$ (small deviation expansion),
in Eq.\eqref{eq:23} $\ms \Omega^{\ms W} $ can be evaluated equivalently at $p$ or $p'$.
Also, in Eq.\eqref{eq:24} the sum of the first and second terms is gauge-invariant, while the third term by itself is gauge-invariant.
An explicitly gauge-invariant formula for $\tfrac 1 2 (\ms \Omega^{\ms W}_{p_\mu p_\nu}-\ms \Omega^{\ms W}_{p_\nu p_\mu})$,
  suitable for numerical evaluation, is provided in App. \ref{sec:expl-form-analyt}. 

The quantity Eq.\eqref{eq:24} may be dubbed for convenience the ``disorder-induced curvature'', although it does not \emph{a priori} qualify as a curvature in the strict mathematical sense.
\{ In the case of fermions in a random scalar potential, with $\ms W(p) \propto \idmatrix$, the two curvatures coincide: then one just has
$$\delta \ms X^\mu_{n,n}(p,p') \approx  (p'-p)_\nu\, (\ms \Omega_{p_\mu p_\nu})_{n,n} ,$$
where $\ms \Omega_{p_\mu p_\nu} = \partial_{p_\mu} \ms A_{p_\nu} - \partial_{p_\nu} \ms A_{p_\mu}$ is the usual Berry curvature.
This recovers the familiar result of \cite{sinitsyn_disorder_2005, sinitsyn_coordinate_2006}. \}

We notice that in general $\ms \Omega^{\ms W}_{p_\mu p_\nu} \neq - \ms \Omega^{\ms W}_{p_\nu p_\mu} $, at odds with the usual Berry curvature.
Meanwhile, from Eq.\eqref{eq:23} it appears that $\ms \Omega^{\ms W}_{p_\mu p_\nu} $ is odd under TR and even under inversion, just like the usual Berry curvature.

Besides our rigorous formal proof of Eq.\eqref{eq:22} (and of the energy current below) including semiclassical disorder corrections,
equations Eqs. \eqref{eq:23}, \eqref{eq:24} are the main new result of this work,
displaying both qualitative and quantitative differences with the known theory for the case of fermions in a random scalar potential.

\subsection{The disorder matrix structure $\hat{\sf W}$}
\label{sec:disord-matr-struct}

We now provide more details on the interpretation and the determination in practice of the disorder matrix  ${\ms W}(p)= \ms \Gamma \hat W(p)$, which plays a crucial role in defining the disorder-induced curvature $\ms \Omega^{\ms W}$ in Eqs.\eqref{eq:24}, \eqref{eq:113}. 

An impurity localized at a given position modifies locally the hamiltonian. 
Should one place such an impurity at each site, translational invariance would be restored and the homogeneous hamiltonian would simply be shifted by a matrix function of momentum, $\ms H(p)\rightarrow \ms H(p)+\hat{W}(p)$.
This defines $\hat{W}(p)$ and provides a way to compute it concretely. Such a disorder matrix can describe any local variations of the parameters of the hamiltonian (magnetic exchanges, hopping amplitudes, elastic moduli, etc), that define the matrix structure and momentum dependence of $\ms H(p)$:
in fact, at this stage $\hat{W}(p)$ could simply be absorbed into a redefinition of  $\ms H(p)$.

Now, more generally, in the inhomogeneous setting, distributing a ``bunch'' of impurities around position $X$ weighted by a smoothly decreasing function of distance
shifts the hamiltonian as $\ms H(X,p)\rightarrow \ms H(X,p)+\tilde{W}(X,p)$, where $\tilde{W}(X,p)$ reduces to $\hat W(p)$ in the homogeneous limit where the impurities ``extend'' over the whole system. 
To describe the realistic case of correlated but dilute impurities, we describe them as a self-averaging, effectively random distribution. We introduce such disorder as a random variable,
$\tilde{W}(X,p) \mapsto \hat{V}(X,p)$, and assume gaussian correlations between the impurity positions, Eq.\eqref{eq:8}, which equivalently in phase-space notations reads
\begin{align}
  \langle \hat{V}_{ab}(X,k) \hat{V}_{cd}(X',k')\rangle = \hat{W}_{ab}(k) \hat{W}_{cd}(k') \; w(X-X'). \nonumber
\end{align}
Thus $\hat{V}(X,p) $ can be simply understood as a random (self-averaging) distribution of impurities with fixed matrix structure $\hat{W}(p)$, whose random positions $X$ have a gaussian correlation function $w(X-X')$.

Although these formal details are needed to establish rigorously our results, 
we emphasize that determining $\hat W(p)$, whence ${\ms W}(p)= \ms \Gamma \hat W(p)$ 
and $\ms \Omega^{\ms W}_{p_\mu p_\nu}$ from Eqs.\eqref{eq:24}, \eqref{eq:113}, in practice is more straightforward and can be done in a fully deterministic (non-random) setting.

\section{Coordinate shift effects in the thermal Hall current}
\label{sec:side-jump-effect}

In the presence of disorder, there arise two new contributions to the thermal Hall current,
originating (1) from a displacement of the equilibrium distribution due to the $O(\hbar^1)$ position shift,
and (2) from a new term of order $O(\hbar^1)$ in the current operator itself.

\subsection{Solving the kinetic equation}
\label{sec:solv-kinet-equat}

We now look for a distribution $\ul{\ms F}_{\rm d} (X,p)$, solution of the kinetic equation Eq.\eqref{eq:103}, in the form
\begin{align}
  \label{eq:25}
  \ul{\ms F}_{\rm d} (X,p) &= f \left (\ul{\ms K}_{\rm d} (X,p) ,T(X) \right ) + \ms g_{\rm d}(X,p) ,
\end{align}
where the first term is the clean equilibrium solution, with $f(\varepsilon)=\varepsilon \left [n_{\rm B}(\varepsilon)+\tfrac 1 2 \right ]$,
and the second term contains both the out-of-equilibrium distribution and the ``compensation term'' to the local equilibrium distribution due to the position shift.

Details are given in Appendix \ref{sec:solv-boltzm-eqn}. One finds
\begin{subequations}
  \begin{align}
  \label{eq:28a}
    \ms g_{\rm d}(X,p) &=  \ms g_{\rm d}^{\rm sj}(X,p) + \ms g_{\rm d}^{\rm rta}(X,p),\\
    \label{eq:28b}
    \ms g_{\rm d}^{\rm sj}(X,p) &= - p_\nu \, \ms \Omega^{\ms W}_{p_\lambda p_\nu}(p)  \; \partial_\lambda T \;\partial_T f \!\left ( \ul{\ms K}_{\rm d} \right ),\\
    \label{eq:28c}
    \ms g_{\rm d}^{\rm rta}(X,p) &=  \partial_{k_\lambda} \ul{\ms K}_{\rm d} \;\partial_{\lambda} T \; \partial_T f(\ul{\ms K}_{\rm d}) \,\stau (X,p),
\end{align}
\end{subequations}
where $f$ is still implicitly a function of $T(X)$, all objects $\ms g_{\rm d}, \ms \Omega^{\ms W}, \ul{\ms K}_{\rm d}$ are diagonal matrices in the basis of energy bands,
the dependence $ \ul{\ms K}_{\rm d} (X , p) $ remains implicit,
Eq.\eqref{eq:28c} relies on the same assumptions as the relaxation time approximation, and
\begin{align}
  \label{eq:26}
 \stau^{-1}_{n,n}(X,p) &\equiv  \frac  {2\pi} {\hbar} \, \int_{p'} \tilde w(p-p')\,   \left | \ms M_{n,n}(p,p') \right |^2\,
    \left ( 1 - \hat{\mb p}\cdot\hat{\mb p}' \right )  \nonumber \\
  &\quad \times   \delta \Big ( (\ul{\ms K}_{\rm d})_{n,n}(X, p') - (\ul{\ms K}_{\rm d})_{n,n} (X , p) \Big ),
\end{align}
where $\hat{\mb p}=\mb p/|{\mb p}|$.

\subsection{The ``jump-forward'' term}
\label{sec:jump-forward-comp}

Plugging Eq.\eqref{eq:28b} into the $O(\hbar^0)$ formula for the energy current density \cite{mangeolle_quantum_2024},
one obtains the first disorder contribution to the thermal Hall current:
\begin{align}
  \label{eq:27}
 J_{\mu}^{\rm gv,sj}(X)&= \frac 1 2 \int_p \sum_n \partial_{p_\mu}(\ul{\ms K}_{\rm d})_{n,n} \,  (\ms g_{\rm d}^{\rm sj})_{n,n}(X,p).
\end{align}
Since we assumed disorder is the only source of scattering, $ J_{\mu}^{\rm gv,sj}$ does not depend on the disorder strength,
in the sense that under a rescaling $\tilde w(k) \rightarrow \zeta \tilde w(k)$, with $\zeta \in \mathbb R$,
$ \ms g_{\rm d}^{\rm sj}$ and thus $ J_{\mu}^{\rm gv,sj}$ are invariant.

\subsection{The ``side-jump'' accumulation term}
\label{sec:side-jump-accum}

We now derive the ``side-jump'' contribution, corresponding intuitively to the accumulation of position shifts over the course of a trajectory.
In the case of charge transport by electrons, it was originally postulated \cite{sinitsyn_disorder_2005, sinitsyn_coordinate_2006} on a phenomenological basis.
The charge current \emph{operator}  may also be derived rigorously using a Ward identity \cite{konig_quantum_2021},
but its expectation value still required a phenomenological input.
\footnote{Namely, one needs to fix whether the distribution function (against which the current operator is integrated)
  should be evaluated before or after the scattering event.}
In the present case of a thermal current carried by nonconserved bosons, we cannot resort to such an approach.
Instead we use the local conservation equation of energy to identify the energy current,
similar to the derivation of Eq.\eqref{eq:5} in the case without disorder.
For details of the derivation in the presence of disorder, we refer the reader to Appendix \ref{sec:first-identification}.
The result is identical to Eq.\eqref{eq:5} with the replacements $\ms H \rightarrow \ms H + V$ and $\ms K \rightarrow \ms K + \ms V$.

Expanding the Moyal product, we find to the order $O(\hbar^1)$ two new contributions in the energy current density.
One of these is due to the perturbative shift of energy levels, which (see discussion below Eq.\eqref{eq:16} and details in App. \ref{sec:first-identification})
falls outside of the semiclassical picture: we do not consider it here.
The other one can be identified as the side-jump current,
\begin{align}
  \label{eq:28}
   J_{\mu}^{\rm sj}&= \frac{1}{2} \int_p {\rm Re}\, {\rm Tr}   \left [\ms S^{-1} (\partial_{p_\mu} \ms V)  \ms S  {\sf F}_{\rm d} \right ].
\end{align}
After some lengthy algebra, very similar to that involved in deriving the kinetic equation Eq. \eqref{eq:22} and detailed in Appendix \ref{sec:now-big-semiclassics},
this can be recast into the form
\begin{align}
  \label{eq:29}
  J_{\mu}^{\rm sj}(X) &\approx - \frac {\pi} {\hbar}  \sum_{n} \int_{p,p'} \tilde w(p-p') \, \left | \ms M_{n,n}(p,p') \right |^2 \\
  &\qquad \qquad \qquad \times (p'-p)_\nu \, (\ms \Omega^{\ms W}_{p_\mu p_\nu})_{n,n}(p)  \nn  \\
                      &\times    \delta \Big ( (\ul{\ms K}_{\rm d})_{n,n}(p') - (\ul{\ms K}_{\rm d})_{n,n} ( p) \Big ) \,  (\ul{\ms F}_{\rm d})_{n,n}(X ,p'). \nonumber
\end{align}

In this expression, only the out-of-equilibrium part $\ms g_{\rm d}^{\rm rta}$ of $\ul{\ms F}_{\rm d}$ may yield a finite current:
this follows from basic thermodynamic principles and may also be easily proven directly (see Appendix \ref{sec:formal-proof-vanish}).
Plugging Eq.\eqref{eq:28c} into Eq.\eqref{eq:29}, and again under the usual assumptions of the relaxation time approximation (see App. \ref{sec:solv-boltzm-eqn}),
we obtain $ J_{\mu}^{\rm sj} \rightarrow  J_{\mu}^{\rm sj,rta}$:
\begin{align}
  \label{eq:30}
 J_{\mu}^{\rm sj,rta}(X)
  &\approx  \frac 1 2 \sum_{n} \int_p p_\nu \, (\ms \Omega^{\ms W}_{p_\mu p_\nu})_{n,n}(p) \;  \partial_{\lambda} T \\
  & \qquad \times \partial_T f( (\ul{\ms K}_{\rm d})_{n,n}(X,p)) \; \partial_{p_\lambda} (\ul{\ms K}_{\rm d})_{n,n}(X,p)\nonumber .
\end{align}
Crucially, the fact that $J_{\mu}^{\rm sj,rta} $ (Eq.\eqref{eq:30}) ultimately does not depend on $\stau^{-1}$ (Eq.\eqref{eq:26})
relies on the assumption that disorder is the only source of scattering for bosons in the system.

\subsection{Total contribution of disorder to the thermal Hall conductivity}
\label{sec:total-contr-disord}

We now consider a temperature gradient in the $x$ direction.
The local energy current in the $y$ direction induced by disorder has two contributions, $ { J}_{y}^{\rm dis} =  { J}_{y}^{\rm sj,rta} + { J}_{y}^{\rm gv,sj}$.
Contrarily to the clean case, where the bulk thermal Hall current is in part compensated by boundary magnetization currents \cite{qin_energy_2011, cooper1997},
the current induced by disorder $ { J}_{y}^{\rm dis} $ is a fully \emph{bulk} current, and contributions from the boundaries vanish in the thermodynamic limit.
From Eqs. \eqref{eq:27},\eqref{eq:30} we may then obtain the thermal Hall conductivity,
\begin{equation}
  \label{eq:108}
  \kappa_{xy}^{\rm dis} =  \frac {1} {\partial_x T} \lim_{L\rightarrow \infty} \frac 1 L \int_{-\infty}^{+\infty} \text dx J_{y}(X) \overset{\rm bulk}=  \frac { J_{y}} {\partial_x T} ,
\end{equation}
which takes the form given in the introduction, Eq.\eqref{eq:32} -- see details in Appendix \ref{sec:details-thc}.

We note that because $\ms \Omega^{\ms W}$ is not antisymmetric a priori, $ { J}_{x}^{\rm dis} \neq 0$ a priori,
so that from Eqs.\ \eqref{eq:27},\eqref{eq:30} there may be a $O(\hbar^1)$ contribution to the \emph{longitudinal} thermal conductivity, which otherwise is mainly $O(\hbar^0)$.
In the limit of weak disorder, the $O(\hbar^0)$ contribution is the dominant one, as it is larger by one power of the mean-free path.

\subsection{Analogous derivation for the fermionic Hall conductivities}
\label{sec:analogous-derivation}

\textbf{\textit{Thermal Hall conductivity of fermions --}}
The same steps can be adapted with almost no change to derive the intrinsic and disorder contributions to the thermal Hall effect of fermions (be they neutral or charged).
Importantly, the normalization of eigenvectors is now $\ms S^\dagger \star \ms S = \idmatrix$, so that $\ul{\ms F}_{\rm d}$ has dimension of a number not an energy.
The equilibrium distribution is then $f(\varepsilon)=n_{\rm F}(\varepsilon)$, with $n_{\rm F}(x)=1/(e^{x/T}+1)$ the Fermi function.
Then Eqs.\eqref{eq:31}, \eqref{eq:32} still hold, up to the changes $n_{\rm B}\rightarrow n_{\rm F}$ and $\rm Tr_+ \rightarrow Tr$, the trace over all fermionic bands.

Here it is worth noting, by inspection of the fermionic equivalents of Eqs.\eqref{eq:31}, \eqref{eq:32},
that $\kappa_{xy}^{\rm clean}$ gets contributions from all filled states (cf the $\Theta(\varepsilon-\ul{\ms K}_d)$ factor),
whereas $\kappa_{xy}^{\rm dis}$ gets contributions solely from states close to the Fermi level (cf the $\delta(\varepsilon-\ul{\ms K}_d)$ factor).
In particular, in a gapped fermionic phase, $\kappa_{xy}^{\rm dis}$ exhibits a thermal activation behavior and vanishes at temperatures well below the fermion gap,
while no such behavior is expected for $\kappa_{xy}^{\rm clean}$. This makes it possible, in principle, to distinguish the two types of contributions from each other.
We note that a similar separation of the contributions is not possible in the case of bosons.

As a direct application, it will be interesting to study thermal Hall transport e.g.\ in the quantum spin liquid phase of Kitaev's honeycomb model with fermionic spinons in the presence of disorder.\\

\textbf{\textit{Electrical Hall conductivity of fermions --}}
The same steps can also be adapted with very little change to derive the electrical Hall effect of electrons.
The equilibrium distribution is then $f(\varepsilon)=n_{\rm F}(\varepsilon - {\ms e}\phi)$,
with  $-\ms e$ the electron charge and $\phi$ the electrical potential ($\bs E=-\bs \nabla \phi$ the electric field). 

Then Eqs.\ \eqref{eq:25}--\eqref{eq:27} still hold, up to the replacement
$$\partial_\lambda T \, \partial_T f \quad \rightarrow \quad \partial_\lambda \phi \, \partial_\phi f = {\ms e}E_\lambda \partial_\varepsilon f .$$
The ``jump-forward'' current Eq.\eqref{eq:27} requires the simple change
$$J_{\mu}^{\rm gv,sj} \quad \rightarrow \quad 2\,(-\ms e) \,J_{\mu}^{\rm gv,sj}.$$
From the local conservation of particle charge density in phase space,
\begin{align}
  \label{eq:112}
 \partial_t {\rm Tr} \,\ms F = \frac 1 \hbar {\rm Im}{\rm Tr}\left [ \ms K + \ms V \;\overset \star ,  \;\ms F \right ] ,
\end{align}
one can derive analogously the electronic side-jump current, 
which is Eqs.\eqref{eq:28},\eqref{eq:29} again up to the simple change
$$J_{\mu}^{\rm sj} \quad \rightarrow \quad 2\,(-\ms e) \,J_{\mu}^{\rm sj},$$
and Eq.\eqref{eq:30} holds up to a factor of $2(-\ms e)$ and the replacement 
$\partial_\lambda T \, \partial_T f  \rightarrow  {\ms e}E_\lambda \partial_\varepsilon f $.

The analogous formula to our main result Eq.\eqref{eq:32} is then
\begin{align}
  \label{eq:74analog}
  \sigma_{xy}^{\rm dis} &= {\ms e^2}\, {\rm Tr} \int_p p_\lambda 
                          \;  \partial_\varepsilon f \!\left ( \ul {\ms K}_{\rm d}^{} (p) \right ) \nonumber \\
                          &\times \left [ {\ms \Omega}^{\ms W}_{p_y p_\lambda}(p)\, \partial_{p_x} \ul {\ms K}_{\rm d}^{}(p)
                          -  {\ms \Omega}^{\ms W}_{p_x p_\lambda}(p)\,  \partial_{p_y} \ul {\ms K}_{\rm d}^{}(p) \right ],
\end{align}
see details in Appendix \ref{sec:details-ehc}. 
This formula generalizes the known result for electron side-jump based on a wavepacket analysis \cite{sinitsyn_disorder_2005},
by allowing for impurities with a local matrix structure $\hat W$. In the limit case of fermions in a random scalar potential (where $\ms \Omega^{\ms W}=\ms \Omega$ reduces to the Berry curvature), assuming a quadratic dispersion and a uniform (i.e. momentum-independent) Berry curvature $\Omega^z $, one recovers immediately the familiar result 
$\sigma_{xy}^{\rm dis} =-2 \sigma_{xy}^{\rm clean} $, where $\sigma_{xy}^{\rm clean}={\ms e}^2 p_{\rm F}^2 \Omega^z /4\pi $ is given by the Karplus-Luttinger formula ($p_{\rm F}$ the Fermi momentum).

\subsection{Generalizations and discussion}
\label{sec:generalizations}

The above calculations can be generalized to the case of several independent sets of impurities, labeled $(\iota)$,
each with its own matrix structure $\hat W^{(\iota)}$ and spatial correlation function $w^{(\iota)}$. 
One can also include the effect of interactions as an extra collision integral,
\begin{align}
    \label{eq:202}
 \left ( \text d_t  \,  \ul{\ms F}_{\rm d} \right )_{\rm coll}^{\rm int} (X,p) 
    = \int_{p'} \Upsilon^{\rm int}_{p,p'} \left ( \ms g_{\rm d}(X,p') - \ms g_{\rm d}(X,p)   \right ) ,
\end{align}
to be added on the right-hand side of Eq.\eqref{eq:103},
and assumed to be diagonal in the (implicit) band index $n$.

Details of this generalization are provided in Appendix \ref{sec:append-generalization}. 
The upshot is that our main result, Eq.\eqref{eq:32}, still holds up to the replacement
\begin{align}
\label{eq:245}
    {\ms \Omega}^{\ms W}_{p_\mu p_\nu}(p)
    \quad \mapsto \quad \frac{\sum_{(\iota)}  {\ms \Omega}^{\ms W,(\iota)}_{p_\mu p_\nu}(p) \,\stau_{(\iota)}^{-1}(p)}
    { \stau_{\rm int}^{-1}(p)  + \sum_{(\iota)} \stau_{(\iota)}^{-1}(p) } ,
\end{align}
where $\stau_{\rm int}^{-1}(p)\equiv \int_{p'}\Upsilon^{\rm int}_{p,p'} $ and ${\ms \Omega}^{\ms W,(\iota)}_{p_\mu p_\nu}(p) ,\stau_{(\iota)}^{-1}(p)$ are defined analogously to ${\ms \Omega}^{\ms W}_{p_\mu p_\nu}(p) ,\stau^{-1}(p)$ just by putting a $(\iota)$ index on each instance of $\hat W,\tilde w$ encountered.

By allowing for complex realistic systems with various types of impurities and a minimal inclusion of many-body interaction effects, 
Eq.\eqref{eq:245} plugged into Eq.\eqref{eq:32} accounts for many pieces of physics that are absent from the clean contribution Eq.\eqref{eq:31}.

That the contribution we derived in Eq.\eqref{eq:32} is non-perturbative in the disorder strength (i.e.~it scales as $n_{\rm imp}^0$) is an intriguing,
yet well-known \cite{nagaosa_anomalous_2010} property of coordinate shift effects in anomalous Hall transport, and may be given a very intuitive explanation.
Whenever heat is carried by well-defined quasiparticles, one may write $\kappa \sim c\,v\,l$ with $c$ the specific heat, $v$ the velocity and $l$ the mean-free path.
This holds for any contribution independently, so focussing on the side-jump Hall conductivity,
\begin{equation}
    \kappa_{xy}^{\rm sj} \sim c\,v_\perp^{\rm sj}\,l ,
\end{equation}
where $v_\perp^{\rm sj}$ is now the side-jump transverse velocity, $v_\perp^{\rm sj} \propto n_{\rm imp}$.  
In Eq.\eqref{eq:32} we have assumed that $l$ is governed by disorder scattering, $l^{-1} \propto n_{\rm imp}$, which entails that $\kappa_{xy}^{\rm sj}$ is independent of $n_{\rm imp}$.
More generally, when interactions also contribute to the mean-free path, $\kappa_{xy}^{\rm sj} \propto l_{\rm net}/l_{\rm imp}$:
the ratio of the net mean-free path (including all effects) to the impurity mean-free path.
Precisely this correction is captured in Eq.\eqref{eq:245}.

We note that a recent effort was made to include the effect of band broadening in the intrinsic thermal Hall effect \cite{koyama_thermal_2024}:
this effect is perturbative in the disorder strength, and thus fully distinct from side-jump which appears as a separate, non-perturbative contribution.

\section{Applications}
\label{sec:applications}

We first provide a reminder of the equivalence between our formalism, introduced in Sec.\ref{sec:remind-about-clean},
and general bosonic lattice models expressed in a momentum-space formulation.
As a direct concrete application, we consider the honeycomb-lattice $K\Gamma\Gamma'$ model in a field, in the presence of disorder.
Then, for a different illustration of our results and a more formal discussion, we adopt a more ``low-energy'' perspective,
and consider an application to three slightly different physical models with disorder, the results of which are displayed in Fig. \ref{fig:Applications}.

\subsection{General bosonic hamiltonians in second-quantized notations}
\label{sec:gener-boson-hamilt}

A homogeneous quadratic theory of bosons, in its most general form in momentum space, reads
\begin{align}
 \label{eq:96}
  H_{\rm hom} = \frac \hbar 2 \int_{\bs k} \left ( {\sf b}^\dagger_{i,\bs k} \, , \, {\sf b}_{i,-\bs k} \right ) \begin{pmatrix}  (A_{\bs k} )_{ij} &  (B_{\bs k} )_{ij}  \\  (B_{-\bs k} )^*_{ij} &  (A^\top_{-\bs k} )_{ij}  \end{pmatrix}
                                                                                                                                                                      \begin{pmatrix}
                                                                                                                                                                         {\sf b}_{j,\bs k} \\ {\sf b}^\dagger_{j,-\bs k} 
                                                                                                                                                                       \end{pmatrix}
\end{align}
where ${\sf b}^\dagger_{i,\bs k}$ is a creation operator for a complex boson with flavor index $i$ and momentum $\bs k$.
Hermiticity imposes $A_{\bs k} =A^\dagger_{\bs k}$
and without loss of generality $B_{\bs k} = B^\top_{-\bs k} $.
Real bosons may then be defined as
\begin{align}
  \label{eq:119}
  \Phi^\top_{i,\bs k}= \tfrac 1 {\sqrt {2\hbar }}\left ( {\sf b}^\dagger_{i,\bs k} +{\sf b}_{i,-\bs k} \, , \, i ({\sf b}^\dagger_{i,\bs k} -{\sf b}_{i,-\bs k} )\right ) ,
\end{align}
with two components $\gamma=1,2$ commuting as $\left [ \Phi^1_{i,\bs k}, \Phi^2_{j,\bs k'} \right ] = i \hbar \,\delta_{i,j} \delta_{\bs k+\bs k'}$.
Defining the joint index $a=(\gamma,i)$ so that $\Phi^\top_{a,\bs k}=\left ( \{ \Phi^1_{i,\bs k} \}_{i} , \{ \Phi^2_{i,\bs k} \}_{i} \right )_{a=(\gamma,i)}$,
upon Fourier transformation $ H_{\rm hom} $ may be rewritten precisely as Eq.\eqref{eq:1} with
\begin{align}
  \label{eq:98}
  \ms H_{ab}(x_1,x_2) &= \int_{\bs k} e^{-i \bs k (\bs x_1-\bs x_2)} \begin{pmatrix}  C^+_{\bs k} & D^+_{\bs k} \\
  D^-_{\bs k}  &   C^-_{\bs k}  \end{pmatrix}_{ab} 
\end{align}
and
\begin{subequations}
\begin{align}
\label{eq:E47a}
  C^\pm_{\bs k} &= \tfrac 1 2 ( A_{\bs k} +A^*_{-\bs k} ) \pm \tfrac 1 2 ( B_{\bs k}+ B^*_{-\bs k}) , \\
  \label{eq:E47b}
   D^\pm_{\bs k} &= - \tfrac i 2 ( B_{\bs k}- B^*_{-\bs k}) \pm \tfrac i 2 ( A_{\bs k} - A^*_{-\bs k} ) .
\end{align}
\end{subequations}
From $(C^\pm_{\bs k})^\top = C^\pm_{-\bs k}$ and $(D^\pm_{\bs k})^\top = D^\mp_{-\bs k}$ one can check ${\sf H}_{ab}(x_1,x_2)={\sf H}_{ba}(x_2,x_1)$.
From $(C^\pm_{\bs k})^* = C^\pm_{-\bs k}$ and $(D^\pm_{\bs k})^* = D^\pm_{-\bs k}$ one can check ${\sf H}_{ab}(x_1,x_2) \in \mathbb R$.
Besides, with this choice of basis, the commutation relation between fields $\Phi_a(x)$ is just given by
\begin{align}
  \label{eq:118}
  {\sf \Gamma}(x_1,x_2) = \begin{pmatrix}
  0 & i \,\idmatrix \\ - i \,\idmatrix & 0
\end{pmatrix}\, \delta(x_1-x_2) .
\end{align}

Thus any quadratic theory written in terms of ${\sf b}^\dagger_{i,\bs k} , {\sf b}_{i,\bs k} $ can equivalently be put into the form introduced in Sec.\ref{sec:remind-about-clean}.
This includes, notably, any spin hamiltonian defined on any lattice and expanded into magnons to the quadratic order.
\footnote{Note that here the theory is homogeneous (i.e.\ $\ms H$ only depends on the relative coordinate $x_1-x_2$),
  but it can be made inhomogeneous \textit{a posteriori} by allowing the coefficients ${\sf H}_{ab}$ to depend on $(x_1+x_2)/{2}$ and reintroducing the Moyal product.}

In the case of canonical bosons, such as phonons or magnons in the spin path integral representation \cite{auerbach_interacting},
the Berry curvature is known (see \cite{qin_berry_2012} and discussion of the $d_i$ coefficients in Eq.\eqref{eq:33})
to arise from diagonal elements of the dynamical matrix $\ms K = \ms \Gamma \ms H$.
These are proportional to the $D^\pm_{\bs k}$ matrix elements: in models where $D^\pm_{\bs k}=0$, no intrinsic THE is possible.
Now, for such canonical bosons, the TR operation is $\ms b _{\bs k}{}^({}^\dagger {}^)\rightarrow \ms b _{-\bs k}{}^({}^\dagger {}^)$:
this implies the symmetries $A_{\bs k} \overset{\rm TR}= A^*_{-\bs k}$ and $B_{\bs k} \overset{\rm TR}= B^*_{-\bs k}$.
Thus $D^\pm_{\bs k}$ involves precisely those combinations of $A_{\bs k} ,B_{\bs k}$ that break TR.
\footnote{The situation is less transparent for non-canonical bosons, for instance the Holstein-Primakoff representation where TR acts as
  $\ms b_{\bs k} \rightarrow -\ms b _{-\bs k}^\dagger , \ms b_{\bs k}^\dagger \rightarrow -\ms b _{-\bs k}$.}
This illustrates the general fact that TR breaking is a necessary condition for a THE.

The same procedure and physical arguments hold for the disorder matrix $\hat W(p)$. 
Namely, the homogeneous disorder hamiltonian as described in Sec.\ref{sec:disord-matr-struct} 
can be written most generally as
\begin{align}
    \label{eq:200}
    H_{\rm dis} = \frac \hbar 2 \int_{\bs k} \left ( {\sf b}^\dagger_{i,\bs k} \, , \, {\sf b}_{i,-\bs k} \right ) \begin{pmatrix}  (\tilde A_{\bs k} )_{ij} &  (\tilde B_{\bs k} )_{ij}  \\  (\tilde B_{-\bs k} )^*_{ij} &  (\tilde A^\top_{-\bs k} )_{ij}  \end{pmatrix}
    \begin{pmatrix}   {\sf b}_{j,\bs k} \\ {\sf b}^\dagger_{j,-\bs k} \end{pmatrix}
\end{align}
where $\tilde A_{\bs k}=\tilde A^\dagger_{\bs k}$ and $\tilde B_{\bs k}=\tilde B^\top_{-\bs k}$ can be chosen independently from $A_{\bs k}$ and $B_{\bs k}$. Then, in these notations
\begin{align}
    \label{eq:201}
    \hat W(k) = \begin{pmatrix}  \tilde C^+_{\bs k} & \tilde D^+_{\bs k} \\
  \tilde D^-_{\bs k}  &   \tilde C^-_{\bs k}  \end{pmatrix}
\end{align}
where $\tilde C^\pm_{\bs k},\tilde D^\pm_{\bs k}$ are defined in analogy to Eqs.\eqref{eq:E47a},\eqref{eq:E47b}.
This determines $\ms W = \ms \Gamma \hat W$.

Once $H_{\rm hom}$ and $H_{\rm dis}$ are specified, it is straightforward to obtain both $\ms \Omega_{p_xp_y}$ and $\tfrac 1 2 (\ms \Omega^{\ms W}_{p_xp_y}-\ms \Omega^{\ms W}_{p_yp_x})$:
we remind the reader that a recipe suitable for numerical evaluation is detailed in App.\ref{sec:expl-form-analyt}.

\subsection{Concrete application: honeycomb $K\Gamma\Gamma'$ model in a magnetic field with disorder}
\label{sec:concr-appl-honeyc}

We now appy the above to a particular case of a bosonic Hamiltonian, the ferromagnetic honeycomb-lattice $K\Gamma\Gamma'$ spin model in a field,
whose dynamics in the field-aligned ordered phase can be described in terms of magnon excitations.
The clean thermal Hall conductivity of magnons in this model was extensively studied in Ref.~\cite{PhysRevLett.126.147201},
which discusses in great detail how the interplay of globally broken TR symmetry and anisotropic magnetic exchange generates a magnon Berry curvature and a sizable THE.
We will not reproduce this discussion here, but investigate the role of disorder as random local variations of the parameters of the model.

The microscopic spin hamiltonian reads
\begin{align}
  \label{eq:117}
  H_{\rm spin} = \sum_{\gamma \in \{x,y,z\}}\sum_{\langle \ms i,\ms j\rangle \in \gamma} \mb S_{\ms i}^\top H_{\gamma}\,\mb S_{\ms j} - \sum_{\ms i}\bs h \cdot \mb S_{\ms i} ,
\end{align}
where $\langle \ms i,\ms j\rangle \in \gamma $ labels neighboring sites $\ms i,\ms j$ sharing a bond of type $\gamma \in \{x,y,z\}$,
the field $\bs h$ points along the $\hat a$ axis (for geometrical conventions see Fig.~1 in Ref.~\cite{PhysRevLett.126.147201}),
$\mb S_{\ms i}$ is the spin operator at site $\ms i$, and $H_x,H_y,H_z$ are $3\times 3$ matrices with elements
\begin{align}
  \label{eq:120}
  (H_{\gamma})_{\mu\nu} &= J \,\delta_{\mu\nu} + K \,\delta_{\mu\gamma} \delta_{\nu\gamma} + \Gamma \, |\epsilon_{\mu\nu\gamma}|\nonumber \\
  & \quad  + \Gamma'\, (\delta_{\mu\gamma} +\delta_{\nu\gamma}  )(1 - \delta_{\mu\gamma} \delta_{\nu\gamma} ) 
\end{align}
indexed by $\mu,\nu \in  \{x,y,z\}$. As in Ref.~\cite{PhysRevLett.126.147201}, we consider the pure $K\Gamma\Gamma'$ model where $J=0$, in the ferromagnetic case where $K<0$.
Upon spin wave expansion of the $\mb S_{\ms i}$ operators around the classical field-polarized state, $H_{\rm spin}$ can be recast into the form of Eq.\eqref{eq:96}
where the $2\times 2$ matrices  $A_{\bs k}, B_{\bs k} $ are provided in Ref.~\cite{PhysRevLett.126.147201}, equations (S3b--c) and (S7a)
-- again we do not reproduce the details here and refer the reader to the original paper.

For real materials, all parameters in the model depend on the chemical environment of the spins $\mb S_{\ms i}$:
this holds for magnetic exchange parameters $J,K,\Gamma,\Gamma'$ as well as for $\bs h = {\sf g}\mu_{\rm B}\bs B$ through the g-tensor $\sf g$ (here $\bs B$ is the applied magnetic field).
Since this chemical environment may be affected by disorder, all parameters $(J,K, \Gamma,\Gamma',{\sf g})$ have a random component,
which can be described by a disorder Hamiltonian in the form of Eq.\eqref{eq:200}. In the following we will consider these various instances of disorder independently from each other,
and call ``$X$-type disorder'' (where $X \in \{J,K, \Gamma,\Gamma',{\sf g}\}$) that instance of $H_{\rm dis}$ where $\tilde A_{\bs k}, \tilde B_{\bs k} $
are derived from $A_{\bs k}, B_{\bs k} $ by setting all parameters to zero except $X \rightarrow 1$.
We note that because the computed side-jump effect does not depend on the disorder strength, any global rescaling of $H_{\rm dis}$ leads to the same results.

\begin{figure}[htbp]
\centering
\includegraphics[width=.95\columnwidth]{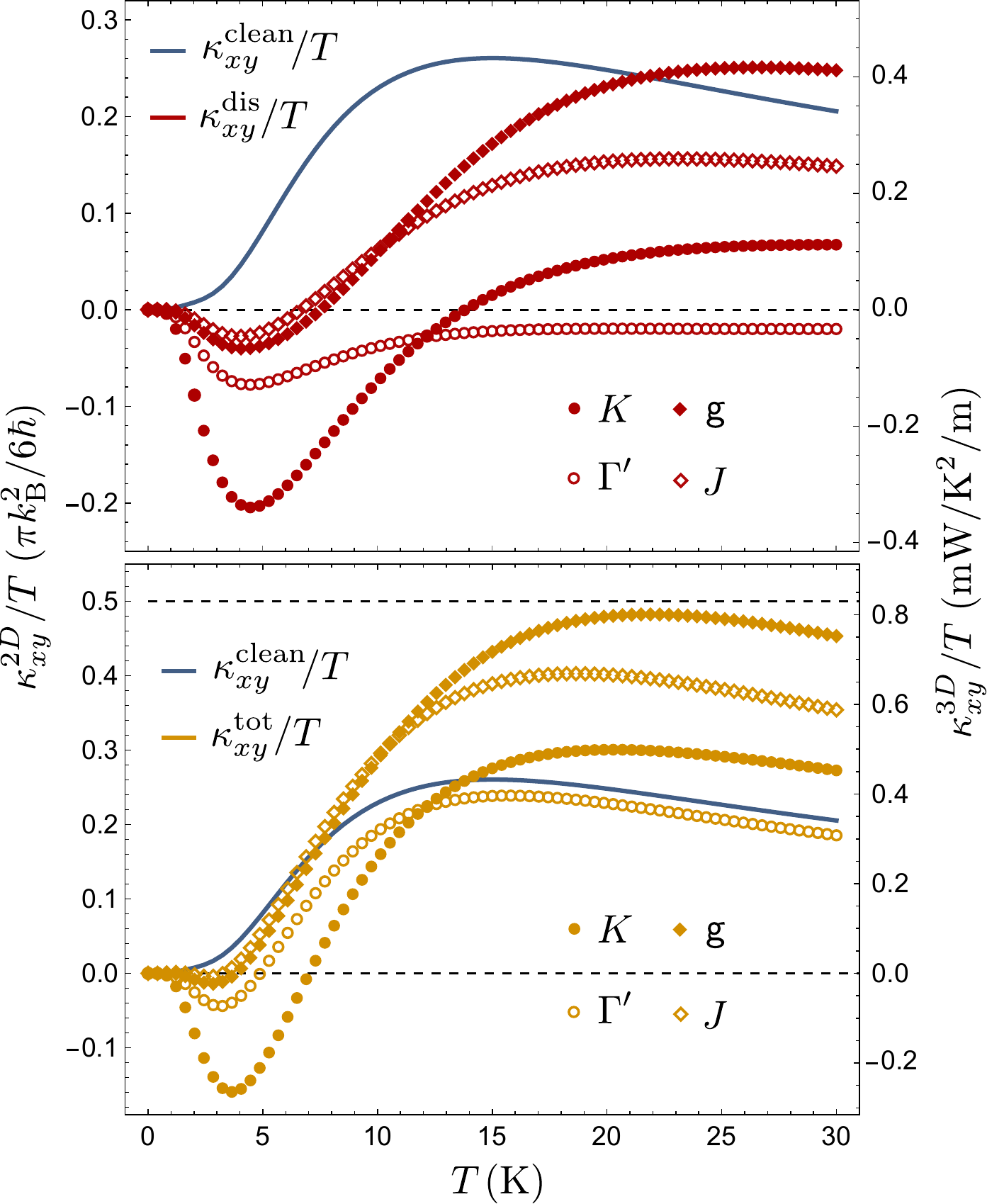}
\caption{ Thermal Hall conductivity of magnons in the disordered $K\Gamma\Gamma'$ model. Blue, red and yellow curves represent the intrinsic, extrinsic and total contribution, respectively.
  Different symbols account for different instances of disorder (for instance filled disks mean $K$-type disorder).
  Right axis: experimental quantity $\kappa_{xy}^{3D}/T$ in SI units. Left axis: equivalent 2D quantity $\kappa_{xy}^{2D}/T$ in units of the thermal Hall quantum.
}
\label{fig:kitaev}
\end{figure}

Using the procedure detailed in App.\ref{sec:expl-form-analyt}, we evaluated the thermal Hall conductivity
given by  Eqs.\eqref{eq:31} ($\kappa_{xy}^{\rm clean}$, intrinsic contribution) and \eqref{eq:32} ($\kappa_{xy}^{\rm dis}$, disorder contribution)
for $K$-type, $\Gamma'$-type \footnote{Because $\Gamma$-type disorder does not satisfy the requirement stated below Eq.\eqref{eq:113},
it cannot be studied within our small-deviation expansion.}, $J$-type and $\sf g$-type disorder.
We used the same relative parameter values $(K,\Gamma,\Gamma')=(-1,0.2,-0.02)$ and $h/S|K|=0.1$ as in Ref.~\cite{PhysRevLett.126.147201} (with $S=1/2$).
We also used the same numerical values in physical units, namely $|K| \approx 80~\rm K$, $\sf g \approx 2.3$ hence $B\approx 3~\rm T$,
and interlayer spacing $\kappa_{xy}^{2D}/\kappa_{xy}^{3D}=5.72 \,\AA$, all inspired by $\alpha$-RuCl$_3$.
The results are displayed in Fig.~\ref{fig:kitaev}. 

We find that different types of disorder can produce very different extrinsic contributions to the THE,
that are typically of the same order of magnitude as the intrinsic contribution, and that can be of the same or the opposite sign (depending in particular on temperature).
In turn, the total thermal Hall conductivity $\kappa_{xy}^{\rm tot} = \kappa_{xy}^{\rm clean} +\kappa_{xy}^{\rm dis}$ is generally quite different from the sole intrinsic contribution,
confirming our qualitative discussions in previous sections.
Qualitative differences appear: notably, we find that $\kappa_{xy}^{\rm tot}$ can undergo a sign change as a function of $T$ that is absent in $\kappa_{xy}^{\rm clean}$.
There are also clear quantitative differences: in some temperature ranges, $\kappa_{xy}$ can be enhanced by a $O(1)$ factor,
that for some types of disorder (especially $\sf g$-type) can be as large as 2 or more.

We note that for the latter type of disorder, our quantitative estimates for $\kappa_{xy}/T$ are comparable to those found in experiments on $\alpha$-RuCl$_3$ \cite{czajka2023planar},
although for different experimental parameters $(B,T)$, and are interestingly close to the half-quantized value that was predicted \cite{PhysRevX.8.031032,PhysRevLett.121.147201} to arise from very different excitations.
We do not claim, from our simple study, any quantitative description of that compound,
which was argued to be rather described by a $JK\Gamma\Gamma'$ \cite{PhysRevB.93.155143} or a $JK\Gamma J_3$ \cite{PhysRevB.93.214431, PhysRevB.96.115103, winter2017breakdown} model.
However, our results clearly show that side-jump effects produce an indispensable contribution to the total thermal Hall effect,
that should always be taken into account in order to reach quantitative agreement with the experimental data.
To achieve this, a realistic description of disorder in the actual compound, that may involve different types of impurities
(that can be incorporated as described in Sec.\ref{sec:generalizations}), will also be necessary.

\subsection{ Formal application and discussion in a low-energy bosonic model}
\label{sec:gener-boson-model}

\subsubsection{Introduction and realization in physical systems}
\label{sec:intr-real-phys}

We now consider a formal application to a less specific instance of a bosonic model,
formulated in the equivalent language of effective field theory.
For the model without disorder, we start from a low-energy perspective and consider a nonlinear sigma-model
for emergent bosonic $\bs m,\bs n$ fields, conjugated with each other. We discuss two physical instances of such a field theory below.
We assume, for simplicity, that the theory is rotationally invariant around the $z$ axis.
Then the most general hamiltonian density preserving this symmetry, 
to the second order in the $\bs m, \bs n$ fields and up to irrelevant terms
\footnote{(in the sense that they contain higher powers of spatial derivatives than another similar term already present in the
hamiltonian density)}, is 
\begin{align}
  \label{eq:33}
  H &= \frac 1 {2\chi} \bs m^2   - \tfrac 1 2 c_1 \bs n \cdot \nabla^2\bs n
      + \tfrac 1 2 c_2 (\nabla \bs n)^2  + \tfrac 1 2\Delta \,\bs n^2 \nonumber \\
    & + \tfrac 1 2 d_1 \bs m \cdot \nabla^2\bs n  + \tfrac 1 2 d_2 (\nabla \bs m) (\nabla \bs n) \nonumber \\
  & + \tfrac 1 2 d_3 \,\bs m \cdot \bs n  -  \tfrac 1 2 d_4\, \hat{\mb z} \cdot (\bs m \times \bs n) ,
\end{align}
where $\chi,\Delta, c_1,c_2,d_1,d_2,d_3,d_4$ are real parameters,
and stability requires $\chi>0, c_1>0, c_1+c_2>0,\Delta \geq 0$).
The coefficients $d_i\, (i=1..4)$, by coupling together conjugated fields $\bs m, \bs n$, will enable the breaking of time reversal symmetry.
Of course, in lattice systems with enforced crystal symmetries, not all of these terms will be allowed
(and possibly other terms, beyond our quadratic low-energy hamiltonian, will turn out to be more relevant).
The actual magnitude of these terms (and of the resulting THE) may also vary vastly from one physical system to the other.
Physical systems described by this hamiltonian Eq.\eqref{eq:33} include:\\

\textbf{\textit{Elasticity with viscosity terms --}} 
In this case, $H$ describes the low-energy lattice dynamics, with $\bs n$ the lattice local displacement field (usually called $\bs u$),
$\bs m$ is the lattice local momentum field (usually called $\bs p$),
$c_1$ and $c_2$ are elasticity moduli \cite{LL7},
$\Delta=0$ is imposed by translational invariance (i.e.\ acoustic phonons are Goldstone bosons),
$\chi^{-1}$ is the lattice mass density (usually called $\rho$), and
$d_i\, (i=1..4)$ are phonon viscosity parameters
\cite{avron_viscosity_1995, barkeshli2012}.
Such terms are known \cite{qin_berry_2012} to generate a phonon Berry curvature and a phonon thermal Hall effect.
Since these viscosity terms break time reversal symmetry, they may only arise in an effective phonon theory beyond pure elasticity, 
for instance from an external magnetic field beyond the Born-Oppenheimer approximation \cite{saito2019berry}
or as a result of spin-phonon interactions in magnetic materials \cite{ye2021}. \\

\textbf{\textit{Magnetic systems --}}
In this case, $H$ describes the low-energy dynamics of spins.
Here we discuss briefly how it can be derived from a microscopic theory, starting from a spin hamiltonian
$H_{\rm spin} = \sum_{\mb r,\mb r'} J_{\mb r,\mb r'}^{ab} S_{\mb r}^a S_{\mb r'}^b$
with $a,b\in\{x,y,z\}$. Assuming a bipartite lattice with Néel antiferromagnetic ordering along $\hat{\mb z}$,
spin fluctuations can be described in terms of the local net magnetization $\bs m$ and local staggered magnetization fluctuations $\bs n$,
which are two-component vectors orthogonal to the Néel order parameter and with each other.
For a concrete illustration one can consider the square lattice and a Kitaev-Heisenberg magnetic exchange
$ J_{\bm r,\bm r'}^{ab} = J \; {\mathbb I}_{\bm r'\in {\rm N}(\bm r)}\delta_{a,b}
                          + K \;  {\mathbb I}_{\bm r'\in {\rm N}_{a}(\bm r)}\delta_{a,b}$
where $ {\rm N}(\bm r)$, respectively $ {\rm N}_{a}(\bm r)$, is the set of nearest neighbors of $\bm r$,
respectively along a bond of type $a \in\{x,y\}$, and $ {\mathbb I}_{\mc E}$ is the indicator of $\mc E$.
To restore rotational symmetry in the effective theory, we select the rotation-invariant component of the hamiltonian,
which then takes the form Eq.\eqref{eq:33} where one identifies $\chi^{-1}=  8 J + 2K ,
c_1 = {\mathtt a}^2 (J+K/8), c_2= {\mathtt a}^2 K/4$, with ${\mathtt a}$ the lattice parameter (see details in Appendix \ref{sec:concrete-spin-model}).
The magnon mass $\Delta$ may result from anisotropic exchange, and non-vanishing $d_i$ coefficients
require the effective breaking of time reversal symmetry in the magnon dynamics, e.g.\ from a ferromagnetic moment
induced by Dzyaloshinskii-Moriya exchange and Zeeman coupling to an exteral magnetic field.
Although these coefficients are essential in the existence of a nonvanishing Hall effect, we do not consider their possible microscopic origin in more detail here,
referring the reader to the literature on magnon Berry curvature \cite{mcclarty2022topological, mook2017topological}.\\

In the following we come back to the low-energy hamiltonian Eq.\eqref{eq:33} without reference to its possible origin from a microscopic spin model.

\subsubsection{The model without disorder}
\label{sec:model-with-disord}

At the stage of defining the commutation relations between $m^a, n^b$, there appears a physical difference between phonons
(where $m^a$ is conjugate with $n^a$) and magnons (where $m^x$ is conjugate with $n^y$ and conversely) at the level of the terms with $d_i$ coefficients.
In the following we focus on the magnon case, and defining $n^1=n^x, n^2=n^y$ and $m^1=m^y, m^2=-m^x$, the commutation relation is
\begin{align}
  \label{eq:34}
  [ n^i, m^j] &= i \delta_{ij} .
\end{align}

Then, up to total derivative terms and to the leading order in $d_1,d_2,d_3$ and gradients,
\begin{align}
  \label{eq:35}
 H_{\rm eff} &=\frac{\chi}{2}\pi^i \pi^i  + \tfrac 1 2 \Delta n^in^i \\
  &\nonumber - \tfrac 1 2 \partial_\mu n^i \left ( c_1 \delta_{\mu,\nu} \delta_{i,j} + c_2  \delta_{i,\mu}\delta_{\nu,j} \right ) \partial_\nu n^j ,
\end{align}
upon introducing the conjugate momentum
\begin{equation}
  \label{eq:36}
  \pi^i= (m^i + \tfrac 1 2 \eta \chi \epsilon_{ij}\partial^2_{\mu\mu}n^j  + \tfrac 1 2 d_3 \epsilon_{ij} n^j   )/\chi 
\end{equation}
with convention $\epsilon_{12}=+$ and where $\eta=d_1-\tfrac 1 2 d_2$. For details see Appendix \ref{sec:effective-low-energy}.

To connect with the general formulation introduced in Sec.\ \ref{sec:remind-about-clean} and \ref{sec:form-with-disord},
we now define the four-vector
\begin{equation}
  \label{eq:37}
  \Phi(r)= \begin{bmatrix} n^1(r), n^2(r), {\pi}^1(r), {\pi}^2(r) \end{bmatrix}^\top  ,
\end{equation}
and considering a homogeneous bulk theory one can readily identify $\ms \Gamma$ and $\ms H$, 
 \begin{subequations}
   \begin{align}
  \label{eq:131a}
  {\sf \Gamma} &=\frac i \chi
  \begin{bmatrix}
    0&\delta_{i,j} \\
    - \delta_{i,j} & - \epsilon_{i,j} (\eta  p^ 2 - d_3/\chi)
  \end{bmatrix}, \\
      \label{eq:131b}
     {\sf H} &=\begin{bmatrix}
 p_\mu \left ( c_1 \delta_{\mu,\nu} \delta_{i,j} + c_2  \delta_{i,\mu}\delta_{\nu,j} \right ) p_\nu +  \Delta \delta_{i,j}&0\\
      0&\chi \delta_{i,j} 
      \end{bmatrix} ,
\end{align}
 \end{subequations}
whence the dynamical matrix $\ms K = \ms \Gamma \ms H$.
 
A boson Berry curvature, leading to an intrinsic Hall effect, is caused by the lower-right block of $ {\sf \Gamma}$, see Eq.\eqref{eq:131a}.
We note that $\eta$ and $d_3$ will play essentially the same role in that regard, but with different momentum dependences. 
Because the $d_3$ term makes the Berry curvature very singular close to zero momentum,
so that physical properties such as the (clean) thermal Hall conductivity may depend strongly on the gap $\Delta$, for convenience we assume in the following $d_3=0$.

These Eqs.\eqref{eq:131a},\eqref{eq:131b} define the clean theory: we now investigate several instances of disorder.

\subsubsection{Three models with disorder}
\label{sec:three-models-disord}

Details of the calculations are provided in App. \ref{sec:two-explicit-sets}, and results for $  \kappa_{xy} ^{\rm dis} $ are displayed in Fig.\ \ref{fig:Applications},
with the numerical values given in Table \ref{tab:parameter-values}.
\begin{table}[htbp]
  \centering
  \begin{tabular}{c|c|c|c|c|c|c|c|c|c}
    \hline\hline
     \; $\chi$ \;  & \;  $c_1$  \; &\;   $c_2$ \;  & \;  $d_1$ \;  &  \; $d_2$\;   & \;  $d_3$ \;  & \;  $\Delta$ \; &  \; $\tilde \eta$ \; & \;  ${\mathtt a}$ \; \\   \hline
   ${\mathtt{1.0}}$ & $\mathtt{1.0}$ & $\mathtt{0.25}$ & $\mathtt{-0.2}$ &$\mathtt{+0.1}$ & $\mathtt{0.0}$ & $\mathtt{0.01}$ &  $\mathtt{0.2}$ & $\mathtt 1.0$ \\
    \hline\hline
  \end{tabular}
  \caption{Parameter values used for the numerical evaluations. Momentum integrals are computed over $\bs p \in [-\pi/{\mathtt a},+\pi/{\mathtt a}]^2$.
We note that in the particular case where $d_1=0$, $d_2=0$, $d_3=0$ and $\tilde \eta=0$, TR symmetry is preserved and no THE is allowed. }
  \label{tab:parameter-values}
\end{table}

\textbf{\textit{Case where $\ms W = \ms K$ --}}
For \emph{any} clean theory $(\ms H, \ms \Gamma)$ with disorder $\hat W = \hat{\ms H}$ and thus $\ms W = \ms K$,
corresponding physically to locally modulating the clean hamiltonian by a random global prefactor,
interesting simplifications occur (see Appendix \ref{sec:case-where-disorder}) and the disorder-induced curvature identifies with the usual Berry curvature:
\begin{align}
  \label{eq:38}
  \ms \Omega^{\ms K}_{p_\mu p_\nu}(p) &=  (\ms \Omega_{p_\mu p_\nu})_{n}(p) .
    \end{align}
 The disorder-induced thermal Hall conductivity Eq.\ \eqref{eq:32} then reads
    \begin{align}
      \label{eq:39}
        \kappa_{xy} ^{\rm dis} &= \frac 1 {T} {\rm Tr}_+ \int_p  {\ms \Omega}^{}_{p_x p_y}(p)\,
                          \left [ p_x\partial_{p_x} {\ms K}_{\rm d}^{}(p)+ p_y  \partial_{p_y} {\ms K}_{\rm d}^{}(p) \right ]\nonumber \\
    &\qquad \times \int \text d \varepsilon \, \delta (\varepsilon-  {\sf K}_{\rm d}^{} ) \,\varepsilon^2
                           \partial_\varepsilon n_{\rm B} \left ( \varepsilon, T\right ),
    \end{align}
 to be compared with Eq.\eqref{eq:31}.
 Crucially, this result is general in that it relies only on the constraint $\ms W = \ms K$ but not on the clean model $(\ms H, \ms \Gamma)$ chosen
 -- in Fig. \ref{fig:Applications} we display the result Eq.\eqref{eq:39} in the particular case of the clean theory Eqs.\eqref{eq:131a},\eqref{eq:131b}. \\

\textbf{\textit{Case with both intrinsic and extrinsic chirality --}}
Starting again from the clean model Eqs.\eqref{eq:131a},\eqref{eq:131b}, which contains ``intrinsic chirality'' for $\eta\neq 0$,
we now consider a different instance of disorder, which will be a source of ``extrinsic chirality''. Namely, 
\begin{align}
  \label{eq:40}
  H_{\rm dis} &= - \tilde \eta \epsilon_{ij} m^i \partial^2_{\mu\mu}n^j = \frac 1 2 \Phi \hat W \Phi, \\
    \label{eq:41}
  \hat W &= \chi \tilde \eta p^2 \begin{bmatrix}
    0& -i\sigma^y\\
   i\sigma^y & 0
  \end{bmatrix} 
\end{align}
where $\tilde \eta$ can be interpreted as a local random fluctuation amplitude of $\eta$.
Let us recall that in the clean case, $\eta$ is responsible for the breaking of TR symmetry and the generation of Berry curvature:
in the disordered case, that role will be shared by $\tilde \eta$.

Details of the calculations are provided in App.\ \ref{sec:application-3:-now}: the third term in Eq.\eqref{eq:24} vanishes, the first one is order $\eta + O(\eta^3)$
and the second one is order $\eta^{-1} + O(\eta)$, singular in the limit $\eta \rightarrow 0$.
As already mentioned in generality, the disorder-induced curvature does not depend on the strength of disorder: $\tilde \eta$ does not appear in the final result.

\textbf{\textit{Case where $\kappa_{xy}^{\rm clean}=0$ --}}
Starting from a clean model which is Eqs.\eqref{eq:131a},\eqref{eq:131b} in the particular case $\eta \rightarrow 0$,
that is to say $\ms\Gamma \rightarrow \ms \Gamma |_{\eta=0}$ whence $\ms K \rightarrow \ms K |_{\eta=0}$,
one finds $\ms \Omega_{p_\mu p_\nu}(p) = 0\;\forall p$ since both inversion and time reversal symmetry are preserved.
This implies the absence of an intrinsic thermal Hall effect, $\kappa^{\rm clean}_{xy}=0$:
any instance of thermal Hall effect will originate solely from extrinsic chirality generated by disorder.
To illustrate this, we consider
\begin{align}
  \label{eq:87}
   \hat{W} &= \ms H + \chi \tilde \eta p^2 \begin{bmatrix} 0& -i \sigma^y\\ i \sigma^y & 0 \end{bmatrix},
\end{align}
where only the second term generates chirality, and the first term ensures regular behavior -- see App.\ \ref{sec:application-3:-now} for details.
Only the second term in $ (\ms \Omega^{\ms W}_{p_\mu p_\nu})_{n}(p) $, Eq.\eqref{eq:24}, is not zero,
and $\ms \Omega^{\ms W} \propto \tilde \eta p^2$.
This causes a finite thermal Hall conductivity to arise, as shown in Fig. \ref{fig:Applications}(c).
We state again that this stems purely from the elastic scattering with impurities of bosons with zero (intrinsic) Berry curvature.

\subsubsection{Results and discussion}
\label{sec:discussion}

These three examples show that the side-jump contribution to $\kappa_{xy}$ gives an order one correction to the intrinsic part.
Since both its amplitude and its sign depend on the specifics of the disorder hamiltonian, it can enhance or reduce $\kappa_{xy}$ and may even lead to sign changes.
In addition, even in the case of a trivial (i.e.~TR preserving) clean system, special forms of disorder like local viscosity terms can lead to a non-vanishing THE.

The temperature dependences of $\kappa_{xy}^{\rm clean}$ and $\kappa_{xy}^{\rm dis}$ are imposed by the distribution in momentum space of $\ms \Omega_{p_xp_y}$ and $\ms \Omega_{p_xp_y}^{\ms W}$,
respectively, and by the bosonic dispersion $\ul{\ms K}_{\rm d}$. An approximately linear dispersion $\ul{\ms K}_{\rm d} \sim k$
with a curvature going as $\ms \Omega^{(\ms W)}\sim 1/k$ yield a thermal Hall conductivity $\kappa_{xy} \sim T^2$, as is seen in all three plots.
The sharp peak in case (b) is governed by the interplay between the momentum distribution of  $\ms \Omega^{\ms W}$ and the dispersion gap $\Delta$, so that its position is approximately $T \sim \Delta$.
The maximum of the ``broad peak'' seen in all three plots corresponds to the saturation of the integral of the Berry curvature,
which in our model is imposed by the Brillouin zone boundaries (i.e. the maximum occurs around $T \approx \theta_{\rm D}$ the Debye temperature of the boson bands).
The orders of magnitude we provide here are purely qualitative, 
given that the quantitative determination of the Berry curvature (and similarly disorder-induced curvatures) of bosonic excitations
in concrete models of materials is a challenge in its own right \cite{saito2019berry}.

Even these qualitative features are of course very model-dependent.
In particular, in our examples, $\ms \Omega_{p_x,p_y}$ and $\ms \Omega^{\ms W}_{p_x,p_y}$ both are concentrated close to the gamma-point $k=0$,
but it would be interesting, as another direct application, to study cases, for instance hybridized magnon-phonon bands,
where one or both of these quantities is largest at a finite momentum, e.g.~around the avoided-crossing point.

\subsection{Experimental aspects}
\label{sec:experimental-aspects}

On the experimental side, the theory of side-jump developed above could help with the interpretation of thermal Hall measurements, 
especially in systems where heat is carried by chiral bosons such as spin excitations in quantum magnets.
It is in principle possible to distinguish side-jump from other mechanisms of THE in cases where sample dependence for one given compound is observed, or by investigating dependence upon doping/substitutions.
One key experimental feature of side-jump THE, due to the relation $\kappa_{xy}^{\rm sj} \propto l_{\rm net}/l_{\rm imp}$,
is that $\kappa_{xy}$ as a function of $\kappa_{xx}$ is, as a rule of thumb, \emph{decreasing} or constant, depending on whether $l_{\rm net}$ is dominated by interactions or disorder, respectively.
This is at odds with skew-scattering mechanisms, where a thermal Hall resistivity $\rho_{xy}$ is generated from either disorder or interactions;
then the thermal Hall conductivity $\kappa_{xy}\approx - \rho_{xy}/ \rho_{xx}^2$ is, as a rule of thumb, \emph{increasing} or constant as a function of $\kappa_{xx} \approx 1/\rho_{xx}$,
depending on whether skew-scattering originates from interactions or disorder, respectively.
This is also at odds with the intrinsic mechanism, where $\kappa_{xy}$ simply does not depend on $\kappa_{xx}$.
These features are summarized in Table \ref{tab:rules-of-thumb}.

\begin{table}[htbp]
  \centering
  \begin{tabular}{c|c}
    \hline\hline
  Mechanism generating THE & Expected $\kappa_{xy}$ vs $\kappa_{xx}$ \\   \hline
    Intrinsic & Constant \\
    Skew-scattering (interactions) & Increasing ($\kappa_{xy} \overset \sim \propto \kappa_{xx}^2$) \\
    Skew-scattering (disorder) & $\sim$ Constant \\
    Side-jump (case $l_{\rm net} \ll l_{\rm imp}$) & Decreasing ($\kappa_{xy} \overset \sim \propto \kappa_{xx}^{-1}$)\\
    Side-jump (case $l_{\rm net}\approx l_{\rm imp}$) & $\sim$ Constant \\
    \hline\hline
  \end{tabular}
  \caption{Compared variations of $\kappa_{xy}, \kappa_{xx}$ upon tuning disorder (expected behavior).}
  \label{tab:rules-of-thumb}
\end{table}

Experimental examples of compounds where $\kappa_{xy}$ is smaller for larger $\kappa_{xx}$ include the kagome disordered antiferromagnet Volborthite \cite{watanabe2016emergence}, where the reported THE was attributed to spin excitations.
Another example are pyrochlore terbium oxides, $\rm Tb_2Ti_2O_7$ \cite{hirschberger2015large}
and $\rm (Tb_{.3}Y_{.7})_2Ti_2O_7$ \cite{hirokane2019}, where upon substituting magnetic $\rm Tb^{3+}$ ions with nonmagnetic $\rm Y^{3+}$ one observes that $\kappa_{xx}$ slightly decreases while $\kappa_{xy}$ noticeably increases.
\footnote{We note that the physics of thermal transport in fully nonmagnetic $\rm Y_2Ti_2O_7$ \cite{sharma2024phonon} appears to be quite different.}
While arguments have been put forward to discard either phonons or spin excitations as heat carriers in this family of compounds, we note that these observations are compatible with a non-negligible role of side-jump.
A $\kappa_{xy}$ smaller for larger $\kappa_{xx}$ was also found \cite{PhysRevMaterials.7.114403} in the Kitaev compound $\alpha \text{-RuCl}_3$
  -- a fact that may be considered in regard with our Sec.\ref{sec:concr-appl-honeyc}.

As a final note, while decreasing $\kappa_{xy}$ vs $\kappa_{xx}$ is in principle a \emph{sufficient} argument supporting a side-jump mechanism, it is not necessary.
Indeed, sample dependence with $\kappa_{xy},\kappa_{xx}$ roughly proportional to each other has been reported in kagome antiferromagnets Ca-kapellasite \cite{doki2018spin} and Cd-kapellasite \cite{akazawa2020thermal}
as well as in honeycomb ferromagnet $\rm VI_3$ \cite{zhang2021anomalous}, and in all three cases was attribued to spurious geometrical factors afflicting the data:
unfortunately, such effects obscure possible signatures of side-jump mechanisms.

\section{Conclusion}
\label{sec:conclusion}

We have derived systematically the side-jump contribution to the bosonic THE, Eq.\eqref{eq:32}. 
As a side product, we also found a generalized formula of the side-jump contribution to the electronic AHE from generic types of impurities
(i.e., any local deformation of the Hamiltonian), see Eq.~\ref{eq:74analog}. This contribution is a sizeable correction to the intrinsic part of the Hall response,
that is non-perturbative i.e.\ independent of the density of impurities (in the absence of interactions). Thus, the prevailing interpretation of thermal Hall experiments with Eq.\eqref{eq:31} alone is always incomplete,
and we have shown quantitatively (both formally and in a realistic spin model for $\alpha\text{-RuCl}_3$) that extrinsic contributions can be of similar magnitude.
This is especially so in bosonic systems, where so far  Eq.\eqref{eq:31} has been most commonly used alone.
Therefore, we argue that any reasonable estimate of the THE in the absence of interactions should involve both Eq.\eqref{eq:31} and Eq.\eqref{eq:32} together
-- the latter can be evaluated straightforwardly, for instance using the procedure detailed in App. \ref{sec:expl-form-analyt}.

  This independence of Eq.\eqref{eq:32} on the impurity density holds \emph{only if} the carriers' mean free path is governed by disorder scattering.
For sufficiently strong interactions, such that the mean free path is independent of impurity density, the side-jump contribution becomes perturbative ($\propto n_{\rm imp}$)
and might be negligible -- the price to pay is that interactions, in turn, will play a role in the Hall conductivity, e.g.\ via many-body skew-scattering mechanisms.
Ultimately, for bosonic systems there is no reason to believe that Eq.\eqref{eq:31} alone should ever provide a good approximation of the thermal Hall conductivity in any experiment.
  In particular, should there be a near-``universal'' mechanism for THE in trivial non-magnetic insulators \cite{behnia2025phononthermalhalllattice}, interactions should be involved to some extent.

  Beyond thermal Hall transport, our gauge invariant formulation of a kinetic equation for bosons including disorder scattering promises many useful applications.
  For instance it naturally allows for the calculation of other transport coefficients, such as the spin Nernst effect. It also points to several concrete directions for future research:
  (1) The kinetic formalism should be extended for calculating nonlinear response functions -- for a recent treatment of the clean case see~\cite{park2025quantum}.
  (2) It can also be generalized to include time-dependent (random) terms in the Hamiltonian, allowing in particular to study in a minimal version
  the dissipative quantum dynamics of systems with nontrivial band geometry.
  (3) Our results indicate that quantitative predictions for the THE in the intrinsic regime will require knowledge of the operatorial form of disorder.
  Thus, it will be useful to interface our theory with quantitative ab-initio efforts to obtain not only the clean features (energies, wavefunctions)
  of the phonons or magnons \cite{dhakal2024theory}, but also input for the effective disorder Hamiltonian, as we describe in Sec.\ref{sec:disord-matr-struct}.
      (4) A challenging next step will be the inclusion of interaction effects between bosons, which will not only provide yet another contribution to the thermal Hall effect from skew-scattering, but will also alter the side-jump contribution from impurities, and may even generate a side-jump-like contribution of their own by mixing many-body dynamics with band geometrical properties.

The promise of THE measurements in magnetic insulators is to learn about the properties of charge neutral low energy excitations, most commonly of bosonic nature. Disentangling the different intrinsic and extrinsic contributions will be crucial for a proper understanding. In this sense our work, 
by providing means to compute extrinsic, disorder-dependent contributions that might otherwise obscure some more ``universal'' features, serves as a rigorous step towards a full quantitative theory.

\section*{Acknowledgements}
\label{sec:acknowledgements}

L.M. thanks Elio König and Joji Nasu for a useful discussion. 
L.M. thanks Lucile Savary and Leon Balents for helpful comments and a previous collaboration that led to this work. We thank Peng Rao for useful discussion and  acknowledge support from the Imperial-TUM flagship
partnership, from the Deutsche Forschungsgemeinschaft 
(DFG, German Research Foundation) under Germany’s Excellence Strategy–EXC– 2111–390814868, DFG grants No. KN1254/1-2, KN1254/2-1, and TRR 360 - 492547816, as well as the Munich Quantum Valley, which is supported by the
Bavarian state government with funds from the Hightech Agenda Bayern Plus.

\bibliography{library}

\begin{thebibliography}{94}%
\makeatletter
\providecommand \@ifxundefined [1]{%
 \@ifx{#1\undefined}
}%
\providecommand \@ifnum [1]{%
 \ifnum #1\expandafter \@firstoftwo
 \else \expandafter \@secondoftwo
 \fi
}%
\providecommand \@ifx [1]{%
 \ifx #1\expandafter \@firstoftwo
 \else \expandafter \@secondoftwo
 \fi
}%
\providecommand \natexlab [1]{#1}%
\providecommand \enquote  [1]{``#1''}%
\providecommand \bibnamefont  [1]{#1}%
\providecommand \bibfnamefont [1]{#1}%
\providecommand \citenamefont [1]{#1}%
\providecommand \href@noop [0]{\@secondoftwo}%
\providecommand \href [0]{\begingroup \@sanitize@url \@href}%
\providecommand \@href[1]{\@@startlink{#1}\@@href}%
\providecommand \@@href[1]{\endgroup#1\@@endlink}%
\providecommand \@sanitize@url [0]{\catcode `\\12\catcode `\$12\catcode
  `\&12\catcode `\#12\catcode `\^12\catcode `\_12\catcode `\%12\relax}%
\providecommand \@@startlink[1]{}%
\providecommand \@@endlink[0]{}%
\providecommand \url  [0]{\begingroup\@sanitize@url \@url }%
\providecommand \@url [1]{\endgroup\@href {#1}{\urlprefix }}%
\providecommand \urlprefix  [0]{URL }%
\providecommand \Eprint [0]{\href }%
\providecommand \doibase [0]{http://dx.doi.org/}%
\providecommand \selectlanguage [0]{\@gobble}%
\providecommand \bibinfo  [0]{\@secondoftwo}%
\providecommand \bibfield  [0]{\@secondoftwo}%
\providecommand \translation [1]{[#1]}%
\providecommand \BibitemOpen [0]{}%
\providecommand \bibitemStop [0]{}%
\providecommand \bibitemNoStop [0]{.\EOS\space}%
\providecommand \EOS [0]{\spacefactor3000\relax}%
\providecommand \BibitemShut  [1]{\csname bibitem#1\endcsname}%
\let\auto@bib@innerbib\@empty
\bibitem [{\citenamefont {Klitzing}\ \emph {et~al.}(1980)\citenamefont
  {Klitzing}, \citenamefont {Dorda},\ and\ \citenamefont
  {Pepper}}]{klitzing1980new}%
  \BibitemOpen
  \bibfield  {author} {\bibinfo {author} {\bibfnamefont {K~v}\ \bibnamefont
  {Klitzing}}, \bibinfo {author} {\bibfnamefont {Gerhard}\ \bibnamefont
  {Dorda}}, \ and\ \bibinfo {author} {\bibfnamefont {Michael}\ \bibnamefont
  {Pepper}},\ }\bibfield  {title} {\enquote {\bibinfo {title} {New method for
  high-accuracy determination of the fine-structure constant based on quantized
  hall resistance},}\ }\href@noop {} {\bibfield  {journal} {\bibinfo  {journal}
  {Physical review letters}\ }\textbf {\bibinfo {volume} {45}},\ \bibinfo
  {pages} {494} (\bibinfo {year} {1980})}\BibitemShut {NoStop}%
\bibitem [{\citenamefont {Thouless}\ \emph {et~al.}(1982)\citenamefont
  {Thouless}, \citenamefont {Kohmoto}, \citenamefont {Nightingale},\ and\
  \citenamefont {den Nijs}}]{thouless1982quantized}%
  \BibitemOpen
  \bibfield  {author} {\bibinfo {author} {\bibfnamefont {David~J}\ \bibnamefont
  {Thouless}}, \bibinfo {author} {\bibfnamefont {Mahito}\ \bibnamefont
  {Kohmoto}}, \bibinfo {author} {\bibfnamefont {M~Peter}\ \bibnamefont
  {Nightingale}}, \ and\ \bibinfo {author} {\bibfnamefont {Marcel}\
  \bibnamefont {den Nijs}},\ }\bibfield  {title} {\enquote {\bibinfo {title}
  {Quantized hall conductance in a two-dimensional periodic potential},}\
  }\href@noop {} {\bibfield  {journal} {\bibinfo  {journal} {Physical review
  letters}\ }\textbf {\bibinfo {volume} {49}},\ \bibinfo {pages} {405}
  (\bibinfo {year} {1982})}\BibitemShut {NoStop}%
\bibitem [{\citenamefont {Nagaosa}\ \emph {et~al.}()\citenamefont {Nagaosa},
  \citenamefont {Sinova}, \citenamefont {Onoda}, \citenamefont {MacDonald},\
  and\ \citenamefont {Ong}}]{nagaosa_anomalous_2010}%
  \BibitemOpen
  \bibfield  {author} {\bibinfo {author} {\bibfnamefont {Naoto}\ \bibnamefont
  {Nagaosa}}, \bibinfo {author} {\bibfnamefont {Jairo}\ \bibnamefont {Sinova}},
  \bibinfo {author} {\bibfnamefont {Shigeki}\ \bibnamefont {Onoda}}, \bibinfo
  {author} {\bibfnamefont {A.~H.}\ \bibnamefont {MacDonald}}, \ and\ \bibinfo
  {author} {\bibfnamefont {N.~P.}\ \bibnamefont {Ong}},\ }\bibfield  {title}
  {\enquote {\bibinfo {title} {Anomalous hall effect},}\ }\href {\doibase
  10.1103/RevModPhys.82.1539} {\bibfield  {journal} {\bibinfo  {journal} {Rev.
  Mod. Phys.}\ }\textbf {\bibinfo {volume} {82}},\ \bibinfo {pages}
  {1539--1592}}\BibitemShut {NoStop}%
\bibitem [{\citenamefont {Luttinger}(1964)}]{luttinger1964theory}%
  \BibitemOpen
  \bibfield  {author} {\bibinfo {author} {\bibfnamefont {J.~M.}\ \bibnamefont
  {Luttinger}},\ }\bibfield  {title} {\enquote {\bibinfo {title} {Theory of
  thermal transport coefficients},}\ }\href {\doibase
  10.1103/PhysRev.135.A1505} {\bibfield  {journal} {\bibinfo  {journal} {Phys.
  Rev.}\ }\textbf {\bibinfo {volume} {135}},\ \bibinfo {pages} {A1505--A1514}
  (\bibinfo {year} {1964})}\BibitemShut {NoStop}%
\bibitem [{\citenamefont {Kitaev}(2006)}]{kitaev2006anyons}%
  \BibitemOpen
  \bibfield  {author} {\bibinfo {author} {\bibfnamefont {Alexei}\ \bibnamefont
  {Kitaev}},\ }\bibfield  {title} {\enquote {\bibinfo {title} {Anyons in an
  exactly solved model and beyond},}\ }\href@noop {} {\bibfield  {journal}
  {\bibinfo  {journal} {Annals of Physics}\ }\textbf {\bibinfo {volume}
  {321}},\ \bibinfo {pages} {2--111} (\bibinfo {year} {2006})}\BibitemShut
  {NoStop}%
\bibitem [{\citenamefont {Kane}\ and\ \citenamefont
  {Fisher}(1997)}]{kane1997quantized}%
  \BibitemOpen
  \bibfield  {author} {\bibinfo {author} {\bibfnamefont {CL}~\bibnamefont
  {Kane}}\ and\ \bibinfo {author} {\bibfnamefont {Matthew~PA}\ \bibnamefont
  {Fisher}},\ }\bibfield  {title} {\enquote {\bibinfo {title} {Quantized
  thermal transport in the fractional quantum hall effect},}\ }\href@noop {}
  {\bibfield  {journal} {\bibinfo  {journal} {Physical Review B}\ }\textbf
  {\bibinfo {volume} {55}},\ \bibinfo {pages} {15832} (\bibinfo {year}
  {1997})}\BibitemShut {NoStop}%
\bibitem [{\citenamefont {Kasahara}\ \emph {et~al.}(2018)\citenamefont
  {Kasahara}, \citenamefont {Ohnishi}, \citenamefont {Mizukami}, \citenamefont
  {Tanaka}, \citenamefont {Ma}, \citenamefont {Sugii}, \citenamefont {Kurita},
  \citenamefont {Tanaka}, \citenamefont {Nasu}, \citenamefont {Motome} \emph
  {et~al.}}]{kasahara2018majorana}%
  \BibitemOpen
  \bibfield  {author} {\bibinfo {author} {\bibfnamefont {Yuichi}\ \bibnamefont
  {Kasahara}}, \bibinfo {author} {\bibfnamefont {Tsuneya}\ \bibnamefont
  {Ohnishi}}, \bibinfo {author} {\bibfnamefont {Yuta}\ \bibnamefont
  {Mizukami}}, \bibinfo {author} {\bibfnamefont {Osamu}\ \bibnamefont
  {Tanaka}}, \bibinfo {author} {\bibfnamefont {Sixiao}\ \bibnamefont {Ma}},
  \bibinfo {author} {\bibfnamefont {Kaori}\ \bibnamefont {Sugii}}, \bibinfo
  {author} {\bibfnamefont {Nobuyuki}\ \bibnamefont {Kurita}}, \bibinfo {author}
  {\bibfnamefont {Hidekazu}\ \bibnamefont {Tanaka}}, \bibinfo {author}
  {\bibfnamefont {Joji}\ \bibnamefont {Nasu}}, \bibinfo {author} {\bibfnamefont
  {Yukitoshi}\ \bibnamefont {Motome}},  \emph {et~al.},\ }\bibfield  {title}
  {\enquote {\bibinfo {title} {Majorana quantization and half-integer thermal
  quantum hall effect in a kitaev spin liquid},}\ }\href@noop {} {\bibfield
  {journal} {\bibinfo  {journal} {Nature}\ }\textbf {\bibinfo {volume} {559}},\
  \bibinfo {pages} {227--231} (\bibinfo {year} {2018})}\BibitemShut {NoStop}%
\bibitem [{\citenamefont {Paul}\ \emph {et~al.}(2024)\citenamefont {Paul},
  \citenamefont {Tiwari}, \citenamefont {Melcer}, \citenamefont {Umansky},\
  and\ \citenamefont {Heiblum}}]{paul2024topological}%
  \BibitemOpen
  \bibfield  {author} {\bibinfo {author} {\bibfnamefont {Arup~Kumar}\
  \bibnamefont {Paul}}, \bibinfo {author} {\bibfnamefont {Priya}\ \bibnamefont
  {Tiwari}}, \bibinfo {author} {\bibfnamefont {Ron}\ \bibnamefont {Melcer}},
  \bibinfo {author} {\bibfnamefont {Vladimir}\ \bibnamefont {Umansky}}, \ and\
  \bibinfo {author} {\bibfnamefont {Moty}\ \bibnamefont {Heiblum}},\ }\bibfield
   {title} {\enquote {\bibinfo {title} {Topological thermal hall conductance of
  even-denominator fractional states},}\ }\href@noop {} {\bibfield  {journal}
  {\bibinfo  {journal} {Physical Review Letters}\ }\textbf {\bibinfo {volume}
  {133}},\ \bibinfo {pages} {076601} (\bibinfo {year} {2024})}\BibitemShut
  {NoStop}%
\bibitem [{\citenamefont {Onose}\ \emph {et~al.}(2010)\citenamefont {Onose},
  \citenamefont {Ideue}, \citenamefont {Katsura}, \citenamefont {Shiomi},
  \citenamefont {Nagaosa},\ and\ \citenamefont
  {Tokura}}]{onose2010observation}%
  \BibitemOpen
  \bibfield  {author} {\bibinfo {author} {\bibfnamefont {Y}~\bibnamefont
  {Onose}}, \bibinfo {author} {\bibfnamefont {T}~\bibnamefont {Ideue}},
  \bibinfo {author} {\bibfnamefont {H}~\bibnamefont {Katsura}}, \bibinfo
  {author} {\bibfnamefont {Y}~\bibnamefont {Shiomi}}, \bibinfo {author}
  {\bibfnamefont {N}~\bibnamefont {Nagaosa}}, \ and\ \bibinfo {author}
  {\bibfnamefont {Y}~\bibnamefont {Tokura}},\ }\bibfield  {title} {\enquote
  {\bibinfo {title} {Observation of the magnon hall effect},}\ }\href@noop {}
  {\bibfield  {journal} {\bibinfo  {journal} {Science}\ }\textbf {\bibinfo
  {volume} {329}},\ \bibinfo {pages} {297--299} (\bibinfo {year}
  {2010})}\BibitemShut {NoStop}%
\bibitem [{\citenamefont {Ideue}\ \emph {et~al.}(2012)\citenamefont {Ideue},
  \citenamefont {Onose}, \citenamefont {Katsura}, \citenamefont {Shiomi},
  \citenamefont {Ishiwata}, \citenamefont {Nagaosa},\ and\ \citenamefont
  {Tokura}}]{ideue2012effect}%
  \BibitemOpen
  \bibfield  {author} {\bibinfo {author} {\bibfnamefont {T}~\bibnamefont
  {Ideue}}, \bibinfo {author} {\bibfnamefont {Y}~\bibnamefont {Onose}},
  \bibinfo {author} {\bibfnamefont {H}~\bibnamefont {Katsura}}, \bibinfo
  {author} {\bibfnamefont {Y}~\bibnamefont {Shiomi}}, \bibinfo {author}
  {\bibfnamefont {S}~\bibnamefont {Ishiwata}}, \bibinfo {author} {\bibfnamefont
  {N}~\bibnamefont {Nagaosa}}, \ and\ \bibinfo {author} {\bibfnamefont
  {Y}~\bibnamefont {Tokura}},\ }\bibfield  {title} {\enquote {\bibinfo {title}
  {Effect of lattice geometry on magnon hall effect in ferromagnetic
  insulators},}\ }\href@noop {} {\bibfield  {journal} {\bibinfo  {journal}
  {Physical Review B—Condensed Matter and Materials Physics}\ }\textbf
  {\bibinfo {volume} {85}},\ \bibinfo {pages} {134411} (\bibinfo {year}
  {2012})}\BibitemShut {NoStop}%
\bibitem [{\citenamefont {Hirschberger}\ \emph
  {et~al.}(2015{\natexlab{a}})\citenamefont {Hirschberger}, \citenamefont
  {Krizan}, \citenamefont {Cava},\ and\ \citenamefont
  {Ong}}]{hirschberger2015large}%
  \BibitemOpen
  \bibfield  {author} {\bibinfo {author} {\bibfnamefont {Max}\ \bibnamefont
  {Hirschberger}}, \bibinfo {author} {\bibfnamefont {Jason~W}\ \bibnamefont
  {Krizan}}, \bibinfo {author} {\bibfnamefont {RJ}~\bibnamefont {Cava}}, \ and\
  \bibinfo {author} {\bibfnamefont {NP}~\bibnamefont {Ong}},\ }\bibfield
  {title} {\enquote {\bibinfo {title} {Large thermal hall conductivity of
  neutral spin excitations in a frustrated quantum magnet},}\ }\href@noop {}
  {\bibfield  {journal} {\bibinfo  {journal} {Science}\ }\textbf {\bibinfo
  {volume} {348}},\ \bibinfo {pages} {106--109} (\bibinfo {year}
  {2015}{\natexlab{a}})}\BibitemShut {NoStop}%
\bibitem [{\citenamefont {Hirschberger}\ \emph
  {et~al.}(2015{\natexlab{b}})\citenamefont {Hirschberger}, \citenamefont
  {Chisnell}, \citenamefont {Lee},\ and\ \citenamefont
  {Ong}}]{hirschberger2015thermal}%
  \BibitemOpen
  \bibfield  {author} {\bibinfo {author} {\bibfnamefont {Max}\ \bibnamefont
  {Hirschberger}}, \bibinfo {author} {\bibfnamefont {Robin}\ \bibnamefont
  {Chisnell}}, \bibinfo {author} {\bibfnamefont {Young~S}\ \bibnamefont {Lee}},
  \ and\ \bibinfo {author} {\bibfnamefont {Nai~Phuan}\ \bibnamefont {Ong}},\
  }\bibfield  {title} {\enquote {\bibinfo {title} {Thermal hall effect of spin
  excitations in a kagome magnet},}\ }\href@noop {} {\bibfield  {journal}
  {\bibinfo  {journal} {Physical review letters}\ }\textbf {\bibinfo {volume}
  {115}},\ \bibinfo {pages} {106603} (\bibinfo {year}
  {2015}{\natexlab{b}})}\BibitemShut {NoStop}%
\bibitem [{\citenamefont {Watanabe}\ \emph {et~al.}(2016)\citenamefont
  {Watanabe}, \citenamefont {Sugii}, \citenamefont {Shimozawa}, \citenamefont
  {Suzuki}, \citenamefont {Yajima}, \citenamefont {Ishikawa}, \citenamefont
  {Hiroi}, \citenamefont {Shibauchi}, \citenamefont {Matsuda},\ and\
  \citenamefont {Yamashita}}]{watanabe2016emergence}%
  \BibitemOpen
  \bibfield  {author} {\bibinfo {author} {\bibfnamefont {Daiki}\ \bibnamefont
  {Watanabe}}, \bibinfo {author} {\bibfnamefont {Kaori}\ \bibnamefont {Sugii}},
  \bibinfo {author} {\bibfnamefont {Masaaki}\ \bibnamefont {Shimozawa}},
  \bibinfo {author} {\bibfnamefont {Yoshitaka}\ \bibnamefont {Suzuki}},
  \bibinfo {author} {\bibfnamefont {Takeshi}\ \bibnamefont {Yajima}}, \bibinfo
  {author} {\bibfnamefont {Hajime}\ \bibnamefont {Ishikawa}}, \bibinfo {author}
  {\bibfnamefont {Zenji}\ \bibnamefont {Hiroi}}, \bibinfo {author}
  {\bibfnamefont {Takasada}\ \bibnamefont {Shibauchi}}, \bibinfo {author}
  {\bibfnamefont {Yuji}\ \bibnamefont {Matsuda}}, \ and\ \bibinfo {author}
  {\bibfnamefont {Minoru}\ \bibnamefont {Yamashita}},\ }\bibfield  {title}
  {\enquote {\bibinfo {title} {Emergence of nontrivial magnetic excitations in
  a spin-liquid state of kagom{\'e} volborthite},}\ }\href@noop {} {\bibfield
  {journal} {\bibinfo  {journal} {Proceedings of the National Academy of
  Sciences}\ }\textbf {\bibinfo {volume} {113}},\ \bibinfo {pages} {8653--8657}
  (\bibinfo {year} {2016})}\BibitemShut {NoStop}%
\bibitem [{\citenamefont {Doki}\ \emph {et~al.}(2018)\citenamefont {Doki},
  \citenamefont {Akazawa}, \citenamefont {Lee}, \citenamefont {Han},
  \citenamefont {Sugii}, \citenamefont {Shimozawa}, \citenamefont {Kawashima},
  \citenamefont {Oda}, \citenamefont {Yoshida},\ and\ \citenamefont
  {Yamashita}}]{doki2018spin}%
  \BibitemOpen
  \bibfield  {author} {\bibinfo {author} {\bibfnamefont {Hayato}\ \bibnamefont
  {Doki}}, \bibinfo {author} {\bibfnamefont {Masatoshi}\ \bibnamefont
  {Akazawa}}, \bibinfo {author} {\bibfnamefont {Hyun-Yong}\ \bibnamefont
  {Lee}}, \bibinfo {author} {\bibfnamefont {Jung~Hoon}\ \bibnamefont {Han}},
  \bibinfo {author} {\bibfnamefont {Kaori}\ \bibnamefont {Sugii}}, \bibinfo
  {author} {\bibfnamefont {Masaaki}\ \bibnamefont {Shimozawa}}, \bibinfo
  {author} {\bibfnamefont {Naoki}\ \bibnamefont {Kawashima}}, \bibinfo {author}
  {\bibfnamefont {Migaku}\ \bibnamefont {Oda}}, \bibinfo {author}
  {\bibfnamefont {Hiroyuki}\ \bibnamefont {Yoshida}}, \ and\ \bibinfo {author}
  {\bibfnamefont {Minoru}\ \bibnamefont {Yamashita}},\ }\bibfield  {title}
  {\enquote {\bibinfo {title} {Spin thermal hall conductivity of a kagome
  antiferromagnet},}\ }\href@noop {} {\bibfield  {journal} {\bibinfo  {journal}
  {Physical review letters}\ }\textbf {\bibinfo {volume} {121}},\ \bibinfo
  {pages} {097203} (\bibinfo {year} {2018})}\BibitemShut {NoStop}%
\bibitem [{\citenamefont {Hirokane}\ \emph {et~al.}(2019)\citenamefont
  {Hirokane}, \citenamefont {Nii}, \citenamefont {Tomioka},\ and\ \citenamefont
  {Onose}}]{hirokane2019}%
  \BibitemOpen
  \bibfield  {author} {\bibinfo {author} {\bibfnamefont {Yuji}\ \bibnamefont
  {Hirokane}}, \bibinfo {author} {\bibfnamefont {Yoichi}\ \bibnamefont {Nii}},
  \bibinfo {author} {\bibfnamefont {Yasuhide}\ \bibnamefont {Tomioka}}, \ and\
  \bibinfo {author} {\bibfnamefont {Yoshinori}\ \bibnamefont {Onose}},\
  }\bibfield  {title} {\enquote {\bibinfo {title} {Phononic thermal hall effect
  in diluted terbium oxides},}\ }\href {\doibase 10.1103/PhysRevB.99.134419}
  {\bibfield  {journal} {\bibinfo  {journal} {Phys. Rev. B}\ }\textbf {\bibinfo
  {volume} {99}},\ \bibinfo {pages} {134419} (\bibinfo {year}
  {2019})}\BibitemShut {NoStop}%
\bibitem [{\citenamefont {Akazawa}\ \emph {et~al.}(2020)\citenamefont
  {Akazawa}, \citenamefont {Shimozawa}, \citenamefont {Kittaka}, \citenamefont
  {Sakakibara}, \citenamefont {Okuma}, \citenamefont {Hiroi}, \citenamefont
  {Lee}, \citenamefont {Kawashima}, \citenamefont {Han},\ and\ \citenamefont
  {Yamashita}}]{akazawa2020thermal}%
  \BibitemOpen
  \bibfield  {author} {\bibinfo {author} {\bibfnamefont {Masatoshi}\
  \bibnamefont {Akazawa}}, \bibinfo {author} {\bibfnamefont {Masaaki}\
  \bibnamefont {Shimozawa}}, \bibinfo {author} {\bibfnamefont {Shunichiro}\
  \bibnamefont {Kittaka}}, \bibinfo {author} {\bibfnamefont {Toshiro}\
  \bibnamefont {Sakakibara}}, \bibinfo {author} {\bibfnamefont {Ryutaro}\
  \bibnamefont {Okuma}}, \bibinfo {author} {\bibfnamefont {Zenji}\ \bibnamefont
  {Hiroi}}, \bibinfo {author} {\bibfnamefont {Hyun-Yong}\ \bibnamefont {Lee}},
  \bibinfo {author} {\bibfnamefont {Naoki}\ \bibnamefont {Kawashima}}, \bibinfo
  {author} {\bibfnamefont {Jung~Hoon}\ \bibnamefont {Han}}, \ and\ \bibinfo
  {author} {\bibfnamefont {Minoru}\ \bibnamefont {Yamashita}},\ }\bibfield
  {title} {\enquote {\bibinfo {title} {Thermal hall effects of spins and
  phonons in kagome antiferromagnet cd-kapellasite},}\ }\href@noop {}
  {\bibfield  {journal} {\bibinfo  {journal} {Physical Review X}\ }\textbf
  {\bibinfo {volume} {10}},\ \bibinfo {pages} {041059} (\bibinfo {year}
  {2020})}\BibitemShut {NoStop}%
\bibitem [{\citenamefont {Zhang}\ \emph {et~al.}(2021)\citenamefont {Zhang},
  \citenamefont {Xu}, \citenamefont {Carnahan}, \citenamefont {Sretenovic},
  \citenamefont {Suri}, \citenamefont {Xiao},\ and\ \citenamefont
  {Ke}}]{zhang2021anomalous}%
  \BibitemOpen
  \bibfield  {author} {\bibinfo {author} {\bibfnamefont {Heda}\ \bibnamefont
  {Zhang}}, \bibinfo {author} {\bibfnamefont {Chunqiang}\ \bibnamefont {Xu}},
  \bibinfo {author} {\bibfnamefont {Caitlin}\ \bibnamefont {Carnahan}},
  \bibinfo {author} {\bibfnamefont {Milos}\ \bibnamefont {Sretenovic}},
  \bibinfo {author} {\bibfnamefont {Nishchay}\ \bibnamefont {Suri}}, \bibinfo
  {author} {\bibfnamefont {Di}~\bibnamefont {Xiao}}, \ and\ \bibinfo {author}
  {\bibfnamefont {Xianglin}\ \bibnamefont {Ke}},\ }\bibfield  {title} {\enquote
  {\bibinfo {title} {Anomalous thermal hall effect in an insulating van der
  waals magnet},}\ }\href@noop {} {\bibfield  {journal} {\bibinfo  {journal}
  {Physical Review Letters}\ }\textbf {\bibinfo {volume} {127}},\ \bibinfo
  {pages} {247202} (\bibinfo {year} {2021})}\BibitemShut {NoStop}%
\bibitem [{\citenamefont {Chen}\ \emph {et~al.}(2022)\citenamefont {Chen},
  \citenamefont {Boulanger}, \citenamefont {Wang}, \citenamefont {Tafti},\ and\
  \citenamefont {Taillefer}}]{chen2022large}%
  \BibitemOpen
  \bibfield  {author} {\bibinfo {author} {\bibfnamefont {Lu}~\bibnamefont
  {Chen}}, \bibinfo {author} {\bibfnamefont {Marie-Eve}\ \bibnamefont
  {Boulanger}}, \bibinfo {author} {\bibfnamefont {Zhi-Cheng}\ \bibnamefont
  {Wang}}, \bibinfo {author} {\bibfnamefont {Fazel}\ \bibnamefont {Tafti}}, \
  and\ \bibinfo {author} {\bibfnamefont {Louis}\ \bibnamefont {Taillefer}},\
  }\bibfield  {title} {\enquote {\bibinfo {title} {Large phonon thermal hall
  conductivity in the antiferromagnetic insulator cu3teo6},}\ }\href@noop {}
  {\bibfield  {journal} {\bibinfo  {journal} {Proceedings of the National
  Academy of Sciences}\ }\textbf {\bibinfo {volume} {119}},\ \bibinfo {pages}
  {e2208016119} (\bibinfo {year} {2022})}\BibitemShut {NoStop}%
\bibitem [{\citenamefont {Lefran{\c{c}}ois}\ \emph {et~al.}(2022)\citenamefont
  {Lefran{\c{c}}ois}, \citenamefont {Grissonnanche}, \citenamefont {Baglo},
  \citenamefont {Lampen-Kelley}, \citenamefont {Yan}, \citenamefont {Balz},
  \citenamefont {Mandrus}, \citenamefont {Nagler}, \citenamefont {Kim},
  \citenamefont {Kim} \emph {et~al.}}]{lefranccois2022evidence}%
  \BibitemOpen
  \bibfield  {author} {\bibinfo {author} {\bibfnamefont {{\'E}}~\bibnamefont
  {Lefran{\c{c}}ois}}, \bibinfo {author} {\bibfnamefont {G}~\bibnamefont
  {Grissonnanche}}, \bibinfo {author} {\bibfnamefont {J}~\bibnamefont {Baglo}},
  \bibinfo {author} {\bibfnamefont {P}~\bibnamefont {Lampen-Kelley}}, \bibinfo
  {author} {\bibfnamefont {J-Q}\ \bibnamefont {Yan}}, \bibinfo {author}
  {\bibfnamefont {C}~\bibnamefont {Balz}}, \bibinfo {author} {\bibfnamefont
  {D}~\bibnamefont {Mandrus}}, \bibinfo {author} {\bibfnamefont
  {SE}~\bibnamefont {Nagler}}, \bibinfo {author} {\bibfnamefont
  {S}~\bibnamefont {Kim}}, \bibinfo {author} {\bibfnamefont {Young-June}\
  \bibnamefont {Kim}},  \emph {et~al.},\ }\bibfield  {title} {\enquote
  {\bibinfo {title} {Evidence of a phonon hall effect in the kitaev spin liquid
  candidate $\alpha$-rucl 3},}\ }\href@noop {} {\bibfield  {journal} {\bibinfo
  {journal} {Physical Review X}\ }\textbf {\bibinfo {volume} {12}},\ \bibinfo
  {pages} {021025} (\bibinfo {year} {2022})}\BibitemShut {NoStop}%
\bibitem [{\citenamefont {Hentrich}\ \emph {et~al.}(2019)\citenamefont
  {Hentrich}, \citenamefont {Roslova}, \citenamefont {Isaeva}, \citenamefont
  {Doert}, \citenamefont {Brenig}, \citenamefont {B{\"u}chner},\ and\
  \citenamefont {Hess}}]{hentrich2019large}%
  \BibitemOpen
  \bibfield  {author} {\bibinfo {author} {\bibfnamefont {Richard}\ \bibnamefont
  {Hentrich}}, \bibinfo {author} {\bibfnamefont {Maria}\ \bibnamefont
  {Roslova}}, \bibinfo {author} {\bibfnamefont {Anna}\ \bibnamefont {Isaeva}},
  \bibinfo {author} {\bibfnamefont {Thomas}\ \bibnamefont {Doert}}, \bibinfo
  {author} {\bibfnamefont {Wolfram}\ \bibnamefont {Brenig}}, \bibinfo {author}
  {\bibfnamefont {Bernd}\ \bibnamefont {B{\"u}chner}}, \ and\ \bibinfo {author}
  {\bibfnamefont {Christian}\ \bibnamefont {Hess}},\ }\bibfield  {title}
  {\enquote {\bibinfo {title} {Large thermal hall effect in $\alpha$-rucl 3:
  Evidence for heat transport by kitaev-heisenberg paramagnons},}\ }\href@noop
  {} {\bibfield  {journal} {\bibinfo  {journal} {Physical Review B}\ }\textbf
  {\bibinfo {volume} {99}},\ \bibinfo {pages} {085136} (\bibinfo {year}
  {2019})}\BibitemShut {NoStop}%
\bibitem [{\citenamefont {Yamashita}\ \emph {et~al.}(2020)\citenamefont
  {Yamashita}, \citenamefont {Gouchi}, \citenamefont {Uwatoko}, \citenamefont
  {Kurita},\ and\ \citenamefont {Tanaka}}]{PhysRevB.102.220404}%
  \BibitemOpen
  \bibfield  {author} {\bibinfo {author} {\bibfnamefont {M.}~\bibnamefont
  {Yamashita}}, \bibinfo {author} {\bibfnamefont {J.}~\bibnamefont {Gouchi}},
  \bibinfo {author} {\bibfnamefont {Y.}~\bibnamefont {Uwatoko}}, \bibinfo
  {author} {\bibfnamefont {N.}~\bibnamefont {Kurita}}, \ and\ \bibinfo {author}
  {\bibfnamefont {H.}~\bibnamefont {Tanaka}},\ }\bibfield  {title} {\enquote
  {\bibinfo {title} {Sample dependence of half-integer quantized thermal hall
  effect in the kitaev spin-liquid candidate
  $\ensuremath{\alpha}\text{\ensuremath{-}}{\mathrm{rucl}}_{3}$},}\ }\href
  {\doibase 10.1103/PhysRevB.102.220404} {\bibfield  {journal} {\bibinfo
  {journal} {Phys. Rev. B}\ }\textbf {\bibinfo {volume} {102}},\ \bibinfo
  {pages} {220404} (\bibinfo {year} {2020})}\BibitemShut {NoStop}%
\bibitem [{\citenamefont {Yokoi}\ \emph {et~al.}(2021)\citenamefont {Yokoi},
  \citenamefont {Ma}, \citenamefont {Kasahara}, \citenamefont {Kasahara},
  \citenamefont {Shibauchi}, \citenamefont {Kurita}, \citenamefont {Tanaka},
  \citenamefont {Nasu}, \citenamefont {Motome}, \citenamefont {Hickey},
  \citenamefont {Trebst},\ and\ \citenamefont
  {Matsuda}}]{10.1126/science.aay5551}%
  \BibitemOpen
  \bibfield  {author} {\bibinfo {author} {\bibfnamefont {T.}~\bibnamefont
  {Yokoi}}, \bibinfo {author} {\bibfnamefont {S.}~\bibnamefont {Ma}}, \bibinfo
  {author} {\bibfnamefont {Y.}~\bibnamefont {Kasahara}}, \bibinfo {author}
  {\bibfnamefont {S.}~\bibnamefont {Kasahara}}, \bibinfo {author}
  {\bibfnamefont {T.}~\bibnamefont {Shibauchi}}, \bibinfo {author}
  {\bibfnamefont {N.}~\bibnamefont {Kurita}}, \bibinfo {author} {\bibfnamefont
  {H.}~\bibnamefont {Tanaka}}, \bibinfo {author} {\bibfnamefont
  {J.}~\bibnamefont {Nasu}}, \bibinfo {author} {\bibfnamefont {Y.}~\bibnamefont
  {Motome}}, \bibinfo {author} {\bibfnamefont {C.}~\bibnamefont {Hickey}},
  \bibinfo {author} {\bibfnamefont {S.}~\bibnamefont {Trebst}}, \ and\ \bibinfo
  {author} {\bibfnamefont {Y.}~\bibnamefont {Matsuda}},\ }\bibfield  {title}
  {\enquote {\bibinfo {title} {Half-integer quantized anomalous thermal hall
  effect in the kitaev material candidate
  $\ensuremath{\alpha}{\mathrm{-rucl}}_{3}$},}\ }\href {\doibase
  10.1126/science.aay5551} {\bibfield  {journal} {\bibinfo  {journal}
  {Science}\ }\textbf {\bibinfo {volume} {373}},\ \bibinfo {pages} {568--572}
  (\bibinfo {year} {2021})}\BibitemShut {NoStop}%
\bibitem [{\citenamefont {Bruin}\ \emph {et~al.}(2022)\citenamefont {Bruin},
  \citenamefont {Claus}, \citenamefont {Matsumoto}, \citenamefont {Kurita},
  \citenamefont {Tanaka},\ and\ \citenamefont {Takagi}}]{bruin2022robustness}%
  \BibitemOpen
  \bibfield  {author} {\bibinfo {author} {\bibfnamefont {JAN}\ \bibnamefont
  {Bruin}}, \bibinfo {author} {\bibfnamefont {RR}~\bibnamefont {Claus}},
  \bibinfo {author} {\bibfnamefont {Y}~\bibnamefont {Matsumoto}}, \bibinfo
  {author} {\bibfnamefont {N}~\bibnamefont {Kurita}}, \bibinfo {author}
  {\bibfnamefont {H}~\bibnamefont {Tanaka}}, \ and\ \bibinfo {author}
  {\bibfnamefont {H}~\bibnamefont {Takagi}},\ }\bibfield  {title} {\enquote
  {\bibinfo {title} {Robustness of the thermal hall effect close to
  half-quantization in $\alpha$-rucl3},}\ }\href@noop {} {\bibfield  {journal}
  {\bibinfo  {journal} {Nature Physics}\ }\textbf {\bibinfo {volume} {18}},\
  \bibinfo {pages} {401--405} (\bibinfo {year} {2022})}\BibitemShut {NoStop}%
\bibitem [{\citenamefont {Kasahara}\ \emph {et~al.}(2022)\citenamefont
  {Kasahara}, \citenamefont {Suetsugu}, \citenamefont {Asaba}, \citenamefont
  {Kasahara}, \citenamefont {Shibauchi}, \citenamefont {Kurita}, \citenamefont
  {Tanaka},\ and\ \citenamefont {Matsuda}}]{PhysRevB.106.L060410}%
  \BibitemOpen
  \bibfield  {author} {\bibinfo {author} {\bibfnamefont {Y.}~\bibnamefont
  {Kasahara}}, \bibinfo {author} {\bibfnamefont {S.}~\bibnamefont {Suetsugu}},
  \bibinfo {author} {\bibfnamefont {T.}~\bibnamefont {Asaba}}, \bibinfo
  {author} {\bibfnamefont {S.}~\bibnamefont {Kasahara}}, \bibinfo {author}
  {\bibfnamefont {T.}~\bibnamefont {Shibauchi}}, \bibinfo {author}
  {\bibfnamefont {N.}~\bibnamefont {Kurita}}, \bibinfo {author} {\bibfnamefont
  {H.}~\bibnamefont {Tanaka}}, \ and\ \bibinfo {author} {\bibfnamefont
  {Y.}~\bibnamefont {Matsuda}},\ }\bibfield  {title} {\enquote {\bibinfo
  {title} {Quantized and unquantized thermal hall conductance of the kitaev
  spin liquid candidate $\ensuremath{\alpha}{\mathrm{-rucl}}_{3}$},}\ }\href
  {\doibase 10.1103/PhysRevB.106.L060410} {\bibfield  {journal} {\bibinfo
  {journal} {Phys. Rev. B}\ }\textbf {\bibinfo {volume} {106}},\ \bibinfo
  {pages} {L060410} (\bibinfo {year} {2022})}\BibitemShut {NoStop}%
\bibitem [{\citenamefont {Zhang}\ \emph {et~al.}(2023)\citenamefont {Zhang},
  \citenamefont {May}, \citenamefont {Miao}, \citenamefont {Sales},
  \citenamefont {Mandrus}, \citenamefont {Nagler}, \citenamefont {McGuire},\
  and\ \citenamefont {Yan}}]{PhysRevMaterials.7.114403}%
  \BibitemOpen
  \bibfield  {author} {\bibinfo {author} {\bibfnamefont {Heda}\ \bibnamefont
  {Zhang}}, \bibinfo {author} {\bibfnamefont {Andrew~F.}\ \bibnamefont {May}},
  \bibinfo {author} {\bibfnamefont {Hu}~\bibnamefont {Miao}}, \bibinfo {author}
  {\bibfnamefont {Brian~C.}\ \bibnamefont {Sales}}, \bibinfo {author}
  {\bibfnamefont {David~G.}\ \bibnamefont {Mandrus}}, \bibinfo {author}
  {\bibfnamefont {Stephen~E.}\ \bibnamefont {Nagler}}, \bibinfo {author}
  {\bibfnamefont {Michael~A.}\ \bibnamefont {McGuire}}, \ and\ \bibinfo
  {author} {\bibfnamefont {Jiaqiang}\ \bibnamefont {Yan}},\ }\bibfield  {title}
  {\enquote {\bibinfo {title} {Sample-dependent and sample-independent thermal
  transport properties of
  $\ensuremath{\alpha}\text{\ensuremath{-}}{\mathrm{rucl}}_{3}$},}\ }\href
  {\doibase 10.1103/PhysRevMaterials.7.114403} {\bibfield  {journal} {\bibinfo
  {journal} {Phys. Rev. Mater.}\ }\textbf {\bibinfo {volume} {7}},\ \bibinfo
  {pages} {114403} (\bibinfo {year} {2023})}\BibitemShut {NoStop}%
\bibitem [{\citenamefont {Zhang}\ \emph
  {et~al.}(2024{\natexlab{a}})\citenamefont {Zhang}, \citenamefont {McGuire},
  \citenamefont {May}, \citenamefont {Chao}, \citenamefont {Zheng},
  \citenamefont {Chi}, \citenamefont {Sales}, \citenamefont {Mandrus},
  \citenamefont {Nagler}, \citenamefont {Miao}, \citenamefont {Ye},\ and\
  \citenamefont {Yan}}]{PhysRevMaterials.8.014402}%
  \BibitemOpen
  \bibfield  {author} {\bibinfo {author} {\bibfnamefont {Heda}\ \bibnamefont
  {Zhang}}, \bibinfo {author} {\bibfnamefont {Michael~A.}\ \bibnamefont
  {McGuire}}, \bibinfo {author} {\bibfnamefont {Andrew~F.}\ \bibnamefont
  {May}}, \bibinfo {author} {\bibfnamefont {Hsin-Yun}\ \bibnamefont {Chao}},
  \bibinfo {author} {\bibfnamefont {Qiang}\ \bibnamefont {Zheng}}, \bibinfo
  {author} {\bibfnamefont {Miaofang}\ \bibnamefont {Chi}}, \bibinfo {author}
  {\bibfnamefont {Brian~C.}\ \bibnamefont {Sales}}, \bibinfo {author}
  {\bibfnamefont {David~G.}\ \bibnamefont {Mandrus}}, \bibinfo {author}
  {\bibfnamefont {Stephen~E.}\ \bibnamefont {Nagler}}, \bibinfo {author}
  {\bibfnamefont {Hu}~\bibnamefont {Miao}}, \bibinfo {author} {\bibfnamefont
  {Feng}\ \bibnamefont {Ye}}, \ and\ \bibinfo {author} {\bibfnamefont
  {Jiaqiang}\ \bibnamefont {Yan}},\ }\bibfield  {title} {\enquote {\bibinfo
  {title} {Stacking disorder and thermal transport properties of
  $\ensuremath{\alpha}\text{\ensuremath{-}}{\mathrm{rucl}}_{3}$},}\ }\href
  {\doibase 10.1103/PhysRevMaterials.8.014402} {\bibfield  {journal} {\bibinfo
  {journal} {Phys. Rev. Mater.}\ }\textbf {\bibinfo {volume} {8}},\ \bibinfo
  {pages} {014402} (\bibinfo {year} {2024}{\natexlab{a}})}\BibitemShut
  {NoStop}%
\bibitem [{\citenamefont {Czajka}\ \emph {et~al.}(2023)\citenamefont {Czajka},
  \citenamefont {Gao}, \citenamefont {Hirschberger}, \citenamefont
  {Lampen-Kelley}, \citenamefont {Banerjee}, \citenamefont {Quirk},
  \citenamefont {Mandrus}, \citenamefont {Nagler},\ and\ \citenamefont
  {Ong}}]{czajka2023planar}%
  \BibitemOpen
  \bibfield  {author} {\bibinfo {author} {\bibfnamefont {Peter}\ \bibnamefont
  {Czajka}}, \bibinfo {author} {\bibfnamefont {Tong}\ \bibnamefont {Gao}},
  \bibinfo {author} {\bibfnamefont {Max}\ \bibnamefont {Hirschberger}},
  \bibinfo {author} {\bibfnamefont {Paula}\ \bibnamefont {Lampen-Kelley}},
  \bibinfo {author} {\bibfnamefont {Arnab}\ \bibnamefont {Banerjee}}, \bibinfo
  {author} {\bibfnamefont {Nicholas}\ \bibnamefont {Quirk}}, \bibinfo {author}
  {\bibfnamefont {David~G}\ \bibnamefont {Mandrus}}, \bibinfo {author}
  {\bibfnamefont {Stephen~E}\ \bibnamefont {Nagler}}, \ and\ \bibinfo {author}
  {\bibfnamefont {N~Phuan}\ \bibnamefont {Ong}},\ }\bibfield  {title} {\enquote
  {\bibinfo {title} {Planar thermal hall effect of topological bosons in the
  kitaev magnet $\alpha$-rucl3},}\ }\href@noop {} {\bibfield  {journal}
  {\bibinfo  {journal} {Nature Materials}\ }\textbf {\bibinfo {volume} {22}},\
  \bibinfo {pages} {36--41} (\bibinfo {year} {2023})}\BibitemShut {NoStop}%
\bibitem [{\citenamefont {Zhang}\ \emph
  {et~al.}(2024{\natexlab{b}})\citenamefont {Zhang}, \citenamefont {Gao},\ and\
  \citenamefont {Chen}}]{zhang2024thermal}%
  \BibitemOpen
  \bibfield  {author} {\bibinfo {author} {\bibfnamefont {Xiao-Tian}\
  \bibnamefont {Zhang}}, \bibinfo {author} {\bibfnamefont {Yong~Hao}\
  \bibnamefont {Gao}}, \ and\ \bibinfo {author} {\bibfnamefont {Gang}\
  \bibnamefont {Chen}},\ }\bibfield  {title} {\enquote {\bibinfo {title}
  {Thermal hall effects in quantum magnets},}\ }\href@noop {} {\bibfield
  {journal} {\bibinfo  {journal} {Physics Reports}\ }\textbf {\bibinfo {volume}
  {1070}},\ \bibinfo {pages} {1--59} (\bibinfo {year}
  {2024}{\natexlab{b}})}\BibitemShut {NoStop}%
\bibitem [{\citenamefont {Matsumoto}\ and\ \citenamefont
  {Murakami}(2011)}]{matsumoto_rotational_2011}%
  \BibitemOpen
  \bibfield  {author} {\bibinfo {author} {\bibfnamefont {Ryo}\ \bibnamefont
  {Matsumoto}}\ and\ \bibinfo {author} {\bibfnamefont {Shuichi}\ \bibnamefont
  {Murakami}},\ }\bibfield  {title} {\enquote {\bibinfo {title} {Rotational
  motion of magnons and the thermal {Hall} effect},}\ }\href {\doibase
  10.1103/PhysRevB.84.184406} {\bibfield  {journal} {\bibinfo  {journal}
  {Physical Review B}\ }\textbf {\bibinfo {volume} {84}},\ \bibinfo {pages}
  {184406} (\bibinfo {year} {2011})}\BibitemShut {NoStop}%
\bibitem [{\citenamefont {Qin}\ \emph {et~al.}(2011)\citenamefont {Qin},
  \citenamefont {Niu},\ and\ \citenamefont {Shi}}]{qin_energy_2011}%
  \BibitemOpen
  \bibfield  {author} {\bibinfo {author} {\bibfnamefont {Tao}\ \bibnamefont
  {Qin}}, \bibinfo {author} {\bibfnamefont {Qian}\ \bibnamefont {Niu}}, \ and\
  \bibinfo {author} {\bibfnamefont {Junren}\ \bibnamefont {Shi}},\ }\bibfield
  {title} {\enquote {\bibinfo {title} {Energy magnetization and the thermal
  hall effect},}\ }\href {\doibase 10.1103/PhysRevLett.107.236601} {\bibfield
  {journal} {\bibinfo  {journal} {Phys. Rev. Lett.}\ }\textbf {\bibinfo
  {volume} {107}},\ \bibinfo {pages} {236601} (\bibinfo {year}
  {2011})}\BibitemShut {NoStop}%
\bibitem [{\citenamefont {Qin}\ \emph {et~al.}()\citenamefont {Qin},
  \citenamefont {Zhou},\ and\ \citenamefont {Shi}}]{qin_berry_2012}%
  \BibitemOpen
  \bibfield  {author} {\bibinfo {author} {\bibfnamefont {Tao}\ \bibnamefont
  {Qin}}, \bibinfo {author} {\bibfnamefont {Jianhui}\ \bibnamefont {Zhou}}, \
  and\ \bibinfo {author} {\bibfnamefont {Junren}\ \bibnamefont {Shi}},\
  }\bibfield  {title} {\enquote {\bibinfo {title} {Berry curvature and the
  phonon hall effect},}\ }\href {\doibase 10.1103/PhysRevB.86.104305}
  {\bibfield  {journal} {\bibinfo  {journal} {Phys. Rev. B}\ }\textbf {\bibinfo
  {volume} {86}},\ \bibinfo {pages} {104305}}\BibitemShut {NoStop}%
\bibitem [{\citenamefont {Matsumoto}\ \emph {et~al.}()\citenamefont
  {Matsumoto}, \citenamefont {Shindou},\ and\ \citenamefont
  {Murakami}}]{matsumoto_thermal_2014}%
  \BibitemOpen
  \bibfield  {author} {\bibinfo {author} {\bibfnamefont {Ryo}\ \bibnamefont
  {Matsumoto}}, \bibinfo {author} {\bibfnamefont {Ryuichi}\ \bibnamefont
  {Shindou}}, \ and\ \bibinfo {author} {\bibfnamefont {Shuichi}\ \bibnamefont
  {Murakami}},\ }\bibfield  {title} {\enquote {\bibinfo {title} {Thermal hall
  effect of magnons in magnets with dipolar interaction},}\ }\href {\doibase
  10.1103/PhysRevB.89.054420} {\bibfield  {journal} {\bibinfo  {journal} {Phys.
  Rev. B}\ }\textbf {\bibinfo {volume} {89}},\ \bibinfo {pages}
  {054420}}\BibitemShut {NoStop}%
\bibitem [{\citenamefont {Shitade}(2014)}]{shitade2014heat}%
  \BibitemOpen
  \bibfield  {author} {\bibinfo {author} {\bibfnamefont {Atsuo}\ \bibnamefont
  {Shitade}},\ }\bibfield  {title} {\enquote {\bibinfo {title} {Heat transport
  as torsional responses and keldysh formalism in a curved spacetime},}\
  }\href@noop {} {\bibfield  {journal} {\bibinfo  {journal} {Progress of
  Theoretical and Experimental Physics}\ }\textbf {\bibinfo {volume} {2014}},\
  \bibinfo {pages} {123I01} (\bibinfo {year} {2014})}\BibitemShut {NoStop}%
\bibitem [{\citenamefont {Zhang}\ \emph {et~al.}(2020)\citenamefont {Zhang},
  \citenamefont {Gao},\ and\ \citenamefont {Xiao}}]{zhang2020thermodynamics}%
  \BibitemOpen
  \bibfield  {author} {\bibinfo {author} {\bibfnamefont {Yinhan}\ \bibnamefont
  {Zhang}}, \bibinfo {author} {\bibfnamefont {Yang}\ \bibnamefont {Gao}}, \
  and\ \bibinfo {author} {\bibfnamefont {Di}~\bibnamefont {Xiao}},\ }\bibfield
  {title} {\enquote {\bibinfo {title} {Thermodynamics of energy
  magnetization},}\ }\href@noop {} {\bibfield  {journal} {\bibinfo  {journal}
  {Physical Review B}\ }\textbf {\bibinfo {volume} {102}},\ \bibinfo {pages}
  {235161} (\bibinfo {year} {2020})}\BibitemShut {NoStop}%
\bibitem [{\citenamefont {Mangeolle}\ \emph {et~al.}()\citenamefont
  {Mangeolle}, \citenamefont {Savary},\ and\ \citenamefont
  {Balents}}]{mangeolle_quantum_2024}%
  \BibitemOpen
  \bibfield  {author} {\bibinfo {author} {\bibfnamefont {Léo}\ \bibnamefont
  {Mangeolle}}, \bibinfo {author} {\bibfnamefont {Lucile}\ \bibnamefont
  {Savary}}, \ and\ \bibinfo {author} {\bibfnamefont {Leon}\ \bibnamefont
  {Balents}},\ }\bibfield  {title} {\enquote {\bibinfo {title} {Quantum kinetic
  equation and thermal conductivity tensor for bosons},}\ }\href {\doibase
  10.1103/PhysRevB.109.235137} {\bibfield  {journal} {\bibinfo  {journal}
  {Phys. Rev. B}\ }\textbf {\bibinfo {volume} {109}},\ \bibinfo {pages}
  {235137}}\BibitemShut {NoStop}%
\bibitem [{\citenamefont {Mori}\ \emph {et~al.}(2014)\citenamefont {Mori},
  \citenamefont {Spencer-Smith}, \citenamefont {Sushkov},\ and\ \citenamefont
  {Maekawa}}]{PhysRevLett.113.265901}%
  \BibitemOpen
  \bibfield  {author} {\bibinfo {author} {\bibfnamefont {Michiyasu}\
  \bibnamefont {Mori}}, \bibinfo {author} {\bibfnamefont {Alexander}\
  \bibnamefont {Spencer-Smith}}, \bibinfo {author} {\bibfnamefont {Oleg~P.}\
  \bibnamefont {Sushkov}}, \ and\ \bibinfo {author} {\bibfnamefont {Sadamichi}\
  \bibnamefont {Maekawa}},\ }\bibfield  {title} {\enquote {\bibinfo {title}
  {Origin of the phonon hall effect in rare-earth garnets},}\ }\href {\doibase
  10.1103/PhysRevLett.113.265901} {\bibfield  {journal} {\bibinfo  {journal}
  {Phys. Rev. Lett.}\ }\textbf {\bibinfo {volume} {113}},\ \bibinfo {pages}
  {265901} (\bibinfo {year} {2014})}\BibitemShut {NoStop}%
\bibitem [{\citenamefont {Flebus}\ and\ \citenamefont
  {MacDonald}(2022)}]{PhysRevB.105.L220301}%
  \BibitemOpen
  \bibfield  {author} {\bibinfo {author} {\bibfnamefont {B.}~\bibnamefont
  {Flebus}}\ and\ \bibinfo {author} {\bibfnamefont {A.~H.}\ \bibnamefont
  {MacDonald}},\ }\bibfield  {title} {\enquote {\bibinfo {title} {Charged
  defects and phonon hall effects in ionic crystals},}\ }\href {\doibase
  10.1103/PhysRevB.105.L220301} {\bibfield  {journal} {\bibinfo  {journal}
  {Phys. Rev. B}\ }\textbf {\bibinfo {volume} {105}},\ \bibinfo {pages}
  {L220301} (\bibinfo {year} {2022})}\BibitemShut {NoStop}%
\bibitem [{\citenamefont {Sun}\ \emph {et~al.}(2022)\citenamefont {Sun},
  \citenamefont {Chen},\ and\ \citenamefont {Kivelson}}]{PhysRevB.106.144111}%
  \BibitemOpen
  \bibfield  {author} {\bibinfo {author} {\bibfnamefont {Xiao-Qi}\ \bibnamefont
  {Sun}}, \bibinfo {author} {\bibfnamefont {Jing-Yuan}\ \bibnamefont {Chen}}, \
  and\ \bibinfo {author} {\bibfnamefont {Steven~A.}\ \bibnamefont {Kivelson}},\
  }\bibfield  {title} {\enquote {\bibinfo {title} {Large extrinsic phonon
  thermal hall effect from resonant scattering},}\ }\href {\doibase
  10.1103/PhysRevB.106.144111} {\bibfield  {journal} {\bibinfo  {journal}
  {Phys. Rev. B}\ }\textbf {\bibinfo {volume} {106}},\ \bibinfo {pages}
  {144111} (\bibinfo {year} {2022})}\BibitemShut {NoStop}%
\bibitem [{\citenamefont {Mangeolle}\ \emph
  {et~al.}(2022{\natexlab{a}})\citenamefont {Mangeolle}, \citenamefont
  {Balents},\ and\ \citenamefont {Savary}}]{PhysRevX.12.041031}%
  \BibitemOpen
  \bibfield  {author} {\bibinfo {author} {\bibfnamefont {L\'eo}\ \bibnamefont
  {Mangeolle}}, \bibinfo {author} {\bibfnamefont {Leon}\ \bibnamefont
  {Balents}}, \ and\ \bibinfo {author} {\bibfnamefont {Lucile}\ \bibnamefont
  {Savary}},\ }\bibfield  {title} {\enquote {\bibinfo {title} {Phonon thermal
  hall conductivity from scattering with collective fluctuations},}\ }\href
  {\doibase 10.1103/PhysRevX.12.041031} {\bibfield  {journal} {\bibinfo
  {journal} {Phys. Rev. X}\ }\textbf {\bibinfo {volume} {12}},\ \bibinfo
  {pages} {041031} (\bibinfo {year} {2022}{\natexlab{a}})}\BibitemShut
  {NoStop}%
\bibitem [{\citenamefont {Mangeolle}\ \emph
  {et~al.}(2022{\natexlab{b}})\citenamefont {Mangeolle}, \citenamefont
  {Balents},\ and\ \citenamefont {Savary}}]{PhysRevB.106.245139}%
  \BibitemOpen
  \bibfield  {author} {\bibinfo {author} {\bibfnamefont {L\'eo}\ \bibnamefont
  {Mangeolle}}, \bibinfo {author} {\bibfnamefont {Leon}\ \bibnamefont
  {Balents}}, \ and\ \bibinfo {author} {\bibfnamefont {Lucile}\ \bibnamefont
  {Savary}},\ }\bibfield  {title} {\enquote {\bibinfo {title} {Thermal
  conductivity and theory of inelastic scattering of phonons by collective
  fluctuations},}\ }\href {\doibase 10.1103/PhysRevB.106.245139} {\bibfield
  {journal} {\bibinfo  {journal} {Phys. Rev. B}\ }\textbf {\bibinfo {volume}
  {106}},\ \bibinfo {pages} {245139} (\bibinfo {year}
  {2022}{\natexlab{b}})}\BibitemShut {NoStop}%
\bibitem [{\citenamefont {Chatzichrysafis}\ and\ \citenamefont
  {Mook}(2025)}]{PhysRevB.111.134405}%
  \BibitemOpen
  \bibfield  {author} {\bibinfo {author} {\bibfnamefont {Dimos}\ \bibnamefont
  {Chatzichrysafis}}\ and\ \bibinfo {author} {\bibfnamefont {Alexander}\
  \bibnamefont {Mook}},\ }\bibfield  {title} {\enquote {\bibinfo {title}
  {Thermal hall effect of magnons from many-body skew scattering},}\ }\href
  {\doibase 10.1103/PhysRevB.111.134405} {\bibfield  {journal} {\bibinfo
  {journal} {Phys. Rev. B}\ }\textbf {\bibinfo {volume} {111}},\ \bibinfo
  {pages} {134405} (\bibinfo {year} {2025})}\BibitemShut {NoStop}%
\bibitem [{Note1()}]{Note1}%
  \BibitemOpen
  \bibinfo {note} {See \cite {nagaosa_anomalous_2010} and especially, in Sec.
  I.B.3: ``\protect \textit {A practical approach, which is followed at present
  for materials in which $\sigma ^{\protect \rm AH}$ seems to be independent of
  $\sigma _{xx}$, is to first calculate the intrinsic contribution to the AHE.
  If this explains the observation (and it appears that it usually does), then
  it is deemed that the intrinsic mechanism dominates. If not, we can take some
  comfort from understanding on the basis of simple model results that there
  can be other contributions to $\sigma ^{\protect \rm AH}$ which are also
  independent of $\sigma _{xx}$ and can for the most part be identified with
  the side-jump mechanism.}''}\BibitemShut {NoStop}%
\bibitem [{\citenamefont {Abrahams}\ \emph {et~al.}(1981)\citenamefont
  {Abrahams}, \citenamefont {Anderson}, \citenamefont {Lee},\ and\
  \citenamefont {Ramakrishnan}}]{abrahams1981quasiparticle}%
  \BibitemOpen
  \bibfield  {author} {\bibinfo {author} {\bibfnamefont {Elihu}\ \bibnamefont
  {Abrahams}}, \bibinfo {author} {\bibfnamefont {PW}~\bibnamefont {Anderson}},
  \bibinfo {author} {\bibfnamefont {PA}~\bibnamefont {Lee}}, \ and\ \bibinfo
  {author} {\bibfnamefont {TV}~\bibnamefont {Ramakrishnan}},\ }\bibfield
  {title} {\enquote {\bibinfo {title} {Quasiparticle lifetime in disordered
  two-dimensional metals},}\ }\href@noop {} {\bibfield  {journal} {\bibinfo
  {journal} {Physical Review B}\ }\textbf {\bibinfo {volume} {24}},\ \bibinfo
  {pages} {6783} (\bibinfo {year} {1981})}\BibitemShut {NoStop}%
\bibitem [{\citenamefont {Abrahams}\ \emph {et~al.}(1979)\citenamefont
  {Abrahams}, \citenamefont {Anderson}, \citenamefont {Licciardello},\ and\
  \citenamefont {Ramakrishnan}}]{abrahams1979scaling}%
  \BibitemOpen
  \bibfield  {author} {\bibinfo {author} {\bibfnamefont {Elihu}\ \bibnamefont
  {Abrahams}}, \bibinfo {author} {\bibfnamefont {Philip~W}\ \bibnamefont
  {Anderson}}, \bibinfo {author} {\bibfnamefont {Donald~C}\ \bibnamefont
  {Licciardello}}, \ and\ \bibinfo {author} {\bibfnamefont {Tiruppattur~V}\
  \bibnamefont {Ramakrishnan}},\ }\bibfield  {title} {\enquote {\bibinfo
  {title} {Scaling theory of localization: Absence of quantum diffusion in two
  dimensions},}\ }\href@noop {} {\bibfield  {journal} {\bibinfo  {journal}
  {Physical Review Letters}\ }\textbf {\bibinfo {volume} {42}},\ \bibinfo
  {pages} {673} (\bibinfo {year} {1979})}\BibitemShut {NoStop}%
\bibitem [{\citenamefont {Prange}\ and\ \citenamefont
  {Girvin}(1987)}]{prange1987quantum}%
  \BibitemOpen
  \bibfield  {author} {\bibinfo {author} {\bibfnamefont {Richard~E}\
  \bibnamefont {Prange}}\ and\ \bibinfo {author} {\bibfnamefont {Steven~M}\
  \bibnamefont {Girvin}},\ }\href@noop {} {\emph {\bibinfo {title} {The quantum
  Hall effect}}}\ (\bibinfo  {publisher} {Springer},\ \bibinfo {year}
  {1987})\BibitemShut {NoStop}%
\bibitem [{Note2()}]{Note2}%
  \BibitemOpen
  \bibinfo {note} {In magnetic systems, long-range interactions also exist such
  as dipolar exchange, though instances where these dominate over other (local)
  magnetic interactions governing the physical properties are rare, especially
  for bulk 3D insulating magnets.}\BibitemShut {Stop}%
\bibitem [{\citenamefont {Chern}\ \emph {et~al.}(2021)\citenamefont {Chern},
  \citenamefont {Zhang},\ and\ \citenamefont {Kim}}]{PhysRevLett.126.147201}%
  \BibitemOpen
  \bibfield  {author} {\bibinfo {author} {\bibfnamefont {Li~Ern}\ \bibnamefont
  {Chern}}, \bibinfo {author} {\bibfnamefont {Emily~Z.}\ \bibnamefont {Zhang}},
  \ and\ \bibinfo {author} {\bibfnamefont {Yong~Baek}\ \bibnamefont {Kim}},\
  }\bibfield  {title} {\enquote {\bibinfo {title} {Sign structure of thermal
  hall conductivity and topological magnons for in-plane field polarized kitaev
  magnets},}\ }\href {\doibase 10.1103/PhysRevLett.126.147201} {\bibfield
  {journal} {\bibinfo  {journal} {Phys. Rev. Lett.}\ }\textbf {\bibinfo
  {volume} {126}},\ \bibinfo {pages} {147201} (\bibinfo {year}
  {2021})}\BibitemShut {NoStop}%
\bibitem [{\citenamefont {Karplus}\ and\ \citenamefont
  {Luttinger}()}]{karplus_hall_1954}%
  \BibitemOpen
  \bibfield  {author} {\bibinfo {author} {\bibfnamefont {Robert}\ \bibnamefont
  {Karplus}}\ and\ \bibinfo {author} {\bibfnamefont {J.~M.}\ \bibnamefont
  {Luttinger}},\ }\bibfield  {title} {\enquote {\bibinfo {title} {Hall effect
  in ferromagnetics},}\ }\href {\doibase 10.1103/PhysRev.95.1154} {\bibfield
  {journal} {\bibinfo  {journal} {Phys. Rev.}\ }\textbf {\bibinfo {volume}
  {95}},\ \bibinfo {pages} {1154--1160}}\BibitemShut {NoStop}%
\bibitem [{\citenamefont {Sundaram}\ and\ \citenamefont
  {Niu}()}]{sundaram_wave-packet_1999}%
  \BibitemOpen
  \bibfield  {author} {\bibinfo {author} {\bibfnamefont {Ganesh}\ \bibnamefont
  {Sundaram}}\ and\ \bibinfo {author} {\bibfnamefont {Qian}\ \bibnamefont
  {Niu}},\ }\bibfield  {title} {\enquote {\bibinfo {title} {Wave-packet
  dynamics in slowly perturbed crystals: Gradient corrections and berry-phase
  effects},}\ }\href {\doibase 10.1103/PhysRevB.59.14915} {\bibfield  {journal}
  {\bibinfo  {journal} {Phys. Rev. B}\ }\textbf {\bibinfo {volume} {59}},\
  \bibinfo {pages} {14915--14925}}\BibitemShut {NoStop}%
\bibitem [{\citenamefont {Wickles}\ and\ \citenamefont
  {Belzig}()}]{wickles_effective_2013}%
  \BibitemOpen
  \bibfield  {author} {\bibinfo {author} {\bibfnamefont {Christian}\
  \bibnamefont {Wickles}}\ and\ \bibinfo {author} {\bibfnamefont {Wolfgang}\
  \bibnamefont {Belzig}},\ }\bibfield  {title} {\enquote {\bibinfo {title}
  {Effective quantum theories for bloch dynamics in inhomogeneous systems with
  nontrivial band structure},}\ }\href {\doibase 10.1103/PhysRevB.88.045308}
  {\bibfield  {journal} {\bibinfo  {journal} {Phys. Rev. B}\ }\textbf {\bibinfo
  {volume} {88}},\ \bibinfo {pages} {045308}}\BibitemShut {NoStop}%
\bibitem [{\citenamefont {Adams}\ and\ \citenamefont
  {Blount}()}]{adams_energy_1959}%
  \BibitemOpen
  \bibfield  {author} {\bibinfo {author} {\bibfnamefont {E.N.}\ \bibnamefont
  {Adams}}\ and\ \bibinfo {author} {\bibfnamefont {E.I.}\ \bibnamefont
  {Blount}},\ }\bibfield  {title} {\enquote {\bibinfo {title} {Energy bands in
  the presence of an external force field—{II}},}\ }\href {\doibase
  10.1016/0022-3697(59)90004-6} {\bibfield  {journal} {\bibinfo  {journal}
  {Journal of Physics and Chemistry of Solids}\ }\textbf {\bibinfo {volume}
  {10}},\ \bibinfo {pages} {286--303}}\BibitemShut {NoStop}%
\bibitem [{\citenamefont {Fivaz}()}]{fivaz_transport_1969}%
  \BibitemOpen
  \bibfield  {author} {\bibinfo {author} {\bibfnamefont {R.~C.}\ \bibnamefont
  {Fivaz}},\ }\bibfield  {title} {\enquote {\bibinfo {title} {Transport theory
  for ferromagnets},}\ }\href {\doibase 10.1103/PhysRev.183.586} {\bibfield
  {journal} {\bibinfo  {journal} {Phys. Rev.}\ }\textbf {\bibinfo {volume}
  {183}},\ \bibinfo {pages} {586--594}}\BibitemShut {NoStop}%
\bibitem [{\citenamefont {Nozières}\ and\ \citenamefont
  {Lewiner}()}]{nozieres_simple_1973}%
  \BibitemOpen
  \bibfield  {author} {\bibinfo {author} {\bibfnamefont {P.}~\bibnamefont
  {Nozières}}\ and\ \bibinfo {author} {\bibfnamefont {C.}~\bibnamefont
  {Lewiner}},\ }\bibfield  {title} {\enquote {\bibinfo {title} {A simple theory
  of the anomalous hall effect in semiconductors},}\ }\href {\doibase
  10.1051/jphys:019730034010090100} {\bibfield  {journal} {\bibinfo  {journal}
  {J. Phys. France}\ }\textbf {\bibinfo {volume} {34}},\ \bibinfo {pages}
  {901--915}}\BibitemShut {NoStop}%
\bibitem [{\citenamefont {Sinitsyn}\ \emph {et~al.}(2005)\citenamefont
  {Sinitsyn}, \citenamefont {Niu}, \citenamefont {Sinova},\ and\ \citenamefont
  {Nomura}}]{sinitsyn_disorder_2005}%
  \BibitemOpen
  \bibfield  {author} {\bibinfo {author} {\bibfnamefont {N.~A.}\ \bibnamefont
  {Sinitsyn}}, \bibinfo {author} {\bibfnamefont {Qian}\ \bibnamefont {Niu}},
  \bibinfo {author} {\bibfnamefont {Jairo}\ \bibnamefont {Sinova}}, \ and\
  \bibinfo {author} {\bibfnamefont {Kentaro}\ \bibnamefont {Nomura}},\
  }\bibfield  {title} {\enquote {\bibinfo {title} {Disorder effects in the
  anomalous hall effect induced by berry curvature},}\ }\href {\doibase
  10.1103/PhysRevB.72.045346} {\bibfield  {journal} {\bibinfo  {journal} {Phys.
  Rev. B}\ }\textbf {\bibinfo {volume} {72}},\ \bibinfo {pages} {045346}
  (\bibinfo {year} {2005})}\BibitemShut {NoStop}%
\bibitem [{\citenamefont {Sinitsyn}\ \emph {et~al.}(2006)\citenamefont
  {Sinitsyn}, \citenamefont {Niu},\ and\ \citenamefont
  {MacDonald}}]{sinitsyn_coordinate_2006}%
  \BibitemOpen
  \bibfield  {author} {\bibinfo {author} {\bibfnamefont {N.~A.}\ \bibnamefont
  {Sinitsyn}}, \bibinfo {author} {\bibfnamefont {Q.}~\bibnamefont {Niu}}, \
  and\ \bibinfo {author} {\bibfnamefont {A.~H.}\ \bibnamefont {MacDonald}},\
  }\bibfield  {title} {\enquote {\bibinfo {title} {Coordinate shift in the
  semiclassical boltzmann equation and the anomalous hall effect},}\ }\href
  {\doibase 10.1103/PhysRevB.73.075318} {\bibfield  {journal} {\bibinfo
  {journal} {Phys. Rev. B}\ }\textbf {\bibinfo {volume} {73}},\ \bibinfo
  {pages} {075318} (\bibinfo {year} {2006})}\BibitemShut {NoStop}%
\bibitem [{\citenamefont {König}\ and\ \citenamefont
  {Levchenko}()}]{konig_quantum_2021}%
  \BibitemOpen
  \bibfield  {author} {\bibinfo {author} {\bibfnamefont {Elio~J.}\ \bibnamefont
  {König}}\ and\ \bibinfo {author} {\bibfnamefont {Alex}\ \bibnamefont
  {Levchenko}},\ }\bibfield  {title} {\enquote {\bibinfo {title} {Quantum
  kinetics of anomalous and nonlinear hall effects in topological
  semimetals},}\ }\href {\doibase 10.1016/j.aop.2021.168492} {\bibfield
  {journal} {\bibinfo  {journal} {Annals of Physics}\ }\bibinfo {series}
  {Special issue on Philip W. Anderson},\ \textbf {\bibinfo {volume} {435}},\
  \bibinfo {pages} {168492}}\BibitemShut {NoStop}%
\bibitem [{\citenamefont {Sinitsyn}\ \emph {et~al.}(2007)\citenamefont
  {Sinitsyn}, \citenamefont {MacDonald}, \citenamefont {Jungwirth},
  \citenamefont {Dugaev},\ and\ \citenamefont
  {Sinova}}]{sinitsyn2007anomalous}%
  \BibitemOpen
  \bibfield  {author} {\bibinfo {author} {\bibfnamefont {NA}~\bibnamefont
  {Sinitsyn}}, \bibinfo {author} {\bibfnamefont {AH}~\bibnamefont {MacDonald}},
  \bibinfo {author} {\bibfnamefont {T}~\bibnamefont {Jungwirth}}, \bibinfo
  {author} {\bibfnamefont {VK}~\bibnamefont {Dugaev}}, \ and\ \bibinfo {author}
  {\bibfnamefont {Jairo}\ \bibnamefont {Sinova}},\ }\bibfield  {title}
  {\enquote {\bibinfo {title} {Anomalous hall effect in a two-dimensional dirac
  band: The link between the kubo-streda formula and the semiclassical
  boltzmann equation approach},}\ }\href@noop {} {\bibfield  {journal}
  {\bibinfo  {journal} {Physical Review B—Condensed Matter and Materials
  Physics}\ }\textbf {\bibinfo {volume} {75}},\ \bibinfo {pages} {045315}
  (\bibinfo {year} {2007})}\BibitemShut {NoStop}%
\bibitem [{\citenamefont {Guo}\ \emph {et~al.}(2022)\citenamefont {Guo},
  \citenamefont {Joshi},\ and\ \citenamefont {Sachdev}}]{guo2022resonant}%
  \BibitemOpen
  \bibfield  {author} {\bibinfo {author} {\bibfnamefont {Haoyu}\ \bibnamefont
  {Guo}}, \bibinfo {author} {\bibfnamefont {Darshan~G}\ \bibnamefont {Joshi}},
  \ and\ \bibinfo {author} {\bibfnamefont {Subir}\ \bibnamefont {Sachdev}},\
  }\bibfield  {title} {\enquote {\bibinfo {title} {Resonant thermal hall effect
  of phonons coupled to dynamical defects},}\ }\href@noop {} {\bibfield
  {journal} {\bibinfo  {journal} {Proceedings of the National Academy of
  Sciences}\ }\textbf {\bibinfo {volume} {119}},\ \bibinfo {pages}
  {e2215141119} (\bibinfo {year} {2022})}\BibitemShut {NoStop}%
\bibitem [{\citenamefont {Kapustin}\ and\ \citenamefont
  {Spodyneiko}(2020)}]{kapustin_thermal_2020}%
  \BibitemOpen
  \bibfield  {author} {\bibinfo {author} {\bibfnamefont {Anton}\ \bibnamefont
  {Kapustin}}\ and\ \bibinfo {author} {\bibfnamefont {Lev}\ \bibnamefont
  {Spodyneiko}},\ }\bibfield  {title} {\enquote {\bibinfo {title} {Thermal
  {Hall} conductance and a relative topological invariant of gapped
  two-dimensional systems},}\ }\href {\doibase 10.1103/PhysRevB.101.045137}
  {\bibfield  {journal} {\bibinfo  {journal} {Physical Review B}\ }\textbf
  {\bibinfo {volume} {101}},\ \bibinfo {pages} {045137} (\bibinfo {year}
  {2020})}\BibitemShut {NoStop}%
\bibitem [{\citenamefont {Kamenev}(2011)}]{kamenev2011field}%
  \BibitemOpen
  \bibfield  {author} {\bibinfo {author} {\bibfnamefont {Alex}\ \bibnamefont
  {Kamenev}},\ }\href@noop {} {\emph {\bibinfo {title} {Field theory of
  non-equilibrium systems}}}\ (\bibinfo  {publisher} {Cambridge University
  Press},\ \bibinfo {year} {2011})\BibitemShut {NoStop}%
\bibitem [{Note3()}]{Note3}%
  \BibitemOpen
  \bibinfo {note} {In \cite {konig_quantum_2021} an extra valley index is
  introduced, but the $(a,b)$ and $(r_1,r_2)$ indices remain decoupled, the
  spatial dependence is still of the $\delta (r_1-r_2)$ type, and within one
  valley disorder is still $\protect \hat V \propto \protect \text {{\protect
  \fontencoding {U}\protect \fontfamily {bbold}\protect \selectfont
  1}}$.}\BibitemShut {Stop}%
\bibitem [{\citenamefont {Culcer}\ \emph {et~al.}()\citenamefont {Culcer},
  \citenamefont {Sekine},\ and\ \citenamefont
  {{MacDonald}}}]{culcer_interband_2017}%
  \BibitemOpen
  \bibfield  {author} {\bibinfo {author} {\bibfnamefont {Dimitrie}\
  \bibnamefont {Culcer}}, \bibinfo {author} {\bibfnamefont {Akihiko}\
  \bibnamefont {Sekine}}, \ and\ \bibinfo {author} {\bibfnamefont {Allan~H.}\
  \bibnamefont {{MacDonald}}},\ }\bibfield  {title} {\enquote {\bibinfo {title}
  {Interband coherence response to electric fields in crystals: Berry-phase
  contributions and disorder effects},}\ }\href {\doibase
  10.1103/PhysRevB.96.035106} {\bibfield  {journal} {\bibinfo  {journal} {Phys.
  Rev. B}\ }\textbf {\bibinfo {volume} {96}},\ \bibinfo {pages}
  {035106}}\BibitemShut {NoStop}%
\bibitem [{\citenamefont {Rammer}(1998)}]{rammerbook}%
  \BibitemOpen
  \bibfield  {author} {\bibinfo {author} {\bibfnamefont {Jorgen}\ \bibnamefont
  {Rammer}},\ }\href@noop {} {\emph {\bibinfo {title} {Quantum transport
  theory}}}\ (\bibinfo  {publisher} {Perseus Books},\ \bibinfo {year}
  {1998})\BibitemShut {NoStop}%
\bibitem [{foo()}]{footnote_perturbative}%
  \BibitemOpen
  \href@noop {} {}\bibinfo {note} {Throughout the paper, each time we use the
  word ``perturbative'' this refers specifically to an expansion in powers of
  the disorder strength.}\BibitemShut {Stop}%
\bibitem [{\citenamefont {Gutzwiller}(1990)}]{gutzwillerbook}%
  \BibitemOpen
  \bibfield  {author} {\bibinfo {author} {\bibfnamefont {Martin~C.}\
  \bibnamefont {Gutzwiller}},\ }\href@noop {} {\emph {\bibinfo {title} {Chaos
  in Classical and Quantum Mechanic}}}\ (\bibinfo  {publisher}
  {Springer-Verlag},\ \bibinfo {year} {1990})\BibitemShut {NoStop}%
\bibitem [{\citenamefont {Ziman}(1960)}]{zimanbook}%
  \BibitemOpen
  \bibfield  {author} {\bibinfo {author} {\bibfnamefont {John~M.}\ \bibnamefont
  {Ziman}},\ }\href@noop {} {\emph {\bibinfo {title} {Electrons and Phonons:
  the theory of transport phenomena in solids}}}\ (\bibinfo  {publisher}
  {Oxford University Press},\ \bibinfo {year} {1960})\BibitemShut {NoStop}%
\bibitem [{\citenamefont {Allen}(1978)}]{PhysRevB.17.3725}%
  \BibitemOpen
  \bibfield  {author} {\bibinfo {author} {\bibfnamefont {P.~B.}\ \bibnamefont
  {Allen}},\ }\bibfield  {title} {\enquote {\bibinfo {title} {New method for
  solving boltzmann's equation for electrons in metals},}\ }\href {\doibase
  10.1103/PhysRevB.17.3725} {\bibfield  {journal} {\bibinfo  {journal} {Phys.
  Rev. B}\ }\textbf {\bibinfo {volume} {17}},\ \bibinfo {pages} {3725--3734}
  (\bibinfo {year} {1978})}\BibitemShut {NoStop}%
\bibitem [{Note4()}]{Note4}%
  \BibitemOpen
  \bibinfo {note} {Namely, one needs to fix whether the distribution function
  (against which the current operator is integrated) should be evaluated before
  or after the scattering event.}\BibitemShut {Stop}%
\bibitem [{\citenamefont {Cooper}\ \emph {et~al.}(1997)\citenamefont {Cooper},
  \citenamefont {Halperin},\ and\ \citenamefont {Ruzin}}]{cooper1997}%
  \BibitemOpen
  \bibfield  {author} {\bibinfo {author} {\bibfnamefont {N.~R.}\ \bibnamefont
  {Cooper}}, \bibinfo {author} {\bibfnamefont {B.~I.}\ \bibnamefont
  {Halperin}}, \ and\ \bibinfo {author} {\bibfnamefont {I.~M.}\ \bibnamefont
  {Ruzin}},\ }\bibfield  {title} {\enquote {\bibinfo {title} {Thermoelectric
  response of an interacting two-dimensional electron gas in a quantizing
  magnetic field},}\ }\href {\doibase 10.1103/PhysRevB.55.2344} {\bibfield
  {journal} {\bibinfo  {journal} {Phys. Rev. B}\ }\textbf {\bibinfo {volume}
  {55}},\ \bibinfo {pages} {2344--2359} (\bibinfo {year} {1997})}\BibitemShut
  {NoStop}%
\bibitem [{\citenamefont {Koyama}\ and\ \citenamefont
  {Nasu}()}]{koyama_thermal_2024}%
  \BibitemOpen
  \bibfield  {author} {\bibinfo {author} {\bibfnamefont {Shinnosuke}\
  \bibnamefont {Koyama}}\ and\ \bibinfo {author} {\bibfnamefont {Joji}\
  \bibnamefont {Nasu}},\ }\bibfield  {title} {\enquote {\bibinfo {title}
  {Thermal hall effect incorporating magnon damping in localized spin
  systems},}\ }\href {\doibase 10.1103/PhysRevB.109.174442} {\bibfield
  {journal} {\bibinfo  {journal} {Phys. Rev. B}\ }\textbf {\bibinfo {volume}
  {109}},\ \bibinfo {pages} {174442}}\BibitemShut {NoStop}%
\bibitem [{Note5()}]{Note5}%
  \BibitemOpen
  \bibinfo {note} {Note that here the theory is homogeneous (i.e.\ $\protect
  \mathsf H$ only depends on the relative coordinate $x_1-x_2$), but it can be
  made inhomogeneous \protect \textit {a posteriori} by allowing the
  coefficients ${\protect \sf H}_{ab}$ to depend on $(x_1+x_2)/{2}$ and
  reintroducing the Moyal product.}\BibitemShut {Stop}%
\bibitem [{\citenamefont {Auerbach}(1994)}]{auerbach_interacting}%
  \BibitemOpen
  \bibfield  {author} {\bibinfo {author} {\bibfnamefont {Assa}\ \bibnamefont
  {Auerbach}},\ }\href@noop {} {\emph {\bibinfo {title} {Interacting electrons
  and quantum magnetism}}}\ (\bibinfo  {publisher} {Springer-Verlag},\ \bibinfo
  {year} {1994})\BibitemShut {NoStop}%
\bibitem [{Note6()}]{Note6}%
  \BibitemOpen
  \bibinfo {note} {The situation is less transparent for non-canonical bosons,
  for instance the Holstein-Primakoff representation where TR acts as $\protect
  \mathsf b_{\protect \bm {k}} \rightarrow -\protect \mathsf b _{-\protect \bm
  {k}}^\dagger , \protect \mathsf b_{\protect \bm {k}}^\dagger \rightarrow
  -\protect \mathsf b _{-\protect \bm {k}}$.}\BibitemShut {Stop}%
\bibitem [{Note7()}]{Note7}%
  \BibitemOpen
  \bibinfo {note} {Because $\Gamma $-type disorder does not satisfy the
  requirement stated below Eq.\protect \eqref {eq:113}, it cannot be studied
  within our small-deviation expansion.}\BibitemShut {Stop}%
\bibitem [{\citenamefont {Vinkler-Aviv}\ and\ \citenamefont
  {Rosch}(2018)}]{PhysRevX.8.031032}%
  \BibitemOpen
  \bibfield  {author} {\bibinfo {author} {\bibfnamefont {Yuval}\ \bibnamefont
  {Vinkler-Aviv}}\ and\ \bibinfo {author} {\bibfnamefont {Achim}\ \bibnamefont
  {Rosch}},\ }\bibfield  {title} {\enquote {\bibinfo {title} {Approximately
  quantized thermal hall effect of chiral liquids coupled to phonons},}\ }\href
  {\doibase 10.1103/PhysRevX.8.031032} {\bibfield  {journal} {\bibinfo
  {journal} {Phys. Rev. X}\ }\textbf {\bibinfo {volume} {8}},\ \bibinfo {pages}
  {031032} (\bibinfo {year} {2018})}\BibitemShut {NoStop}%
\bibitem [{\citenamefont {Ye}\ \emph {et~al.}(2018)\citenamefont {Ye},
  \citenamefont {Hal\'asz}, \citenamefont {Savary},\ and\ \citenamefont
  {Balents}}]{PhysRevLett.121.147201}%
  \BibitemOpen
  \bibfield  {author} {\bibinfo {author} {\bibfnamefont {Mengxing}\
  \bibnamefont {Ye}}, \bibinfo {author} {\bibfnamefont {G\'abor~B.}\
  \bibnamefont {Hal\'asz}}, \bibinfo {author} {\bibfnamefont {Lucile}\
  \bibnamefont {Savary}}, \ and\ \bibinfo {author} {\bibfnamefont {Leon}\
  \bibnamefont {Balents}},\ }\bibfield  {title} {\enquote {\bibinfo {title}
  {Quantization of the thermal hall conductivity at small hall angles},}\
  }\href {\doibase 10.1103/PhysRevLett.121.147201} {\bibfield  {journal}
  {\bibinfo  {journal} {Phys. Rev. Lett.}\ }\textbf {\bibinfo {volume} {121}},\
  \bibinfo {pages} {147201} (\bibinfo {year} {2018})}\BibitemShut {NoStop}%
\bibitem [{\citenamefont {Kim}\ and\ \citenamefont
  {Kee}(2016)}]{PhysRevB.93.155143}%
  \BibitemOpen
  \bibfield  {author} {\bibinfo {author} {\bibfnamefont {Heung-Sik}\
  \bibnamefont {Kim}}\ and\ \bibinfo {author} {\bibfnamefont {Hae-Young}\
  \bibnamefont {Kee}},\ }\bibfield  {title} {\enquote {\bibinfo {title}
  {Crystal structure and magnetism in
  $\ensuremath{\alpha}\ensuremath{-}{\mathrm{rucl}}_{3}$: An ab initio
  study},}\ }\href {\doibase 10.1103/PhysRevB.93.155143} {\bibfield  {journal}
  {\bibinfo  {journal} {Phys. Rev. B}\ }\textbf {\bibinfo {volume} {93}},\
  \bibinfo {pages} {155143} (\bibinfo {year} {2016})}\BibitemShut {NoStop}%
\bibitem [{\citenamefont {Winter}\ \emph {et~al.}(2016)\citenamefont {Winter},
  \citenamefont {Li}, \citenamefont {Jeschke},\ and\ \citenamefont
  {Valent\'{\i}}}]{PhysRevB.93.214431}%
  \BibitemOpen
  \bibfield  {author} {\bibinfo {author} {\bibfnamefont {Stephen~M.}\
  \bibnamefont {Winter}}, \bibinfo {author} {\bibfnamefont {Ying}\ \bibnamefont
  {Li}}, \bibinfo {author} {\bibfnamefont {Harald~O.}\ \bibnamefont {Jeschke}},
  \ and\ \bibinfo {author} {\bibfnamefont {Roser}\ \bibnamefont
  {Valent\'{\i}}},\ }\bibfield  {title} {\enquote {\bibinfo {title} {Challenges
  in design of kitaev materials: Magnetic interactions from competing energy
  scales},}\ }\href {\doibase 10.1103/PhysRevB.93.214431} {\bibfield  {journal}
  {\bibinfo  {journal} {Phys. Rev. B}\ }\textbf {\bibinfo {volume} {93}},\
  \bibinfo {pages} {214431} (\bibinfo {year} {2016})}\BibitemShut {NoStop}%
\bibitem [{\citenamefont {Wang}\ \emph {et~al.}(2017)\citenamefont {Wang},
  \citenamefont {Dong}, \citenamefont {Yu},\ and\ \citenamefont
  {Li}}]{PhysRevB.96.115103}%
  \BibitemOpen
  \bibfield  {author} {\bibinfo {author} {\bibfnamefont {Wei}\ \bibnamefont
  {Wang}}, \bibinfo {author} {\bibfnamefont {Zhao-Yang}\ \bibnamefont {Dong}},
  \bibinfo {author} {\bibfnamefont {Shun-Li}\ \bibnamefont {Yu}}, \ and\
  \bibinfo {author} {\bibfnamefont {Jian-Xin}\ \bibnamefont {Li}},\ }\bibfield
  {title} {\enquote {\bibinfo {title} {Theoretical investigation of magnetic
  dynamics in $\ensuremath{\alpha}\ensuremath{-}{\mathrm{rucl}}_{3}$},}\ }\href
  {\doibase 10.1103/PhysRevB.96.115103} {\bibfield  {journal} {\bibinfo
  {journal} {Phys. Rev. B}\ }\textbf {\bibinfo {volume} {96}},\ \bibinfo
  {pages} {115103} (\bibinfo {year} {2017})}\BibitemShut {NoStop}%
\bibitem [{\citenamefont {Winter}\ \emph {et~al.}(2017)\citenamefont {Winter},
  \citenamefont {Riedl}, \citenamefont {Maksimov}, \citenamefont {Chernyshev},
  \citenamefont {Honecker},\ and\ \citenamefont
  {Valent{\'\i}}}]{winter2017breakdown}%
  \BibitemOpen
  \bibfield  {author} {\bibinfo {author} {\bibfnamefont {Stephen~M}\
  \bibnamefont {Winter}}, \bibinfo {author} {\bibfnamefont {Kira}\ \bibnamefont
  {Riedl}}, \bibinfo {author} {\bibfnamefont {Pavel~A}\ \bibnamefont
  {Maksimov}}, \bibinfo {author} {\bibfnamefont {Alexander~L}\ \bibnamefont
  {Chernyshev}}, \bibinfo {author} {\bibfnamefont {Andreas}\ \bibnamefont
  {Honecker}}, \ and\ \bibinfo {author} {\bibfnamefont {Roser}\ \bibnamefont
  {Valent{\'\i}}},\ }\bibfield  {title} {\enquote {\bibinfo {title} {Breakdown
  of magnons in a strongly spin-orbital coupled magnet},}\ }\href@noop {}
  {\bibfield  {journal} {\bibinfo  {journal} {Nature communications}\ }\textbf
  {\bibinfo {volume} {8}},\ \bibinfo {pages} {1152} (\bibinfo {year}
  {2017})}\BibitemShut {NoStop}%
\bibitem [{Note8()}]{Note8}%
  \BibitemOpen
  \bibinfo {note} {(in the sense that they contain higher powers of spatial
  derivatives than another similar term already present in the hamiltonian
  density)}\BibitemShut {NoStop}%
\bibitem [{\citenamefont {Landau}\ and\ \citenamefont {Lifshitz}(1970)}]{LL7}%
  \BibitemOpen
  \bibfield  {author} {\bibinfo {author} {\bibfnamefont {Lev~D.}\ \bibnamefont
  {Landau}}\ and\ \bibinfo {author} {\bibfnamefont {Evgeny~M.}\ \bibnamefont
  {Lifshitz}},\ }\href@noop {} {\emph {\bibinfo {title} {Theory of
  Elasticity}}},\ \bibinfo {edition} {2nd}\ ed.\ (\bibinfo  {publisher}
  {Pergamon Press},\ \bibinfo {year} {1970})\BibitemShut {NoStop}%
\bibitem [{\citenamefont {Avron}\ \emph {et~al.}(1995)\citenamefont {Avron},
  \citenamefont {Seiler},\ and\ \citenamefont {Zograf}}]{avron_viscosity_1995}%
  \BibitemOpen
  \bibfield  {author} {\bibinfo {author} {\bibfnamefont {J.~E.}\ \bibnamefont
  {Avron}}, \bibinfo {author} {\bibfnamefont {R.}~\bibnamefont {Seiler}}, \
  and\ \bibinfo {author} {\bibfnamefont {P.~G.}\ \bibnamefont {Zograf}},\
  }\bibfield  {title} {\enquote {\bibinfo {title} {Viscosity of quantum hall
  fluids},}\ }\href {\doibase 10.1103/PhysRevLett.75.697} {\bibfield  {journal}
  {\bibinfo  {journal} {Physical Review Letters}\ }\textbf {\bibinfo {volume}
  {75}},\ \bibinfo {pages} {697--700} (\bibinfo {year} {1995})}\BibitemShut
  {NoStop}%
\bibitem [{\citenamefont {Barkeshli}\ \emph {et~al.}(2012)\citenamefont
  {Barkeshli}, \citenamefont {Chung},\ and\ \citenamefont
  {Qi}}]{barkeshli2012}%
  \BibitemOpen
  \bibfield  {author} {\bibinfo {author} {\bibfnamefont {Maissam}\ \bibnamefont
  {Barkeshli}}, \bibinfo {author} {\bibfnamefont {Suk~Bum}\ \bibnamefont
  {Chung}}, \ and\ \bibinfo {author} {\bibfnamefont {Xiao-Liang}\ \bibnamefont
  {Qi}},\ }\bibfield  {title} {\enquote {\bibinfo {title} {Dissipationless
  phonon {Hall} viscosity},}\ }\href {\doibase 10.1103/PhysRevB.85.245107}
  {\bibfield  {journal} {\bibinfo  {journal} {Phys. Rev. B}\ }\textbf {\bibinfo
  {volume} {85}},\ \bibinfo {pages} {245107} (\bibinfo {year}
  {2012})}\BibitemShut {NoStop}%
\bibitem [{\citenamefont {Saito}\ \emph {et~al.}(2019)\citenamefont {Saito},
  \citenamefont {Misaki}, \citenamefont {Ishizuka},\ and\ \citenamefont
  {Nagaosa}}]{saito2019berry}%
  \BibitemOpen
  \bibfield  {author} {\bibinfo {author} {\bibfnamefont {Takuma}\ \bibnamefont
  {Saito}}, \bibinfo {author} {\bibfnamefont {Kou}\ \bibnamefont {Misaki}},
  \bibinfo {author} {\bibfnamefont {Hiroaki}\ \bibnamefont {Ishizuka}}, \ and\
  \bibinfo {author} {\bibfnamefont {Naoto}\ \bibnamefont {Nagaosa}},\
  }\bibfield  {title} {\enquote {\bibinfo {title} {Berry phase of phonons and
  thermal {Hall} effect in nonmagnetic insulators},}\ }\href {\doibase
  10.1103/PhysRevLett.123.255901} {\bibfield  {journal} {\bibinfo  {journal}
  {Phys. Rev. Lett.}\ }\textbf {\bibinfo {volume} {123}},\ \bibinfo {pages}
  {255901} (\bibinfo {year} {2019})}\BibitemShut {NoStop}%
\bibitem [{\citenamefont {Ye}\ \emph {et~al.}(2021)\citenamefont {Ye},
  \citenamefont {Savary},\ and\ \citenamefont {Balents}}]{ye2021}%
  \BibitemOpen
  \bibfield  {author} {\bibinfo {author} {\bibfnamefont {Mengxing}\
  \bibnamefont {Ye}}, \bibinfo {author} {\bibfnamefont {Lucile}\ \bibnamefont
  {Savary}}, \ and\ \bibinfo {author} {\bibfnamefont {Leon}\ \bibnamefont
  {Balents}},\ }\href {https://arxiv.org/abs/2103.04223} {\enquote {\bibinfo
  {title} {Phonon hall viscosity in magnetic insulators},}\ } (\bibinfo {year}
  {2021}),\ \Eprint {http://arxiv.org/abs/2103.04223} {arXiv:2103.04223
  [cond-mat.str-el]} \BibitemShut {NoStop}%
\bibitem [{\citenamefont {McClarty}(2022)}]{mcclarty2022topological}%
  \BibitemOpen
  \bibfield  {author} {\bibinfo {author} {\bibfnamefont {Paul~A}\ \bibnamefont
  {McClarty}},\ }\bibfield  {title} {\enquote {\bibinfo {title} {Topological
  magnons: A review},}\ }\href@noop {} {\bibfield  {journal} {\bibinfo
  {journal} {Annual Review of Condensed Matter Physics}\ }\textbf {\bibinfo
  {volume} {13}},\ \bibinfo {pages} {171--190} (\bibinfo {year}
  {2022})}\BibitemShut {NoStop}%
\bibitem [{\citenamefont {Mook}\ \emph {et~al.}(2017)\citenamefont {Mook} \emph
  {et~al.}}]{mook2017topological}%
  \BibitemOpen
  \bibfield  {author} {\bibinfo {author} {\bibfnamefont {Alexander}\
  \bibnamefont {Mook}} \emph {et~al.},\ }\emph {\bibinfo {title} {Topological
  Magnon Materials and Transverse Magnon Transport}},\ \href@noop {} {Ph.D.
  thesis},\ \bibinfo  {school} {Universit{\"a}ts-und Landesbibliothek
  Sachsen-Anhalt} (\bibinfo {year} {2017})\BibitemShut {NoStop}%
\bibitem [{Note9()}]{Note9}%
  \BibitemOpen
  \bibinfo {note} {We note that the physics of thermal transport in fully
  nonmagnetic $\protect \rm Y_2Ti_2O_7$ \cite {sharma2024phonon} appears to be
  quite different.}\BibitemShut {Stop}%
\bibitem [{\citenamefont {Behnia}(2025)}]{behnia2025phononthermalhalllattice}%
  \BibitemOpen
  \bibfield  {author} {\bibinfo {author} {\bibfnamefont {Kamran}\ \bibnamefont
  {Behnia}},\ }\bibfield  {title} {\enquote {\bibinfo {title} {Phonon thermal
  hall as a lattice aharonov-bohm effect},}\ }\href@noop {} {\bibfield
  {journal} {\bibinfo  {journal} {arXiv preprint arXiv:2502.18236}\ } (\bibinfo
  {year} {2025})}\BibitemShut {NoStop}%
\bibitem [{\citenamefont {Park}\ \emph {et~al.}(2025)\citenamefont {Park},
  \citenamefont {Huang}, \citenamefont {Savary},\ and\ \citenamefont
  {Balents}}]{park2025quantum}%
  \BibitemOpen
  \bibfield  {author} {\bibinfo {author} {\bibfnamefont {Takamori}\
  \bibnamefont {Park}}, \bibinfo {author} {\bibfnamefont {Xiaoyang}\
  \bibnamefont {Huang}}, \bibinfo {author} {\bibfnamefont {Lucile}\
  \bibnamefont {Savary}}, \ and\ \bibinfo {author} {\bibfnamefont {Leon}\
  \bibnamefont {Balents}},\ }\bibfield  {title} {\enquote {\bibinfo {title}
  {Quantum geometry from the moyal product: quantum kinetic equation and
  non-linear response},}\ }\href@noop {} {\bibfield  {journal} {\bibinfo
  {journal} {arXiv preprint arXiv:2504.10447}\ } (\bibinfo {year}
  {2025})}\BibitemShut {NoStop}%
\bibitem [{\citenamefont {Dhakal}\ \emph {et~al.}(2024)\citenamefont {Dhakal},
  \citenamefont {Kaib}, \citenamefont {Choi}, \citenamefont {Biswas},
  \citenamefont {Valenti},\ and\ \citenamefont {Winter}}]{dhakal2024theory}%
  \BibitemOpen
  \bibfield  {author} {\bibinfo {author} {\bibfnamefont {Ramesh}\ \bibnamefont
  {Dhakal}}, \bibinfo {author} {\bibfnamefont {David~AS}\ \bibnamefont {Kaib}},
  \bibinfo {author} {\bibfnamefont {Kate}\ \bibnamefont {Choi}}, \bibinfo
  {author} {\bibfnamefont {Sananda}\ \bibnamefont {Biswas}}, \bibinfo {author}
  {\bibfnamefont {Roser}\ \bibnamefont {Valenti}}, \ and\ \bibinfo {author}
  {\bibfnamefont {Stephen~M}\ \bibnamefont {Winter}},\ }\bibfield  {title}
  {\enquote {\bibinfo {title} {Theory of intrinsic phonon thermal hall effect
  in $\ensuremath{\alpha}{\mathrm{-rucl}}_{3}$},}\ }\href@noop {} {\bibfield
  {journal} {\bibinfo  {journal} {arXiv preprint arXiv:2407.00660}\ } (\bibinfo
  {year} {2024})}\BibitemShut {NoStop}%
\bibitem [{\citenamefont {Sharma}\ \emph {et~al.}(2024)\citenamefont {Sharma},
  \citenamefont {Valldor},\ and\ \citenamefont {Lorenz}}]{sharma2024phonon}%
  \BibitemOpen
  \bibfield  {author} {\bibinfo {author} {\bibfnamefont {Rohit}\ \bibnamefont
  {Sharma}}, \bibinfo {author} {\bibfnamefont {Martin}\ \bibnamefont
  {Valldor}}, \ and\ \bibinfo {author} {\bibfnamefont {Thomas}\ \bibnamefont
  {Lorenz}},\ }\bibfield  {title} {\enquote {\bibinfo {title} {Phonon thermal
  hall effect in nonmagnetic y 2 ti 2 o 7},}\ }\href@noop {} {\bibfield
  {journal} {\bibinfo  {journal} {Physical Review B}\ }\textbf {\bibinfo
  {volume} {110}},\ \bibinfo {pages} {L100301} (\bibinfo {year}
  {2024})}\BibitemShut {NoStop}%
\bibitem [{\citenamefont {Mahan}(2000)}]{mahan}%
  \BibitemOpen
  \bibfield  {author} {\bibinfo {author} {\bibfnamefont {G.D.}\ \bibnamefont
  {Mahan}},\ }\href@noop {} {\emph {\bibinfo {title} {Many-Particle
  Physics}}},\ \bibinfo {edition} {3rd}\ ed.\ (\bibinfo  {publisher} {Kluwer
  Academic Plenum Publishers, New York},\ \bibinfo {year} {2000})\BibitemShut
  {NoStop}%
\end{thebibliography}%

\appendix

\begin{widetext}
  
\section{Derivation of the kinetic equation: calculatory details}
\label{sec:deriv-kinet-equat-calc-det}

We briefly consider Eq.\eqref{eq:14}.
Defining the Fourier transform as $\ms O(x)=\int_p e^{i p x/\hbar} \, \tilde{\ms O}(p)$,
and henceforth omitting to write the tildes (except on $w(r) \overset{\rm FT}\rightarrow \tilde w(k)$ for clarity), one obtains

\begin{align}
  \label{eq:42}
  \langle (\ms V_{\rm d}^s)_{n,n'}(z_1,z_2)  (\ms V^{s'}_{\rm d})_{n',n}(z_3,z_4) \rangle
  &= \int_{p,p',k} \tilde w(k) \; e^{is(p+k)z_1} e^{-is pz_2}  e^{is'p' z_3} e^{-is'(p'+k)z_4} \\
  &\qquad \qquad \times \left [ \ms S^{-1}(p+k)  {\ms W}(p+k/2)  \ms S(p) \right ]^s_{n,n'} \;  \left [  \ms S^{-1}(p') {\ms W}(p'+k/2) \ms S(p'+k) \right ]^{s'}_{n',n} . \nonumber   
\end{align}
Here and in the following we omit to write the position variable of $\ms S$ which is irrelevant to the order considered: the main contribution to the side-jump
effect happens at the order $\hbar^1$ and inhomogeneities of the clean theory may only contribute to $O(\hbar^2)$.

We now start from Eq.\eqref{eq:13}.
Changing variables to highlight the common structure of all four terms, and omitting the symbol for disorder average as explained in the main text, one obtains
\begin{align}
  \label{eq:43}
& i \hbar  \partial_t({\sf F}_{\rm d})_{n,n}(x,y,t) - [{\sf K}_{\rm d}\,\overset{\circ},\,{\sf F}_{\rm d}]_{n,n}(x,y,t) \\
  &=- \frac i \hbar \int \text d \tau \,\theta(t-\tau)\int_{\xi,\xi',u,v,\zeta,\zeta'}\,  (\mc U_0)_{l,l}(u,t|\xi,\tau)  (\ms F_{\rm d})_{m,m}(\zeta,\zeta',\tau) (\mc U_0)_{l',l'}(\xi',\tau|v,t) \nonumber \\
  &\times \Big [  \langle (\ms V_{\rm d})_{n,n'}(x,u)  (\ms V_{\rm d})_{n',n}(\xi,\zeta) \rangle \delta_{(\zeta'-\xi')} \delta_{(v-y)} \delta_{l,n'}\delta_{l',n} \delta_{m,n}
    - \langle (\ms V_{\rm d})_{n,n'} (\xi,\zeta)   (\ms V_{\rm d}^\dagger)_{n',n} (v,y) \rangle \delta_{(x-u)}   \delta_{(\zeta'-\xi')} \delta_{l,n}\delta_{l',n'} \delta_{m,n'}\nonumber \\
  &-  \langle (\ms V_{\rm d})_{n,n'} (x,u)    (\ms V_{\rm d}^\dagger)_{n',n} (\zeta',\xi') \rangle \delta_{(\xi-\zeta)} \delta_{(v-y)} \delta_{l,n'}\delta_{l',n} \delta_{m,n'}
    + \langle (\ms V_{\rm d}^\dagger)_{n,n'} (\zeta',\xi')  (\ms V_{\rm d}^\dagger)_{n',n} (v,y) \rangle  \delta_{(x-u)} \delta_{(\xi-\zeta)} \delta_{l,n}\delta_{l',n'} \delta_{m,n}  \rangle    \Big ] \nonumber,
\end{align}
where the disorder-free evolution operator is
\begin{align}
  \label{eq:44}
  \mc U_0 (r,t|r',t') &= -i \; {\rm sign}(t-t')\int_p e^{\frac i \hbar p(r-r')} \,e^{-\frac i \hbar \ms K_{\rm d}(\frac{r+r'}{2},p)\; |t-t' |} .
\end{align}

For convenience in intermediate calculations, we now perform a Fourier transform on the position variable of Wigner-transformed quantities:
\begin{align}
  \label{eq:45}
 {\ms F}_{\rm d}(\zeta,\zeta',\tau) &= \int_{p}  e^{+\frac i \hbar p (\zeta-\zeta')} \, \ms F_{\rm d}(\tfrac{\zeta+\zeta'}2 , p ,\tau)
                                     \equiv \int_{p,\Pi}  e^{+\frac i \hbar p (\zeta-\zeta')} \, e^{i \Pi \frac{\zeta+\zeta'}2} \,  \breve{\ms F}_{\rm d} (\Pi, p ,\tau) ,
\end{align}
where the first ${\ms F}_{\rm d}$ is in real space, the second ${\ms F}_{\rm d}$ is its Wigner transform, and $\breve{\ms F}_{\rm d}$ is the Fourier transform of the latter.
Similarly, 
\begin{align}
  \label{eq:46}
  (\mc U_0)_{l,l}(u,t|\xi,\tau)  (\mc U_0)_{l',l'}(\xi',\tau|v,t)
 & = \int_{q,q',\Pi,\Pi'} e^{i \Pi\frac{u+\xi}{2}} e^{i \Pi' \frac{v+\xi'}{2}}  e^{\frac i \hbar q (u-\xi)}  e^{- \frac i \hbar q' (v-\xi')}
    e^{-\frac i \hbar (t-\tau) ((\ms K_{\rm d})_{l,l}(\Pi,q)-(\ms K_{\rm d})_{l',l'}(\Pi',q'))} .
\end{align}

Doing the replacements Eqs.\ \eqref{eq:44}, \eqref{eq:45}, \eqref{eq:46}, into Eq.\eqref{eq:43},
and computing the integrals over variables $\xi,\xi',u,v,\zeta,\zeta'$, one obtains 
\begin{align}
  \label{eq:47}
  & i \hbar  \partial_t({\sf F}_{\rm d})_{n,n}(x,y,t) - [{\sf K}_{\rm d}\,\overset{\circ},\,{\sf F}_{\rm d}]_{n,n}(x,y,t) \nonumber \\
  &= - \frac i \hbar \int \text d \tau \theta(t-\tau) \int_{p,p',K,\Pi,\Pi'}  \tilde w(p-p') \,  e^{i(K+\Pi+\Pi')\frac{x+y}2}\\
  & \Bigg \{ e^{i(p+(\Pi-\Pi')/2) (x-y)} \,e^{-\frac i \hbar (t-\tau) ((\breve{\ms K}_{\rm d})_{n',n'}(\Pi, p'+K/2-\Pi/2)-(\breve{\ms K}_{\rm d})_{n,n} (\Pi',p-K/2-\Pi'/2))} (\breve{\ms F}_{\rm d})_{n,n}(K,p ,\tau) \nonumber \\
  & \times  \left [ \ms S^{-1} (p+K/2) \ms W( \tfrac{p+p'}2+K/2) \ms S(p'+K/2) \right ]_{n,n'}
    \left [ \ms S^{-1} (p'+K/2-\Pi) \ms W(\tfrac{p+p'}2+K/2-\Pi) \ms S(p+K/2-\Pi) \right ]_{n',n} \nonumber \\
  &-  e^{i(p +(\Pi-\Pi')/2) (x-y)} \,e^{-\frac i \hbar (t-\tau) ((\breve{\ms K}_{\rm d})_{n,n} (\Pi,p+K/2+\Pi/2)-(\breve{\ms K}_{\rm d})_{n',n'} (\Pi',p'-K/2-\Pi'/2))}(\breve{\ms F}_{\rm d})_{n',n'} (K, p',\tau) \nonumber \\
  & \times  \left [ \ms S^{-1} (p+K/2) \ms W(\tfrac{p+p'}2+K/2) \ms S (p'+K/2) \right ]_{n,n'}
  \left [ \ms S^{-1} (p-K/2-\Pi') \ms W(\tfrac{p+p'}2-K/2-\Pi')  \ms S (p'-K/2-\Pi') \right ]^\dagger_{n',n} \nonumber \\
  & -   e^{i(p +(\Pi-\Pi')/2) (x-y)} \, e^{-\frac i \hbar (t-\tau) ((\breve{\ms K}_{\rm d})_{n',n'} (\Pi,p'+K/2+\Pi/2)-(\breve{\ms K}_{\rm d})_{n,n} (\Pi',p-K/2-\Pi'/2))}(\breve{\ms F}_{\rm d})_{n',n'} (K,p',\tau) \nonumber \\
  &  \times   \left [\ms S^{-1} (p+K/2) \ms W(\tfrac{p+p'}2+K/2)\ms S (p'+K/2) \right ]_{n,n'}
  \left [ \ms S^{-1} (p-K/2-\Pi) \ms W (\tfrac{p+p'}2-K/2-\Pi)   \ms S (p'-K/2-\Pi) \right ]^\dagger_{n',n} \nonumber \\
  &+  e^{i(p+(\Pi-\Pi')/2) (x-y)} \, e^{-\frac i \hbar (t-\tau) ((\breve{\ms K}_{\rm d})_{n,n} (\Pi, p+K/2+\Pi/2)-(\breve{\ms K}_{\rm d})_{n',n'} (\Pi',p'-K/2+\Pi'/2))}(\breve{\ms F}_{\rm d})_{n,n} (K,p ,\tau) \nonumber \\
  & \times \left [ \ms S^{-1} (p'-K/2) \ms W (\tfrac{p+p'}2-K/2)   \ms S (p-K/2) \right ]^\dagger_{n,n'}
   \left [\ms S^{-1}  (p-K/2-\Pi')  \ms W (\tfrac{p+p'}2-K/2-\Pi')   \ms S(p'-K/2-\Pi') \right ]^\dagger_{n',n} \Bigg \} . \nonumber
\end{align}

We now perform the Wigner transform with respect to $(x,y)$ on both sides, by writing $X=\frac{x+y}2$ and performing the Fourier transform with respect to $x-y$.
We further perform one momentum integral, resolve one momentum conservation delta-function, identify the matrix elements $\ms M_{n,n'}(p,p')$,
perform momentum shifts by amounts $\pm\Pi,\pm\Pi'$ (different for each term in the sum)
and neglect such $K,\Pi,\Pi'$ dependence in the second (i.e. fast) variable of $\breve{\ms F}_{\rm d}, \breve{\ms K}_{\rm d}$ as explained in the main text, Sec.\ \ref{sec:phase-space-form}.

One then recognizes -- up to changes of variables $\Pi \leftrightarrow \Pi'$ and $(K,\Pi,\Pi') \rightarrow (K,\Pi,\Pi')$ -- that complex-conjugated terms appear pairwise on the right-hand side,
so that one eventually obtains
\begin{align}
  \label{eq:48}
   & i \hbar  \partial_t({\sf F}_{\rm d})_{n,n}(X,p,t) - [{\sf K}_{\rm d}\,\overset{\star},\,{\sf F}_{\rm d}]_{n,n}(X,p,t) \nonumber \\
   &= - \frac i \hbar \int \text d \tau \theta(t-\tau) \; 2{\rm Re} \int_{p',K,\Pi,\Pi'}  \tilde w(p-p') \,  e^{i(K+\Pi+\Pi')X}  \\
  & \times \sum_{n'}\Big \{ e^{-\frac i \hbar (t-\tau) ((\breve{\ms K}_{\rm d})_{n',n'}(\Pi, p')-(\breve{\ms K}_{\rm d})_{n,n} (\Pi',p))}\nonumber \\
   & \qquad \qquad \qquad\times \ms M_{n,n'}(p+\tfrac K2+\tfrac \Pi 2,p'+\tfrac K 2+ \tfrac \Pi 2)\; \ms M_{n',n} (p'+\tfrac K2-\tfrac \Pi2, p+\tfrac K 2- \tfrac \Pi 2) \;  (\breve{\ms F}_{\rm d})_{n,n}(K,p ,\tau)
                      \vphantom{  ._{ \substack{p \rightarrow p - (\Pi-\Pi')/2 ,\\ p' \rightarrow p' - (\Pi-\Pi')/2} } }  \nonumber \\
  &\qquad \qquad-  e^{-\frac i \hbar (t-\tau) ((\breve{\ms K}_{\rm d})_{n,n} (\Pi,p)-(\breve{\ms K}_{\rm d})_{n',n'} (\Pi',p'))}\nonumber \\
   & \qquad \qquad \qquad \times \ms M_{n,n'}(p+\tfrac K 2+ \tfrac{\Pi'}2,p'+\tfrac K 2+\tfrac{\Pi'}2 )\;  \ms M_{n,n'} (p-\tfrac K 2- \tfrac{\Pi'} 2,p'- \tfrac K 2-\tfrac{\Pi'} 2)^* \;
     (\breve{\ms F}_{\rm d})_{n',n'} (K, p',\tau)  \Big \}_{ \substack{p \rightarrow p - (\Pi-\Pi')/2 ,\\ p' \rightarrow p' - (\Pi-\Pi')/2} }. \nonumber
\end{align}

We now make the approximation discussed in the main text, Eq.\eqref{eq:16}, then perform the $K,\Pi,\Pi'$ integrations and take the semiclassical limit of time evolution, Eq.\eqref{eq:15}: one obtains
\begin{align}
  \label{eq:49}
  i \hbar  \partial_t & ({\sf F}_{\rm d})_{n,n}(X,p,t) - [{\sf K}_{\rm d}\,\overset{\star},\,{\sf F}_{\rm d}]_{n,n}(X,p,t) \\
  &= -i2 \pi \sum_{n'}\int_{p'} \tilde w(p-p')  \bigg [ (\ms F_{\rm d})_{n,n}(X+ \delta {\ms r}_{n,n'}^{+}(p,p') ,p ,t)\;  {\rm Re} \left [ \ms M_{n,n'}(p,p')\, \ms M_{n',n}(p',p) \right ] \nonumber \\
                    &\qquad \qquad \qquad \qquad\qquad  \qquad \times \delta \left ( (\ms K_{\rm d})_{n',n'}(X + \delta {\ms r}_{n,n'}^{-}(p,p') - \delta {\ms r}_{n,n'}^{+}(p,p')  , p') - (\ms K_{\rm d})_{n,n} (X + \delta {\ms r}_{n,n'}^{+}(p,p')  , p) \right )\nonumber \\
  & \qquad \qquad \qquad\qquad - (\ms F_{\rm d})_{n',n'} (X +\delta {\ms r}_{n,n'}^{+}(p,p') +\delta {\ms r}_{n,n'}^{-}(p,p') , p',t) \;  {\rm Re} \left [ \ms M_{n,n'}(p,p') \,\ms M_{n,n'}(p,p')^* \right ]  \nonumber \\
   & \qquad \qquad \qquad\qquad \qquad \times \delta \left ( (\ms K_{\rm d})_{n',n'}(X +\delta {\ms r}_{n,n'}^{+}(p,p') +\delta {\ms r}_{n,n'}^{-}(p,p') ,p') - (\ms K_{\rm d})_{n,n} (X, p) \right ) \bigg ] ,\nonumber
\end{align}
where one defined $\delta \ms r_{n,n'}^{\pm}(p,p') \equiv  \tfrac 1 2 [\delta \ms r_{n,n'}(p,p') \pm\delta \ms r_{n',n}(p',p)] $ for convenience.
Further neglecting $\delta \ms r_{n,n'}^+(p,p')$, as argued in Appendix \ref{sec:argument-1},  we obtain Eq.\ \eqref{eq:97} in the main text.

\section{Small momentum expansions}
\label{sec:kinet-coord-comm}

\subsection{Obtaining the disorder-induced curvature}
\label{sec:obta-disord-berry}

At small deviations $|p-p'| \ll p,p'$, such that $\ms S(p') \approx \ms S(p)+(p'_\nu-p_\nu)\partial_{p_\nu} \ms S(p)$, we obtain
\begin{align}
  \label{eq:92}
 \Big [ \ms S^{-1}(p) \ms W(\tfrac{p+p'}2) &\ms S(p') \Big ]_{n,n'}
           \approx  \left [ \ms S^{-1}(p) \ms W(p) \ms S(p) \right ]_{n,n'}\times \nonumber \\
  &    \times  \bigg ( 1 + (p'_\nu-p_\nu) \frac{ \left [ \ms S^{-1}(p) \ms W(p) \partial_{p_\nu} \ms S(p)  \right ]_{n,n'}}{ \left [ \ms S^{-1}(p) \ms W(p) \ms S(p) \right ]_{n,n'} }
              +  \tfrac 1 2 (p'_\nu-p_\nu) \frac{\left [ \ms S^{-1}(p) \partial_{p_\nu} \ms W(p) \ms S(p)  \right ]_{n,n'} }{ \left [ \ms S^{-1}(p) \ms W(p) \ms S(p) \right ]_{n,n'} } \bigg ),
\end{align}
which is only a valid expansion provided $\left [ \ms S^{-1}(p) \ms W(p) \ms S(p) \right ]_{n,n'} \neq 0$.
We then take the argument and use ${\rm Arg}(z_1 z_2)={\rm Arg}(z_1)+{\rm Arg}(z_2)$ and ${\rm Arg}(1+z)\approx {\rm Im}(z)$ for $|z| \ll 1$.
We then apply $ (\partial_p+\partial_{p'}) $, which to the order $O(p-p')$ yields
\begin{align}
  \label{eq:93}
\delta \ms r_{n,n'}^{-,\mu}(p,p') 
  &= \tfrac 1 2  \partial_{p_\mu} {\rm Arg} \left [  \ms S^{-1}(p) \ms W(p) \ms S(p)  \right ]_{n,n'}  + \tfrac 1 2 \delta x^\mu_{n,n'}(p)  + \tfrac 1 2 \delta x^\mu_{n',n}(p)   \nonumber \\
    & -  \tfrac 1 2  \partial_{p'_\mu} {\rm Arg} \left [  \ms S^{-1}(p') \ms W(p') \ms S(p')  \right ]_{n',n} -  \tfrac 1 2 \delta x^\mu_{n',n}(p') -  \tfrac 1 2 \delta x^\mu_{n,n'}(p') , \\
  \delta x^\mu_{n,n'}(p) &= - p_\nu \partial_{p_\mu} \,  {\rm Im} \left ( \frac{[\ms S^{-1}(p) \ms W(p) \partial_{p_\nu}\ms S(p)]_{n,n'}}{[\ms S^{-1}(p) \ms W(p) \ms S(p)]_{n,n'}}   \right )
                            - \tfrac 1 2 p_\nu \partial_{p_\mu} \,  {\rm Im} \left ( \frac{[\ms S^{-1}(p) \partial_{p_\nu}\ms W(p) \ms S(p)]_{n,n'}}{[\ms S^{-1}(p) \ms W(p) \ms S(p)]_{n,n'}}   \right )  .
\end{align}

Here is is easy to check gauge invariance under $\ms S \rightarrow \ms S e^{i\stheta}$ of  $ \delta \ms X^\mu_{n,n'}(p,p')$ defined in Eq.\eqref{eq:21}.
Indeed, the gauge transformation rules are $\ms  A^\mu_n(p)  \rightarrow \ms  A^\mu_n(p) + \partial_{p_\mu}\theta_n(p) $ and $\delta x^\mu_{n,n'}(p)  \rightarrow \delta x^\mu_{n,n'}(p) - p_\nu \partial^2_{\mu\nu} \theta_{n'}$
(note the second term in $\delta x^\mu_{n,n'}(p) $ is gauge-invariant by itself), and
\begin{align}
  \label{eq:100}
 \tfrac 1 2  \partial_{p_\mu} {\rm Arg} \left [  \ms S^{-1}(p) \ms W(p) \ms S(p)  \right ]_{n,n'}
& \rightarrow  {\rm idem} -  \tfrac 1 2  \partial_{p_\mu}\theta_{n}(p) + \tfrac 1 2  \partial_{p'_\mu}\theta_{n'}(p') +  \tfrac 1 2 (p_\nu-p'_\nu) \partial^2_{\mu \nu}\theta_{n'}  ,\\
  -  \tfrac 1 2  \partial_{p'_\mu} {\rm Arg} \left [  \ms S^{-1}(p') \ms W(p') \ms S(p')  \right ]_{n',n}
                  & \rightarrow  {\rm idem} +  \tfrac 1 2  \partial_{p'_\mu}\theta_{n'}(p') -  \tfrac 1 2  \partial_{p_\mu}\theta_{n}(p) -  \tfrac 1 2 (p'_\nu-p_\nu) \partial^2_{\mu \nu}\theta_{n} , 
\end{align}
where $\partial^2_{\mu \nu}\theta $ means $\partial^2_{p_\mu p_\nu}\theta(p) \approx \partial^2_{p'_\mu p'_\nu}\theta(p') $, since apostrophes are irrelevant to the order considered.
This shows the gauge invariance of Eq.\eqref{eq:21}.
Now considering specifically diagonal elements $n=n'$ and collecting all terms, one obtains Eq.\eqref{eq:23} in the main text.

\subsection{Case where disorder has the same matrix structure as the clean hamiltonian}
\label{sec:case-where-disorder}

Here we investigate the particular case where $\hat W = \hat{\ms H}$ and thus $\ms W = \ms K$. Then one simply has
       \begin{align}
         \label{eq:83}
         (\ms \Omega^{\ms K}_{p_\mu p_\nu})_{n,n}(p) &=  (\ms \Omega_{p_\mu p_\nu})_{n,n}(p)
                                                     + \tfrac 1 2   \partial_{p_\mu} \,  {\rm Im} \left ( [\ms S^{-1}(p) \partial_{p_\nu} \ms K(p) \ms S(p) ] _{n,n} / [\ms K_{\rm d} ] _{n,n}   \right )  \nonumber \\
         &=  (\ms \Omega_{p_\mu p_\nu})_{n,n}(p)
           + \tfrac 1 2   \partial_{p_\mu}  \,  {\rm Im} \left ( [\partial_{p_\nu} \ms K_{\rm d} + [\ms \Lambda_{p_\nu} , \ms K_{\rm d} ] ] _{n,n} / [\ms K_{\rm d} ] _{n,n}   \right )
           =  (\ms \Omega_{p_\mu p_\nu})_{n,n}(p) ,
       \end{align}
          where $\ms \Lambda_{p_\nu} =  \ms S^{-1} \partial_{p_\nu}\ms S $,   which is Eq.\eqref{eq:38} in the main text.
We also note that in this case,
       \begin{align}
         \label{eq:95}
         \ms S^{-1}(p)\ms K(\tfrac{p+p'}2)\ms S(p') &= \ms K_{\rm d}(\tfrac{p+p'}2) 
                                                      + \tfrac 1 2 (p'_\mu-p_\mu) \left \{ \ms K_{\rm d}(\tfrac{p+p'}2), \ms \Lambda_{p_\mu}(\tfrac{p+p'}2) \right \} +\dots .
       \end{align}
 
     \end{widetext}

     \section{Collection of technical arguments}
\label{sec:coll-techn-argum}

\subsection{Neglecting $\delta \ms r^+$:}
\label{sec:argument-1}

We first note that $\delta \ms r^+$ is gauge-invariant, therefore taking $\delta \ms r^+ \rightarrow 0$ is a well-defined approximation.
Note that in the case of fermions in a random scalar potential, $\delta \ms r^+=0$ is rigorous.
We now argue why it can also be neglected in the cases we are considering in this paper.

In Eq.\eqref{eq:49}, one sees that $\delta \ms r^+$ appears with the same sign in the position coordinate of both $({\sf F}_{\rm d})_{n,n}$ and $({\sf F}_{\rm d})_{n',n'}$ on the right-hand side,
so that it can be absorbed in a global shift of $X$ without any physical consequences (to the order $\hbar^1$).

By contrast, its occurrences in the position coordinate of ${\sf K}_{\rm d}$ on the right-hand side cannot be absorbed in such a shift.
However, their contribution to the collision integral is proportional to $\hbar^1 \partial_X\ms K_{\rm d}$, which can be safely neglected in the homogeneous limit of the bulk theory.
This justifies the assumption $\delta \ms r^+ \rightarrow 0$.

\subsection{Neglecting interband $(n \neq n')$ terms:}
\label{sec:argument-2}

Here we argue, based on the matrix elements $\ms M_{n,n'}$, that interband effects can be neglected,
at least in the particular case $\ms W = \ms K$ -- for a more generally valid proof based on elastic scattering
and the small-deviation approximation, see the main text.

For $\ms W = \ms K$, one sees from Eqs.\eqref{eq:17a},\eqref{eq:95} that within the small-deviation approximation,
\begin{align}
  \label{eq:116}
  \ms M_{n,n'}(p,p') = (\ms K_{\rm d})_{n,n'}(\tfrac{p+p'}2)  + O(p'-p) ,
\end{align}
where we recall that $\ms K_{\rm d}$ is a diagonal matrix.
Therefore, the moduli of off-band-diagonal elements $\left |\ms M_{n,n'}(p,p') \right |^2$ with $n \neq n'$
are smaller by a factor of the order $|p'-p|$ than those of band-diagonal elements.
Now, the important quantity whose diagonal and off-diagonal elements need be compared is
\begin{align}
  \label{eq:115}
  \delta \ms X_{n,n'}(p,p') \; \left | \ms M_{n,n'}(p,p')\right |^2 ,
\end{align}
as is visible e.g.\ in Eqs.\eqref{eq:103},\eqref{eq:56}.
From Eq.\eqref{eq:23} one finds $\delta \ms X_{n,n} = O(p'-p)$ while a priori $\delta \ms X_{n,n'} = O(1)$ for $n\neq n'$.
Thus, ultimately Eq.\eqref{eq:115} is smaller by one power of $|p'-p|$ for off-diagonal elements than for diagonal ones.
This concludes the argument.

\subsection{Replacing $\ms M_{n,n'}(p,p') \rightarrow \left | \ms M_{n,n'}(p,p') \right |$:}
\label{sec:repl-msm-absmsm}

From the small-deviation expansion Eq.\eqref{eq:92}, it is clear that replacing $\ms M_{n,n'}(p,p')$ by its modulus will be justified, to the leading order, provided that
\begin{equation}
\label{eq:1189}
  \left [  \ms S^{-1}(p) \ms W(p) \ms S(p)  \right ]_{n,n'} \; \overset ? \in \; \mathbb R ,
\end{equation}
at least for $n=n'$ (see App.\ref{sec:argument-2}) and for all $p$. We note that because matrix elements $\ms M_{n,n'}(p,p')$ appear pairwise in the collision integral, the condition ``$\in  \mathbb R $'' (instead of ``$>0$'') is sufficient.

In the particular case $\ms W = \ms K$, the validity of Eq.\eqref{eq:1189} follows directly from Eq.\eqref{eq:116}. More generally, all the cases we consider for applications (see Appendix \ref{sec:applications-details}) satisfy this constraint, although this is not \emph{a priori} a general truth. This justifies the replacement here.

\subsection{Validity of the approximation Eq.\eqref{eq:89}:}
\label{sec:valid-appr-eq.eqr}

In the particular case $\ms W = \ms K$, the approximation Eq.\eqref{eq:89} is justified.
Indeed, focusing on band-diagonal terms $n=n'$ as justified above,
it follows immediately from Eq.\eqref{eq:116} to the zeroth order in $|p'-p|$, using the fact that  $(\ms K_{\rm d})_{n,n'} \in \mathbb R$.

\section{Derivation of the accumulation term in the energy current}
\label{sec:deriv-accum-term}

\subsection{Identifying the side-jump current}
\label{sec:first-identification}

We derive the energy current from the conservation equation in phase space, following the same steps as \cite{mangeolle_quantum_2024}.
The phase-space energy density in the presence of disorder $V$ is
\begin{align}
  \label{eq:50}
    \mathcal{H}(X,p)= \frac{1}{4}{\rm Tr}\left\{{\sf H}+ V \,\overset \star ,\, {\sf F}\right\},
\end{align}
whose time evolution is given by
\begin{align}
  \label{eq:51}
   \partial_t\mathcal{H}=\frac{1}{2\hbar}{\rm Im}{\rm Tr}\left([{\sf K} + {\ms V}\,\overset{\star},\,{\sf F}\star ({\sf H} + V  ) ]\right) .
\end{align}
Then after a very short calculation, the phase space energy current, up to a total phase-space derivative and to the order $O(\hbar^1)$, is 
\begin{align}
  \label{eq:52}
  \mathcal{J}_\alpha(X,p)=\frac{1}{2}\omega_{\alpha\beta}{\rm Re}\, {\rm Tr}\left [
  \big( \partial_\beta({\sf K}+{\ms V}) \big ) \big  ({\sf F}\star ({\sf H} + { V} )  \big )\right), 
\end{align}
where $\alpha,\beta=X_\mu,p_\nu$ and $\omega_{X_\mu,p_\nu}=\delta_{\mu,\nu}= - \omega_{p_\nu,X_\mu}$.

Then to the first order in $\hbar$, it will be sufficient to replace the star-product by a ``normal'' product
in those terms of the current that contain disorder operators $\ms V$:
\begin{subequations}
\begin{align}
  \label{eq:53a}
   {\mc J}_{\alpha}(X,p)  &=  \frac{1}{2}\omega_{\alpha\beta}{\rm Re}\, {\rm Tr}
                            \left [ (\partial_\beta {\sf K})   \big  (  {\sf F} \star {\sf H} \big ) \right ] \\
  \label{eq:53b}
  & +  \frac{1}{2}\omega_{\alpha\beta}{\rm Re}\, {\rm Tr}
    \left [ (\partial_\beta \ms V ) \big  ( {\sf F} {\sf H} \big )\right ]  \\
  \label{eq:53c}
    &+   \frac{1}{2}\omega_{\alpha\beta}{\rm Re}\, {\rm Tr}
      \left [ \partial_\beta ( {\sf K} + \ms V ) \big  ({\sf F} V \big )\right ] + \dots .
\end{align}
\end{subequations}

These three terms can be understood as follows:
\begin{itemize}
\item The first term Eq.\eqref{eq:53a} contains both the clean classical energy current (order $\hbar^0$) and its anomalous velocity correction (order $\hbar^1$),
  responsible for the clean anomalous thermal Hall effect Eq.\eqref{eq:31}, as shown in \cite{mangeolle_quantum_2024};
\item The second term Eq.\eqref{eq:53b}, as we will show, has the interpretation of the side-jump accumulation current;
\item The third term Eq.\eqref{eq:53c} features $V$ where ${\sf H} $ was in Eq.\eqref{eq:53a}, whose interpretation was the normalization of eigenvectors or, physically and more loosely speaking,
  the amount of energy carried by a bosonic excitation. Thus Eq.\eqref{eq:53c} corresponds to a local modulation of the normalization of wavepackets,
  or ``equivalently'' to a renormalization of the bosonic energies by disorder. In both cases, as argued in Sec.\ref{sec:phase-space-form},
  this falls outside of the semiclassical picture as it is usually defined: thus we will discard this contribution in the following.
\end{itemize}

Introducing $\ms K_{\rm d},\ms F_{\rm d}$ defined in Sec.\ref{sec:equal-time-formalism} and using the normalization choice $\ms S^\dagger \star \ms H \star \ms S = \idmatrix$,
which as shown in \cite{mangeolle_quantum_2024} guarantees that the clean anomalous thermal Hall current density is band-diagonal, Eq.\eqref{eq:53b} becomes
\begin{align}
  \label{eq:54}
    {\mc J}_{X_\mu}^{\rm sj}(X,p) &= \frac{1}{2}{\rm Re}\, {\rm Tr}
                            \left [\ms S^{-1} (\partial_{p_\mu} \ms V) \ms S  {\sf F}_{\rm d} \right ],
\end{align}
and its integrated version Eq.\eqref{eq:28} in the main text.
\vspace{0.5cm}

\subsection{Semiclassical expansion}
\label{sec:now-big-semiclassics}

We can now  use the fact that in real-space notations
\begin{align}
  \label{eq:55}
  (\partial_{p_\mu} {\ms V})(x,y) \equiv  - i(x_\mu-y_\mu) {\ms V} (x,y) ,
\end{align}
then carry out the same technical derivation as in Sec. \ref{sec:equal-time-formalism} and App. \ref{sec:deriv-kinet-equat-calc-det}.
We do not reproduce the intermediate steps here, since they are very similar. The upshot is
\begin{widetext}
\begin{align}
  \label{eq:56}
 {\mc J}_{X_\mu}^{\rm sj} (X,p) = - \frac \pi \hbar \sum_{n,n'} \int_{p'}  & \, \delta \left ( (\ul{\ms K}_{\rm d})_{n',n'}(X+\delta \ms X_{n,n'}(p,p') , p') - (\ul{\ms K}_{\rm d})_{n,n} (X , p) \right ) \nonumber \\
                                                                           &\times  \delta \ms X_{n,n'}(p,p') \;  (\ul{\ms F}_{\rm d})_{n',n'} (X+ \delta \ms X_{n,n'}(p,p') , p',t)
                                                                             \; \left | \ms M_{n,n'}(p,p') \right |^2\;\tilde w(p-p') ,
\end{align}
\end{widetext}

In Eq.\eqref{eq:56} one already took  $\delta {\ms r}_{n,n'}^{+}(p,p') \rightarrow 0$, as justified in App.\ref{sec:argument-1}, for notational clarity.
Now focusing on band-diagonal processes ($n=n'$), as justified in App.\ref{sec:argument-2} and in the main text, 
and dropping $\delta \ms X_{n,n}(p,p')$ within the position arguments of energies and distribution functions since this is irrelevant to the order $O(\hbar^1)$,
one obtains the result Eq.\eqref{eq:29} in the main text.

It might not be obvious why the same technical steps appear in the derivation of the \emph{kinetic equation} and the \emph{current}. A proof based on a Ward identity is given in \cite{konig_quantum_2021}.
In the present case, here is an intuitive explanation at the technical level. One can rewrite Eq.\eqref{eq:54} as
\begin{align}
  \label{eq:57}
  {\mc J}_{X_\mu}^{\rm sj}&=  \frac{1}{2}{\rm Re}\, {\rm Tr}      \left ( (\partial_{p_\mu} \ms V_{\rm d})  {\sf F}_{\rm d} \right )
                          +  \frac{1}{2}{\rm Re}\, {\rm Tr}  \left ( \ms \Lambda_{p_\mu} \left [\ms V_{\rm d} , {\sf F}_{\rm d} \right ]\right ) ,
\end{align}
with $\ms \Lambda_{p_\mu} = \ms S^{-1} \partial_{p_\mu} \ms S$. In the case where interband processes are neglected (meaning only diagonal entries of $\ms V_{\rm d} $ are considered),
the commutator in the second term vanishes and $ {\mc J}_{\mu}^{\rm sj}$ consists only of the first term.
Using Eq.\eqref{eq:55} and the fact that $x-\xi=2\left ( \tfrac{x+y}2 - \tfrac{y+\xi}2 \right )$, one gets
\begin{align}
  \label{eq:58}
    {\mc J}_{X_\mu}^{\rm sj}(x,y) \approx  {\rm Im}\, {\rm Tr} \int_\xi \left [ \ms V_{\rm d} (x,\xi) {\sf F}_{\rm d}(\xi,y) \right ] \, (\tfrac{x+y}2 - \tfrac{y+\xi}2).
\end{align}

Now it is key to notice (e.g. from Eq.\eqref{eq:9}) that Appendix \ref{sec:deriv-kinet-equat-calc-det} is precisely devoted to computing
\begin{align}
  \label{eq:59}
   \ms V_{\rm d} \circ {\sf F}_{\rm d} - {\sf F}_{\rm d} \circ \ms V_{\rm d}^\dagger =  2i\, {\rm Im}\,\left [ \ms V_{\rm d} \circ {\sf F}_{\rm d}\right ] ,
\end{align}
which  is the right-hand side of the kinetic equation.

Thus $ {\mc J}_{X_\mu}^{\rm sj}$ (Eq.\eqref{eq:58}) can be easily obtained from the calculation of App. \ref{sec:deriv-kinet-equat-calc-det},
by in Eq.\eqref{eq:47} or Eq.\eqref{eq:48} replacing $$ {\sf F}_{\rm d}(K) \rightarrow (X_\mu -i\partial_{K_\mu}) \, {\sf F}_{\rm d}(K),$$
and dividing by $2i$. Assuming $\delta \ms r_{n,n'}^+(p,p')\rightarrow 0$ as always, one thus obtains (cf e.g. Eq.\eqref{eq:49}) 
\begin{widetext}
  \begin{align}
    \label{eq:60}
    {\mc J}_{X_\mu}^{\rm sj} (X,p)&= \frac 1{2i} \sum_{n,n'} \frac {2\pi} {\hbar} (-i) \, \int_{p'} \tilde w(p-p')  \, \delta \left ( (\ul{\ms K}_{\rm d})_{n',n'}(X+\delta X_{n,n'}(p,p') , p') - (\ul{\ms K}_{\rm d})_{n,n} (X , p) \right ) \nonumber \\
                             &\times \bigg [ \big ( X-X \big ) \times (\ul{\ms F}_{\rm d})_{n,n}(X ,p ,t)\; \left | \ms M_{n,n'}(p,p') \right |\, \left |\ms M_{n',n}(p',p) \right |  \nonumber \\
                               &- \big ( X-X-\delta \ms X_{n,n'}(p,p') \big ) \times (\ul{\ms F}_{\rm d})_{n',n'} (X+ \delta \ms X_{n,n'}(p,p') , p',t) \; \left |\ms M_{n,n'}(p,p')  \right |\,\left | \ms M_{n,n'}(p,p')  \right |\, \bigg ] ,
  \end{align}
  which is the correct result Eq.\eqref{eq:56}.

  \vspace{2cm}
  
\end{widetext}

\subsection{Formal proof that the side-jump current vanishes in thermodynamic equilibrium}
\label{sec:formal-proof-vanish}

We now consider its value at equilibrium, $  {\mc J}_{X_\mu}^{\rm sj}(X,p) \big |_{\rm eq}$, obtained by replacing
$$(\ul{\ms F}_{\rm d})_{n,n}(X ,p')  \rightarrow f \big (\ul{\ms K}_{\rm d})_{n,n}(X ,p') ,T(X) \big )$$
within Eq.\eqref{eq:29}, where $f(\varepsilon)=\varepsilon \left [n_{\rm B}(\varepsilon)+\tfrac 1 2 \right ]$.
The deviation ${\mathtt g}_{\rm d}^{\rm sj} = O(\hbar^1)$, which yields the corrected equilibrium distribution,
need not be included to this order: we will only verify that
\begin{align}
  \label{eq:102}
   J_\mu^{\rm sj}(X) \big |_{\rm eq}  &\equiv \int_p {\mc J}_{X_\mu}^{\rm sj}(X,p) \big |_{\rm eq} = 0 + O(\hbar^2).
\end{align}

We know on general grounds (see \cite{mangeolle_quantum_2024} especially Appendix B.3 therein) that the boson spectrum is symmetric,
in the sense that there exists a bijection between bands $n\leftrightarrow \overline n$ such that
\begin{align}
  \label{eq:104a}
  (\ul{ \ms K}_{\rm d})_{n,n}(-p)&=-  (\ul{ \ms K}_{\rm d})_{\overline n, \overline n} (p),\\
  \label{eq:104b}
 (\ms \Omega^{\ms W}_{p_\mu p_\nu})_{n,n} (-p) &=-  (\ms \Omega^{\ms W}_{p_\mu p_\nu})_{\overline n, \overline n} (p) .
\end{align}
Since $f(-\varepsilon)=-f(\varepsilon)$, one finds the relation 
\begin{align}
  \label{eq:61}
  {\mc J}_{X_\mu}^{\rm sj}(X,p) \big |_{\rm eq} &= - {\mc J}_{X_\mu}^{\rm sj}(X,-p) \big |_{\rm eq} ,
\end{align}
therefore its momentum integral vanishes, which proves Eq.\eqref{eq:102}.

\section{Solving Boltzmann's equation and the displaced local equilibrium current}
\label{sec:solv-boltzm-eqn}

Here we derive the solution Eq.\eqref{eq:28a} to the equation Eq.\eqref{eq:103}, to the first order in $\hbar$, equivalently in coordinate shifts $\delta \ms X$.
To this order the ``compensation term'' $\ms g_{\rm d}^{\rm sj}(X,p)$ and out-of-equilibrium part $\ms g_{\rm d}^{\rm rta}(X,p)$ can be solved for independently,
which we do following the steps of Ref.\cite{sinitsyn_disorder_2005} adapted to our generalized bosonic problem and notations.

\subsection{Displaced local equilibrium current}
\label{sec:find-an-equilibrium}

We write the time-independent \emph{local equilibrium} solution to the kinetic equation as
\begin{align}
  \label{eq:62}
  \ul{\ms F}_{\rm d} (X,p) &= f \left (\ul{\ms K}_{\rm d} (X,p) ,T(X) \right ) + \ms g_{\rm d}^{\rm sj} (X,p)
\end{align}
where $\ul{\ms F}_{\rm d}$ is assumed to satisfy local equilibrium, ${\rm D}_t \,\ul{\ms F}_{\rm d}(X,p) =0$.
In the collision integral of Eq.\eqref{eq:103}, we expand to the order of linear response,
\begin{align}
  \label{eq:63}
 & f \! \left ( E,T(X) \right ) - f \! \left (E ,T(X+ \delta \ms X_{n,n'}(p,p')) \right ) \\
  &= - \delta X_{n,n'}(p,p') \;\partial_XT \;\partial_T  f \! \left ( E,T(X) \right ) \nonumber ,
\end{align}
where
\begin{align}
  \label{eq:64}
  E &= (\ul{\ms K}_{\rm d})_{n,n}(X ,p) \simeq  (\ul{\ms K}_{\rm d})_{n',n'}(X ,p').
\end{align}
Here the ``$\simeq$'' approximate equality is enforced by the energy conservation delta-function in Eq.\eqref{eq:22};
below, this is enforced only to the order $O(\hbar^0)$ which is sufficient in that context. Then Eq.\eqref{eq:103} becomes

\begin{widetext}
\begin{align}
  \label{eq:65}
\forall n,\quad  0 &= - \frac {2\pi} {\hbar} \,\sum_{n'} \int_{p'} \tilde w(p-p')\; \left |  \ms M_{n,n'}(p,p')  \right |^2
      \; \delta \Big ( (\ul{\ms K}_{\rm d})_{n',n'}(X , p') - (\ul{\ms K}_{\rm d})_{n,n} (X , p) \Big ) \\
    &\qquad \qquad \times \bigg [ (\ms g_{\rm d})_{n,n}(X,p) -  (\ms g_{\rm d})_{n',n'}(X,p')
      - \delta \ms X^\mu_{n,n'}(p,p') \,\mathsf \partial_{X_\mu} T \;\partial_T f \!\left ( (\ul{\ms K}_{\rm d})_{n^({}'{}^), n^({}'{}^)} (X , p^({}'{}^)) \right ) \bigg ] + \dots \nonumber 
\end{align}
where ``$+ \dots  $'' contains terms higher-order in powers of $\hbar$ or beyond linear response,
and the optional set of apostrophes in $\partial_T f $ follows from Eq.\eqref{eq:64}.

We now use the fact that Eq.\eqref{eq:21}, up to the approximations discussed in App.\ref{sec:kinet-coord-comm}, takes the form Eq.\eqref{eq:23}.
Then, a solution to the collision integral is given by Eq.\eqref{eq:28b} in the main text, which is clearly gauge invariant.

\subsection{Response in the relaxation time approximation (RTA)}
\label{sec:drude-response}

We now consider the $O(\hbar^0)$ out-of-(local-)equilibrium distribution $\ms g_{\rm d}^{\rm rta}$, by again defining
\begin{align}
  \label{eq:67}
 \ul{\ms F}_{\rm d} (X,p) =f(\ul{\ms K}_{\rm d} (X,p),T(X))+ \ms g_{\rm d}^{\rm rta}(X,p) 
\end{align}
but now only requiring that both the left-hand side and the right-hand side of Eq.\eqref{eq:103} be $O(\hbar^0)$.
We focus on the bulk currents of the homogeneous theory, so that spatial gradients $\partial_X \ul{\ms K}_{\rm d} (X,p)$ are neglected, and the left-hand side reads
\begin{align}
  \label{eq:68}
  {\rm D}_t \,\ul{\ms F}_{\rm d}\Big |_{n,n} (X,p) &= \partial_{p_\mu} \ul{\ms K}_{\rm d} \;\partial_{X_\mu} T \; \partial_T f(\ul{\ms K}_{\rm d} (X,p),T(X)) \Big |_{n,n}  
\end{align}
Meanwhile, in the right-hand side, all the position shifts can be discarded at $O(\hbar^0)$, and one obtains 
\begin{align}
  \label{eq:69}
\partial_{k_\mu} \ul{\ms K}_{\rm d} \;\partial_{X_\mu} T \; \partial_T f(\ul{\ms K}_{\rm d} (X,p)) \Big |_{n,n} 
    &= - \frac {2\pi} {\hbar} \, \sum_{n'}\int_{p'} \tilde w(p-p') \times \delta \Big ( (\ul{\ms K}_{\rm d})_{n',n'}(X, p') - (\ul{\ms K}_{\rm d})_{n,n} (X , p) \Big ) \\
    & \qquad \qquad \times \left |  \ms M_{n,n'}(p,p')  \right|^2 
      \times \bigg [ (\ms g_{\rm d}^{\rm rta})_{n,n} (X,p)  -  (\ms g_{\rm d}^{\rm rta})_{n',n'}(X,p') \bigg ] . \nonumber
\end{align}

While this is still in principle a hard problem, it reduces to the relaxation time approximation \cite{mahan},
provided that scattering is elastic (which is the case here), that dispersion is isotropic (meaning $\ul{\ms K}_{\rm d}(|p|)$ only) and non-degenerate,
and that scattering is isotropic i.e.\ $ \ms M_{n,n'}(p,p') $ only depends on the angle between $p$ and $p'$
(i.e.\ on $\hat{\rm p}\cdot\hat{\rm p}' $ where we define $\hat{\rm p} \equiv p/|p|$).

Assuming all this for the sake of analytical tractability, one obtains Eqs.\eqref{eq:28c},\eqref{eq:26} in the main text.

\subsection{Explicit writing of the side-jump current in the RTA}
\label{sec:explicit-writing}

Inserting Eq.\eqref{eq:28c} into Eq.\eqref{eq:29}, one obtains
\begin{align}
  \label{eq:70}
   {\mc J}_{X_\mu}^{\rm sj,rta}(X,p)&\approx - \frac {\pi} {\hbar}  \sum_{n} \int_{p'} \tilde w(p-p')\; \left | \ms M_{n,n}(p,p')  \right |^2 \,
                                      \delta \Big ( (\ul{\ms K}_{\rm d})_{n,n}(X, p') - (\ul{\ms K}_{\rm d})_{n,n} (X , p) \Big ) \; \nonumber\\
                                    &\qquad \qquad \times  (p'-p)_\nu \; (\ms \Omega^{\ms W}_{p_\mu p_\nu})_{n,n}(p)
                                      \times  \partial_{k_\rho} (\ul{\ms K}_{\rm d})_{n,n}(X,p') \;\partial_{X_\rho} T \;
                                      \partial_T f( (\ul{\ms K}_{\rm d})_{n,n}(X,p')) \,\stau_{n,n} (X,p') \nonumber\\
                                    &\approx - \frac 1 2 \sum_{n} (\ms \Omega^{\ms W}_{p_\mu p_\nu})_{n,n}(p) \,  \partial_{k_\rho} (\ul{\ms K}_{\rm d})_{n,n}(X,p) \;
                                      \partial_{X_\rho} T \; \partial_T f( (\ul{\ms K}_{\rm d})_{n,n}(X,p)) \;\stau_{n,n}  (X,p) \; \Xi^\nu_n(X,p) .
\end{align}
In going to the second line we used the fact that $\stau_{n,n} $, defined in Eq.\eqref{eq:26}, only depends on the modulus $|p|=|p'|$,
and we used the temporary notation
\begin{align}
  \label{eq:71}
  \Xi^\nu_n(X,p) &= \frac {2\pi} {\hbar} \int_{p'}  (p'-p)_\nu \; \tilde w(p-p')\;  \left | \ms M_{n,n}(p,p')  \right |^2
                   \, \delta \Big ( (\ul{\ms K}_{\rm d})_{n,n}(X, p') - (\ul{\ms K}_{\rm d})_{n,n} (X , p) \Big ) \\
  &=  \frac  {2\pi} {\hbar} \times (- p_\nu) \int_{p'} \left ( 1 - \hat{\rm p}\cdot\hat{\rm p}' \right ) \;  \tilde w(p-p')\; \left | \ms M_{n,n}(p,p')  \right |^2 \,
    \delta \Big ( (\ul{\ms K}_{\rm d})_{n,n}(X, p') - (\ul{\ms K}_{\rm d})_{n,n} (X , p) \Big ) = - p_\nu \stau_{n,n}^{-1} (X,p) . \nonumber
\end{align}

The cancellation $\stau_{n,n} \stau_{n,n} ^{-1}=1$ leads to Eq.\eqref{eq:30} in the main text.

\subsection{Generalizations}
\label{sec:append-generalization}

We now look for the equilibrium and out-of-equilibrium solutions $\ms g_{\rm d}^{\rm sj}$ and $\ms g_{\rm d}^{\rm rta}$ to the generalized kinetic equation, where we now assume explicitly only interband scattering ($n=n'$ enforced): 
\begin{align}
     {\rm D}_t \,  \ul{\ms F}_{\rm d} (X,p,t) \big |_{n,n}
  &= - \sum_{(\iota)}\int_{p'} (\ms t^{(\iota)}_{n,n})^{-1}(p,p')\;
  \bigg \{ \big [ p_\nu-p'_\nu \big ]\, (\ms \Omega^{\ms W,(\iota)}_{p_\mu p_\nu})_{n,n}(p{}^({}'{}^)) \,
  \partial_{\mu}T\,\partial_T f\!\left ( (\ul{\ms K}_{\rm d})_{n,n} (p{}^({}'{}^)) \right )   \nonumber \\
  & \quad + (\ms g_{\rm d})_{n,n}(p) - (\ms g_{\rm d})_{n,n}(p')   \bigg \}+ \int_{p'} \Upsilon^{\rm int}_{p,p'} \Big [ (\ms g_{\rm d})_{n,n}(p') - (\ms g_{\rm d})_{n,n}(p)   \Big ] ,\\
  (\ms t^{(\iota)}_{n,n})^{-1}(p,p') &\equiv \frac {2\pi} {\hbar} \, \tilde w^{(\iota)}(p-p')  \; \delta \Big ( (\ul{\ms K}_{\rm d})_{n,n}(p') - (\ul{\ms K}_{\rm d})_{n,n} (p) \Big ) \; \big |\ms M^{(\iota)}_{n,n}(p,p') \big |^2 . 
\end{align}
First looking for the local equilibrium solution,where the lhs and the rhs vanish separately, one finds
\begin{align}
   (\ms g_{\rm d}^{\rm sj})_{n,n}(p) &= - \Big ( p_\nu \,
  \partial_{\mu}T\,\partial_T f\!\left ( (\ul{\ms K}_{\rm d})_{n,n} (p) \right )  \;
  \sum_{(\iota)} (\ms \Omega^{\ms W,(\iota)}_{p_\mu p_\nu})_{n,n}(p) \;
  \int_{p'} (\ms t^{(\iota)}_{n,n})^{-1}(p,p')  \Big ) \Big / 
  \Big ( \int_{p'} \Big \{ \Upsilon^{\rm int}_{p,p'} 
  + \sum_{(\iota)} (\ms t^{(\iota)}_{n,n})^{-1}(p,p') \Big \}  \Big ) .
\end{align}
Then taking the order $O(\hbar^0)$ on the rhs and restoring the gradient term on the lhs, one finds
\begin{align}
     (\ms g_{\rm d}^{\rm rta})_{n,n}(p) &=  \partial_{k_\lambda} \ul{\ms K}_{\rm d}(p) \;\partial_{\lambda} T \; \partial_T f(\ul{\ms K}_{\rm d}(p)) 
     \Big / \Big ( \int_{p'} \left ( 1 - \hat {\rm p} \cdot \hat {\rm p}'\right )\Big \{ \Upsilon^{\rm int}_{p,p'} 
  + \sum_{(\iota)} (\ms t^{(\iota)}_{n,n})^{-1}(p,p') \Big \}  \Big ) .
\end{align}
 Assuming that $\Upsilon^{\rm int}_{p,p'}$ just like $(\ms t^{(\iota)}_{n,n})^{-1}(p,p')$ satisfies the usual assumptions of RTA 
 (namely it only depends on the angle $\hat {\rm p} \cdot \hat {\rm p}'$ and on the moduli $|p|=|p'|$ constrained to be equal), one obtains Eq.\eqref{eq:245} in the main text.

\section{Other calculatory details on Hall currents}
\label{sec:other-calc-deta}

\subsection{Thermal Hall conductivity}
\label{sec:details-thc}

We consider for concreteness the case where a temperature gradient is imposed along $x$ and the (local) energy current is measured along $y$.
Then, explicitly, Equations \eqref{eq:27} and \eqref{eq:30} read

\begin{align}
  \label{eq:72}
    { J}_{y}^{\rm sj,rta}(X)
  &=  \partial_{x} T \times  \frac 1 2 \sum_{n}\int_p  p_\lambda \,  \partial_{p_x} (\ul{\ms K}_{\rm d})_{n,n}(X,p) \; (\ms \Omega^{\ms W}_{p_y p_\lambda})_{n,n}(X,p)\,
    \partial_T f \!\left ( (\ul{\ms K}_{\rm d})_{n,n} (X , p) \right )  , \\
   {J}_{y}^{\rm gv,sj}(X)&=  - \partial_{x} T \times \frac 1 2 \sum_n \int_p  p_\lambda \,   \partial_{p_y}(\ul{\ms K}_{\rm d})_{n,n}(X,p) \;  (\ms \Omega^{\ms W}_{p_x p_\lambda})_{n,n}(X,p) \,
                                  \partial_T f \!\left ( (\ul{\ms K}_{\rm d})_{n,n} (X , p) \right ) .
\end{align}

Contrarily to the case of fermions in a random scalar potential, because $\ms \Omega^{\ms W}_{p_\mu p_\nu}$ is not antisymmetric a priori, these two currents are not identical a priori.
The local current due to disorder, $  { J}_{y}^{\rm dis} \equiv   { J}_{y}^{\rm sj,rta}+  { J}_{y}^{\rm gv,sj}$, is thus
\begin{align}
  \label{eq:73}
  { J}_{y}^{\rm dis}(X) &= \partial_{x} T \times  \frac 1 2 \sum_{n}\int_p p_\lambda \;
                          \left [ (\ms \Omega^{\ms W}_{p_y p_\lambda})_{n,n}(p)\, \partial_{p_x} (\ul {\ms K}_{\rm d}^{})_{n,n}(p)
                          -  (\ms \Omega^{\ms W}_{p_x p_\lambda})_{n,n}(p)\,  \partial_{p_y} (\ul {\ms K}_{\rm d}^{})_{n,n}(p) \right ]
                          \; \partial_T f \!\left ( (\ul {\ms K}_{\rm d}^{})_{n,n} ( p) \right )  .
\end{align}

Now one can use the spectrum symmetry relations Eqs.\eqref{eq:104a},\eqref{eq:104b} to replace $\frac 1 2 \sum_n \rightarrow {\rm Tr}_+$, which is
the sum over positive-eigenvalue eigenspaces of the dynamical matrix i.e.\ over bosonic \emph{bands} in the strict physical sense.

We obtain the disorder-induced thermal Hall conductivity
\begin{align}
  \label{eq:74}
  \kappa_{xy}^{\rm dis} &= {\rm Tr}_+ \int_p p_\lambda \;
                          \left [ {\ms \Omega}^{\ms W}_{p_y p_\lambda}(p)\, \partial_{p_x} {\ms K}_{\rm d}^{}(p)
                          -  {\ms \Omega}^{\ms W}_{p_x p_\lambda}(p)\,  \partial_{p_y} {\ms K}_{\rm d}^{}(p) \right ]
                          \;  \partial_T f \!\left ( {\ms K}_{\rm d}^{} (p) ,T\right ) ,
\end{align}
and using $\partial_T  f(\varepsilon,T) = - (\varepsilon^2/T)\, \partial_\varepsilon n_{\rm B}(\varepsilon,T)$ we obtain the final result Eq.\eqref{eq:32} in the main text.

\subsection{Electrical Hall effect}
\label{sec:details-ehc}

The two disorder contributions to the \emph{charge} current are
\begin{align}
  \label{eq:72analog}
    { J}_{y}^{\rm sj,rta}(X)
  &=   -\ms e^2 E_x \sum_{n}\int_p  p_\lambda \,  \partial_{p_x} (\ul{\ms K}_{\rm d})_{n,n}(X,p) \; (\ms \Omega^{\ms W}_{p_y p_\lambda})_{n,n}(X,p)\,
    \partial_\varepsilon f \!\left ( (\ul{\ms K}_{\rm d})_{n,n} (X , p) \right )  , \\
   {J}_{y}^{\rm gv,sj}(X)&=  +\ms e^2 E_x  \sum_n \int_p  p_\lambda \,   \partial_{p_y}(\ul{\ms K}_{\rm d})_{n,n}(X,p) \;  (\ms \Omega^{\ms W}_{p_x p_\lambda})_{n,n}(X,p) \,
                                  \partial_\varepsilon f \!\left ( (\ul{\ms K}_{\rm d})_{n,n} (X , p) \right ) .
\end{align}

The local \emph{charge} current due to disorder, ${ J}_{y}^{\rm dis} \equiv   { J}_{y}^{\rm sj,rta}+  { J}_{y}^{\rm gv,sj}$, is 
\begin{align}
  \label{eq:73analog}
  { J}_{y}^{\rm dis}(X) &=  -\ms e^2 E_x \sum_{n}\int_p p_\lambda \;
                          \left [ (\ms \Omega^{\ms W}_{p_y p_\lambda})_{n,n}(p)\, \partial_{p_x} ({\ms K}_{\rm d}^{})_{n,n}(p)
                          -  (\ms \Omega^{\ms W}_{p_x p_\lambda})_{n,n}(p)\,  \partial_{p_y} ({\ms K}_{\rm d}^{})_{n,n}(p) \right ]
                          \; \partial_\varepsilon f \!\left ( ({\ms K}_{\rm d}^{})_{n,n} ( p) \right )  .
\end{align}
At odds with the bosonic case, here one simply has $\sum_n \rightarrow {\rm Tr}$, which is the sum over fermionic bands.
We obtain the disorder-induced electrical Hall conductivity, Eq.\eqref{eq:74analog} in the main text.

\section{Explicit formulae for numerical evaluation}
\label{sec:expl-form-analyt}

Here we recapitulate how to evaluate Eqs.\eqref{eq:31},\eqref{eq:32} straightforwardly from a disordered bosonic theory of the form of Sec.\ref{sec:gener-boson-hamilt}.
Once the clean hamiltonian Eq.\eqref{eq:96} and the disorder Hamiltonian Eq.\eqref{eq:200} have been specified, Sec.\ref{sec:gener-boson-hamilt} provides explicit expressions
for the dynamical matrix $\sf K$ and its disorder counterpart $\sf W$, both of which are $2N\times 2N$ matrices.
Ultimately  $\sf K$ and $\sf W$ fully determine the thermal Hall conductivities Eqs.\eqref{eq:31},\eqref{eq:32}: we now detail the procedure to derive the latter.

The dynamical matrix is diagonalized as $\ms S_0^{-1}\ms K \ms S_0=\ul{\ms K}_{\rm d}$.
Note that $\ms S_0$, whose columns are the bosonic eigenmodes, is generally \emph{not} unitary (and is of course not uniquely defined).
Note also that $\ul{\ms K}_{\rm d}$ is a diagonal matrix whose $2N$ entries are the $N$ bosonic energies (all positive) and their opposites (all negative).
While Eqs.\eqref{eq:31},\eqref{eq:32} only involve summation over the $N$ physical bands,
actual calculations of the curvatures $\ms \Omega_{p_xp_y}, \ms \Omega^{\ms W}_{p_xp_y}$ can involve all $2N$ eigenvalues and eigenvectors.

For numerical evaluation, it is useful to resort to formulae for the Berry curvature ${\sf \Omega}_{p_xp_y}$
and its disorder counterpart $\tfrac 1 2 (\ms \Omega^{\ms W}_{p_x p_y}-\ms \Omega^{\ms W}_{p_y p_x})$ that avoid computing momentum derivatives of gauge-dependent quantities.
Such gauge-invariant formulae can be derived using the identity
\begin{align}
  \label{eq:53}
  \left [ \ms S_0^{-1}(\partial_{p_x} \ms K) \ms S_0 \right ]_{n,n'}
  =  \left [ \ms S_0^{-1}\partial_{p_x}\ms S_0 \right ]_{n,n'} \left ({[\ul{\ms K}_{\rm d}]_{n',n'} - [\ul{\ms K}_{\rm d}]_{n,n} } \right ),
\end{align}
which can be inverted for $n\neq n'$. For instance it is well known that the Berry curvature can be recast into the form
\begin{align}
  \label{eq:107}
  ({\sf \Omega}_{p_xp_y})_{n,n} &=  {\rm Im}\left [ \ms S_0^{-1}\partial_{p_y} \ms S_0 \ms S_0^{-1}\partial_{p_x} \ms S_0 -(x \leftrightarrow y)\right ]_{n,n} \\
                                &= \sum_{n' \neq n}\frac{{\rm Im} \left (\left [ \ms S_0^{-1}(\partial_{p_x} \ms K) \ms S_0 \right ]_{n,n'}\left [ \ms S_0^{-1}(\partial_{p_y} \ms K) \ms S_0 \right ]_{n',n}\right )}
                              {\left ( [\ul{\ms K}_{\rm d}]_{n',n'} - [\ul{\ms K}_{\rm d}]_{n,n}\right )^2} -(x \leftrightarrow y).
\end{align}
Although only those indices $n$ corresponding to eigenmodes with positive eigenvalues are useful to compute Eq.\eqref{eq:31},
note that the summation $\sum_{n'\neq n}$ involves all $2N-1$ other indices.

A similar formula can be obtained for the antisymmetric part of ${\sf \Omega}^{\sf W}_{p_xp_y}$ (defined in Eq.\eqref{eq:24}): one finds
\begin{align}
  \label{eq:111}
 & \tfrac 1 2  (\ms \Omega^{\ms W}_{p_x p_y}-\ms \Omega^{\ms W}_{p_y p_x})_{n,n}
  = \tfrac 1 2  ({\sf \Omega}_{p_xp_y})_{n,n} \\
                                          &+ \frac 1 2 \sum_{n'\neq n}\sum_{n''\neq n}{\rm Im} \bigg \{
                                            \Big (  \left [ \ms S_0^{-1}\partial_{p_x}\ms S_0 \right ]_{n'',n}\left [ \ms S_0^{-1}\partial_{p_y}\ms S_0 \right ]_{n,n'}   -(x \leftrightarrow y) \Big )
                                            \Big ( \frac{ \left [ \ms S_0^{-1}\ms W \ms S_0 \right ]_{n',n''}}{\left [ \ms S_0^{-1}\ms W \ms S_0 \right ]_{n,n}}
                                            - \frac{\left [ \ms S_0^{-1}\ms W \ms S_0 \right ]_{n',n}\left [ \ms S_0^{-1}\ms W \ms S_0 \right ]_{n,n''}}{\left [ \ms S_0^{-1}\ms W \ms S_0 \right ]_{n,n}^2}\Big ) \bigg \} \nonumber \\
  & + \frac 1 4  \sum_{n'\neq n} {\rm Im} \bigg \{ \frac{1}{ \left [ \ms S_0^{-1}\ms W \ms S_0 \right ]_{n,n}}
    \Big ( \left [ \ms S_0^{-1}\partial_{p_y}\ms S_0 \right ]_{n',n}  \left [ \ms S_0^{-1}(\partial_{p_x}\ms W) \ms S_0\right ]_{n,n'}
    + \left [ \ms S_0^{-1}\partial_{p_y}\ms S_0 \right ]_{n,n'} \left [ \ms S_0^{-1}(\partial_{p_x}\ms W) \ms S_0\right ]_{n',n} \Big )  -(x \leftrightarrow y)  \bigg \} \nonumber \\
                                        & + \frac 1 4 {\rm Im} \bigg \{ \frac{\left [ \ms S_0^{-1}(\partial_{p_y}\ms W) \ms S_0\right ]_{n,n}}{\left [ \ms S_0^{-1}\ms W \ms S_0 \right ]_{n,n}^2}
                                           \sum_{n'\neq n}\Big (  \left [ \ms S_0^{-1}\partial_{p_x}\ms S_0 \right ]_{n,n'}\left [ \ms S_0^{-1}\ms W \ms S_0 \right ]_{n',n}
                                          +   \left [ \ms S_0^{-1}\partial_{p_x}\ms S_0 \right ]_{n',n}\left [ \ms S_0^{-1}\ms W \ms S_0 \right ]_{n,n'} \Big )  -(x \leftrightarrow y)  \bigg \} , \nonumber
\end{align}
which together with Eq.\eqref{eq:53} provides a gauge-invariant formula suitable for numerical evaluation.
This is the one we used to obtain our results in Sec.\ref{sec:concr-appl-honeyc}.
Again, although only those indices $n$ corresponding to eigenmodes with positive eigenvalues are useful to compute Eq.\eqref{eq:32},
note that the summations $\sum_{n'\neq n}$ and $\sum_{n''\neq n}$ involve all $2N-1$ other indices.

\end{widetext}

\section{Application to a low-energy bosonic theory -- Details}
\label{sec:applications-details}

\subsection{Effective low-energy theory for magnons in an antiferromagnet}
\label{sec:effective-low-energy}

For an antiferromagnet on a bipartite lattice, the spin operator may be expressed locally as
\begin{align}
  \label{eq:sigma13}
  \bs S_{\mb r} &= \sum_{a=x,y,z} \hat{\mb u}_a \left [ (-1)^{\mb r} {\rm n}_a(\mb r) + {\rm m}_a(\mb r)\right ]
\end{align}
where the vectors $\hat{\mb u}_{a}$ define an orthonormal basis, $ (-1)^{\mb r}=\pm$ alternates between the two sublattices,
and ${\rm n}_a, {\rm m}_a$ are the staggered and the net magnetization fluctuations, respectively,
constrained by
\begin{align}
  \label{eq:sigma62}
  \mb m^2 + \mb n^2 = 1 ,\qquad \mb m \cdot \mb n = 0 .
\end{align}
Assuming collinear N\'eel order along the $z$ axis, to the cubic order in the transverse components one may replace
\begin{subequations}
  \begin{align}
  \label{eq:sigma63}
    {\rm m}_z &\simeq -  {\rm m}_x  {\rm n}_x -  {\rm m}_y  {\rm n}_y ,\\
    {\rm n}_z &\simeq 1 - \tfrac 1 2 \left (  {\rm m}_x^2+  {\rm n}_x^2 +  {\rm m}_y^2+  {\rm n}_y^2 \right ) .
\end{align}
\end{subequations}
This yields an effective theory in terms of the low-energy fields $n_x,n_y,m_x,m_y$,
defined in the collinear case simply as $n_a= {\rm n}_a, m_a= {\rm m}_a$ for $a \in \{x,y\}$.
These low-energy fields satisfy the commutation relation
\begin{align}
  \label{eq:75}
  [ m_y,n_x] = i \hbar = -  [ m_x,n_y] 
\end{align}
which is Eq.\eqref{eq:34}.

In terms of $m_i,n_i$ with $i=1,2$, and writing $\partial_1\equiv \partial_x, \partial_2\equiv \partial_y$, the hamiltonian Eq.\eqref{eq:33} reads 
\begin{align}
  \label{eq:76}
  H &= \frac 1 {2\chi} m^im^i + \Delta n_in_i \\
  &- \tfrac 1 2 c_1 n^i \partial_{\mu\mu}^2n^i  + \tfrac 1 2 c_2 \partial_in^i \partial_jn^j 
    +   \tfrac 1 2 d_4 m^i \partial_{\mu\mu}^2n^i \nonumber \\
  & + \tfrac 1 2 d_1 \epsilon_{ij} m^i \partial_{\mu\mu}^2n^j    - \tfrac 1 2 d_2 \epsilon_{i \overline i}\partial_im^{\overline i} \partial_jn^j
    + \tfrac 1 2 d_3 \epsilon_{ij} m^i n^j  . \nonumber
\end{align}

Because to leading order in spatial gradients $$ m^i \approx \chi \partial_t n^i,$$
up to integration by parts the $d_4$ term is a total time derivative and does not appear in the effective theory.
It is then possible to write
$$\epsilon_{i \overline i}\partial_im^{\overline i} \partial_jn^j \rightarrow \tfrac 1 2 \epsilon_{ij}m^i\partial^2_{\mu\mu}n^j ,$$
so that an equivalent effective hamiltonian density is
\begin{align}
  \label{eq:77}
  H_{\rm eff} &= \frac 1 {2\chi} m^im^i - \tfrac 1 2 c_1 n^i \partial_{\mu\mu}^2n^i  + \tfrac 1 2 c_2 \partial_in^i \partial_jn^j \\
             &+ \tfrac 1 2 d_3 \epsilon_{ij} m^i n^j     + \tfrac 1 2 (d_1-\tfrac 1 2 d_2) \epsilon_{ij} m^i \partial_{\mu\mu}^2n^j  + \Delta n_in_i  . \nonumber
\end{align}
One can rewrite this, to the first order in $d_1,d_2,d_3$, as Eq.\eqref{eq:35} in the main text.\\

\textit{\textbf{Note --}} In the following, for notational clarity we write all the expressions in the $d_3=0$ and $\Delta=0$ case.
While for numerical evaluations we indeed take $d_3=0$, they are still performed at finite $\Delta$ (with value given in Table \ref{tab:parameter-values}).

\subsection{Diagonalization of the clean problem}
\label{sec:diagonalization}

The dynamical matrix is
\begin{equation}
  \label{eq:78}
\ms K 
  = i \begin{bmatrix}
0 & 1\\
- p_\mu \left ( \frac{c_1}\chi \delta_{\mu,\nu} \delta_{i,j} + \frac{c_2}\chi \delta_{i,\mu}\delta_{\nu,j} \right ) p_\nu
& - \eta \epsilon_{ij} p^2
  \end{bmatrix}.
\end{equation}

 One finds the energy bands 
     \begin{align}
       \label{eq:79}
  {\sf K}^{}_{\rm d} &= {\rm diag}\left ( \varepsilon_1 , -\varepsilon_1,
       \varepsilon_2, -\varepsilon_2 \right ) ,\\
    \varepsilon_1 &= [(c_1+c_2)\bs p^2/\chi]^{\frac 1 2}, \; \quad \varepsilon_2 = [ c_1 \bs p^2/\chi]^{\frac 1 2}, \nonumber 
     \end{align}
     and the Berry curvatures 
       \begin{align}
         \label{eq:82}
     {\ms \Omega}_{p_x p_y}^{(1)}&= - \eta \frac 1 {\varepsilon_1} \frac {4c_1+3c_2} {2 c_2} \quad +O(\eta^3),\nonumber\\
          {\ms \Omega}_{p_x p_y}^{(2)}&= \eta \frac 1 {\varepsilon_2 } \frac {4c_1+c_2} {2 c_2} \quad +O(\eta^3).
       \end{align}
  The normalized eigenvectors are  $\Psi_i, \Psi_i^*$ such that
     \begin{align}
       \label{eq:80}
       \ms S^{} &=
                       \begin{bmatrix}    \Psi_1 \;\vline\; \Psi_1^* \;\vline\;  \Psi_2 \;\vline\;  \Psi_2^*  \end{bmatrix}.
     \end{align}
\begin{widetext}
  
  A suitable choice of properly normalized eigenvectors (in the sense $\ms S^\dagger \ms H \ms S = \text{\textbb 1}$) is
       \begin{align}
         \label{eq:81}
         \Psi_1 &= \frac 1 {\sqrt {2}} \left ( \sqrt{\chi} \frac{\eta }{c_2}\frac{|\bs k|}{|k_2|}  +\frac{i}{\sqrt{\chi} \varepsilon_1} \frac{k_1}{k_2}\frac{|k_2|}{|\bs k|} , \frac{i}{\sqrt{\chi} \varepsilon_1} \frac{|k_2|}{|\bs k|},
        \frac{k_1}{k_2}  \frac 1 {\sqrt {\chi}} \frac{|k_2|}{|\bs k|}  - i \frac{\eta }{c_2} \frac{|\bs k|}{|k_2|} \sqrt {\chi}\varepsilon_1 , \frac 1 {\sqrt {\chi}} \frac{|k_2|}{|\bs k|} \right )^\top \; + O(\eta^2),  \nonumber\\
        \Psi_2 &= -\frac 1 {\sqrt {2}} \left ( \sqrt{\chi} \frac{\eta }{c_2}\frac{|\bs k|}{|k_1|}  +\frac{i}{\sqrt {\chi}\varepsilon_2} \frac{k_2}{k_1}\frac{|k_1|}{|\bs k|} , -\frac{i}{\sqrt {\chi}\varepsilon_2} \frac{|k_1|}{|\bs k|} \frac 1 {\sqrt {\chi}} ,
        \frac{k_2}{k_1}  \frac 1 {\sqrt {\chi}} \frac{|k_1|}{|\bs k|}  - i \frac{\eta }{c_2} \frac{|\bs k|}{|k_1|} \sqrt {\chi}\varepsilon_2 , -\frac 1 {\sqrt {\chi}} \frac{|k_1|}{|\bs k|} \right )^\top \; + O(\eta^2).
       \end{align}
    
  Here we provide $\Psi_1,\Psi_2$ to the first order in $\eta$ only. 
     This is sufficient to compute the leading contribution to the Berry curvature $\ms \Omega_{p_x p_y}$ given above, Eq.\eqref{eq:82},
     which is odd under time reversal, and thus odd in powers of $\eta$.

\subsection{Explicit calculations in two given instances of disorder (b,c)}
\label{sec:two-explicit-sets}
     
\subsubsection{Case (b): a case with both intrinsic and extrinsic chirality}
\label{sec:application-3:-now}

We consider the local disorder hamiltonian density 
\begin{align}
  \label{eq:84}
  H_{\rm dis} &= - \tilde \eta \epsilon_{ij} m^i \partial^2_{\mu\mu}n^j   =  - \tilde \eta \left (m^1 \partial^2_{\mu\mu}n^2 - m^2 \partial^2_{\mu\mu}n^1\right ) 
= \text{Eq.\eqref{eq:41}} , 
\end{align}
such that
\begin{align}
  \label{eq:85}
  \ms W &=   {\sf \Gamma} \hat{W} = - \tilde \eta p^2 \begin{bmatrix}
\sigma^y &0\\
 - i \eta p^2 \text{\textbb 1} & \sigma^y
  \end{bmatrix}.
\end{align}

One finds
\begin{align}
  \label{eq:101}
\left [  \ms S^{-1}(p) \begin{bmatrix} \sigma^y &0\\
    - i \eta p^2 \text{\textbb 1} & \sigma^y \end{bmatrix} \ms S(p) \right ]_{n,n}
                                  &  =  2\frac \chi {c_2} \eta \left ( \varepsilon_1 \; ; \; -  \varepsilon_1 \; ; \;
                 \varepsilon_2 \; ; \; - \varepsilon_2\right ) \;+O(\eta^3) ,
\end{align}
implying that the third term in Eq.\eqref{eq:24} vanishes. As for the first and second terms, they are determined by
  \begin{align}
  \label{eq:86}
{\rm Im} \left [ \ms S^{-1}(p) \begin{bmatrix} \sigma^y &0\\
 - i \eta p^2 \text{\textbb 1} & \sigma^y \end{bmatrix} \partial_{p_\mu}\ms S(p) \right ]_{n,n}
    &=  \epsilon_{\mu,\overline \mu} (p_{\overline \mu}/p^2) \;+O(\eta^2), \\
      \label{eq:109}
    {\rm Im} \left [ \ms S^{-1}(p) \partial_{p_\mu}\ms S(p) \right ]_{n,n}
    &= \frac \eta {c_2} \left ( \frac 1 {\varepsilon_1} \#_\mu(p)  \; ; \; - \frac 1 {\varepsilon_1} \#_\mu(p)  \; ; \;
      \frac 1 {\varepsilon_2} \#'_\mu(p)  \; ; \; - \frac 1 {\varepsilon_2} \#'_\mu(p) \right ) \;+O(\eta^3), 
    \end{align}
    where $\#_\mu(p)$ and $\#'_\mu(p)$ are real quantities which we do not write explicitly here.
\end{widetext}

\subsubsection{Case (c): a case where the clean thermal Hall conductivity is zero}
\label{sec:application-2:-now}

Because
  \begin{align}
  \label{eq:88}
   {\rm Im} \left [ \ms S^{-1} (p) \partial_{p_\mu}\ms S (p) \right ]_{n,n} &= 0 ,
\end{align}
where implicitly $\ms S$ here means $\ms S |_{\eta=0} $, the first term in Eq.\eqref{eq:24} vanishes.

We note that using the disorder $\hat{W}$ of the previous case (Eqs.\eqref{eq:40},\eqref{eq:41}) one would find
$$  [\ms S^{-1}(p) \ms W(p) \ms S(p)]_{n,n} = 0 ,$$
implying that the small-momentum expansion in App. \ref{sec:obta-disord-berry} would be ill-defined.
Here instead we consider $\hat{W} $ as given in Eq.\eqref{eq:87}, whence
\begin{align}
  \label{eq:90}
  \ms W =   {\sf \Gamma}\big |_{\eta=0} \hat{W} =  {\sf K}\big |_{\eta=0}
  - \tilde \eta p^2 \begin{bmatrix}\sigma^y &0\\ 0 & \sigma^y\end{bmatrix}.
\end{align}

We obtain
\begin{align}
  \label{eq:66}
\left [ \ms S^{-1} (p) \ms W(p) \ms S(p) \right ]_{n,n} = \left ( \varepsilon_1   \; ; \; -\varepsilon_1  \; ; \; \varepsilon_2  \; ; \; -\varepsilon_2 \right ) ,
\end{align}
and 
\begin{align}
  \label{eq:94}
 & \left [ \ms S^{-1}(p) \partial_{p_\mu}\ms W(p) \ms S(p) \right ]_{n,n} \\
  &\quad = - \tilde \eta p^2 (p_\mu/p^2)  \left ( \varepsilon_1   \; ; \; -\varepsilon_1  \; ; \; \varepsilon_2 \; ; \; -\varepsilon_2 \right )_{n,n} \nonumber ,
\end{align}
which is real so the third term in Eq.\eqref{eq:24} vanishes. There only subsists the second term, which is determined by
\begin{align}
  \label{eq:91}
 & {\rm Im} \left [ \ms S^{-1}(p) \ms W(p) \partial_{p_\mu}\ms S(p) \right ]_{n,n} \nonumber \\
 &\quad = - \tilde \eta p^2 \, {\rm Im} \left [ \ms S^{-1}(p) \begin{bmatrix} \sigma^y &0\\
 0 & \sigma^y \end{bmatrix} \partial_{p_\mu}\ms S(p) \right ]_{n,n}  \nonumber \\
    &\quad = + \tilde \eta p^2 \left ( \epsilon_{\mu,\overline \mu} (p_{\overline \mu}/p^2) +O(\eta^2) \right )  .
\end{align}

\vspace{0.5cm}

 \subsection{Derivation from a microscopic spin model}
\label{sec:concrete-spin-model}

Here we show how to derive $\chi, c_1,c_2$ from the hamiltonian
\begin{align}
  \label{eq:99}
   H_{\rm spin} &= J \sum_{\langle \mb r,\mb r'\rangle}  \bs S_{\mb r} \cdot \bs S_{\mb r'}
+ K \sum_{\langle \mb r,\mb r'\rangle_a} S^a_{\mb r} S^a_{\mb r'} .
\end{align}

Performing the expansion Eqs. \eqref{eq:sigma13}--\eqref{eq:75} and assuming smooth variations of the emergent $\bs m, \bs n$ fields to derive a continuous low-energy theory,
Eq.\eqref{eq:99} becomes, on the square lattice,
\begin{widetext}
  
\begin{align}
  \label{eq:104}
   H_{\rm mag} &= \frac  J 2 \sum_{\mb r } \sum_{\bs \eta}
                \bigg ( 2 m_a(\mb r) m_a(\mb r) + \tfrac 1 2  \eta^\mu \eta^\nu \partial_\mu n_a(\mb r)  \partial_\nu n_a(\mb r) \bigg )
                 + K  \sum_{\mb r } m_a(\mb r) m_a(\mb r) \nn \\
  & \quad - \frac K 2 \sum_{\mb r }  n_a(\mb r) n_a(\mb r)  
          + \frac K 4 {{\mathtt a}}^2  \sum_{\mb r }
                \bigg (  \partial_x n_x(\mb r)  \partial_x n_x(\mb r)   + \partial_y n_y(\mb r)  \partial_y n_y(\mb r)     \bigg )
                + \dots,
\end{align}
where ${\mathtt a}$ is the lattice parameter and ``$   + \dots$'' contains terms that are higher order in gradients or beyond quadratic order in $\bs m, \bs n$.
Since the purpose of this appendix is to derive the magnon velocities, determined by parameters $\chi,c_1,c_2$, and not the magnon gap which will depend sensitively on other (anisotropic) magnetic exchanges not included in Eq.\eqref{eq:99}, we henceforth drop the mass term in Eq.\eqref{eq:104}.

The last term is not rotationally invariant, therefore to recover the symmetry of Eq.\eqref{eq:33} we select its rotation-invariant component.
Technically this is simply done by writing $\bs n = R_\theta \tilde{\bs n}$ and $\bs\partial = R_\theta \tilde{\bs \partial}$, where $R_\theta$ is the 2D rotation matrix by an angle $\theta$
and $\tilde{\bs n}, \tilde{\bs \partial}$ are the expression of ${\bs n}, {\bs \partial}$ in a rotated orthogonal frame, coinciding with $(\hat{\mb x}, \hat{\mb y})$ at $\theta=0$.
We then average over the angle, $\frac 1 {2\pi}\int \text d \theta$. This yields
\begin{align}
  \label{eq:110}
  \partial_x n_x \partial_x n_x+\partial_y n_y \partial_y n_y
  &\rightarrow \tfrac 1 2 (\bs \nabla \bs n)^2 - \tfrac 1 4 \bs n \cdot \nabla^2 \bs n.
\end{align}

One thus obtains
\begin{align}
  \label{eq:106}
  H_{\rm mag} &= \frac  1 2 \sum_{\mb r } ( 8 J + 2K)  m_a(\mb r) m_a(\mb r) 
                - \frac 1 2 {\mathtt a}^2 J\sum_{\mb r }\bigg ( n_a(\mb r)  \partial^2_{xx} n_a(\mb r) +  n_a(\mb r)  \partial^2_{yy} n_a(\mb r) \bigg ) \nn \\
            &\quad  + \frac K 4 {{\mathtt a}}^2  \sum_{\mb r }
                \bigg (  \frac 1 2 (\bs \nabla \bs n)^2 - \frac 1 4 \bs n \cdot \nabla^2 \bs n   \bigg ) + \dots,
\end{align}
allowing for the identification $\chi^{-1}=  8 J + 2K , c_1 = {\mathtt a}^2 (J+K/8), c_2= {\mathtt a}^2 K/4$ as reported in the main text.\\

\end{widetext}

\end{document}